\documentclass[12pt,a4paper]{article} 
\usepackage{amsmath}
\usepackage{a4}
\usepackage[dvips]{graphics}
\usepackage[dvips]{epsfig}
\usepackage{amsfonts} 
\usepackage{amssymb} 

\begin{document}

\begin{titlepage}
\begin{center}
\begin{huge}\textbf{Kaon and Pion production in relativistic
                    heavy-ion collisions%
\footnote{Supported by GSI Darmstadt}
}\end{huge}

\vspace{0.5cm}

{\large
M. Wagner, 
A.B. Larionov\footnote{On leave from RRC "I.V. Kurchatov Institute",
                       123182 Moscow, Russia}
and U. Mosel}

\vspace{0.5cm}

{\large
Institut f\"ur Theoretische Physik, Universit\"at Giessen,
              D-35392 Giessen, Germany }

\vspace{3cm}
\textbf{Abstract}\\
\vspace{0.5cm}
\end{center}

The BUU transport model is applied to study strangeness
and pion production in nucleus-nucleus collisions.
Starting from proton induced reactions, we further investigate Si+Au, 
Au+Au and Pb+Pb collisions in the energy range between 2 and 40 A$\cdot$GeV and compare with data and with other transport calculations. The $q \bar q$ annihilation, or resonance, channel simulated by the string model in meson-nucleon collisions at $\sqrt{s} > 2$ GeV is 
introduced. The importance of this channel for a good description of the 
proton-nucleus data on $K^+$ production is demonstrated. We, furthermore, show that the meson-meson collisions contribute significantly to the $K \overline K$ production in heavy-ion collisions above 5 A GeV and improve an agreement with data on the $K^+/\pi^+$ ratio. Finally, we study the influence of the in-medium modifications of 
the FRITIOF model on the pion and kaon production.

\end{titlepage} 

\section{Introduction}

High-energy heavy-ion collisions offer a unique possibility to study nuclear 
matter at high densities and temperatures under laboratory conditions. The 
maximum compression is expected at the beam energy around 30 A$\cdot$GeV. 
The most intriguing phenomenon which can happen in highly compressed nuclear 
matter is the transition to the quark-gluon plasma. Pioneering work at large 
baryon densities was done at the AGS in  Brookhaven where the energy range up 
to 15 A$\cdot$GeV was explored (see \cite{siau1} and refs. therein).  
The future facility at GSI will provide beams from 2 A$\cdot$GeV up to 40 
A$\cdot$GeV. One believes that an indirect signal for the quark-gluon plasma 
is the strangeness enhancement, which was first suggested by Rafelski and 
M\"uller \cite{mueller}. The enhancement should then be seen in the most 
abundant strange particles, the kaons. At the AGS and the SPS  energies the 
$K^+/\pi^+$ ratio was studied and, indeed, a maximum in the ratio was 
found at about 30 A$\cdot$GeV \cite{na491}.  Theoretical calculations with 
different transport codes --- RQMD \cite{RQMD,na491}, HSD \cite{cass} and UrQMD 
\cite{strange} --- have recently been performed. 
These calculations were not able to reproduce the peak in the ratio, which 
was either at a wrong energy (RQMD, UrQMD) or not present at all (HSD).  
In the case of  HSD and UrQMD this discrepancy  was due to  
overpredicted  pion production, while the kaon yield was well described 
\cite{strange}. With an eye on the planned CBM experiment at GSI it is therefore important to check whether the mentioned difficulties are genuine difficulties in the transport approach or consequences of a particular numerical implementation. 

In the present work we study pion and kaon production at the energies 
2 - 40 A$\cdot$GeV within the BUU model \cite{BUU,MEPhD}; we stress that this is a numerical implementation independent of those employed in \cite{na491,RQMD,strange}. First,  we 
systematically increase the system size and show its effect on particle 
production.  In particular, an analysis of the centrality dependence of 
the pion and kaon production from Au+Au collisions at 10.7 A$\cdot$GeV 
is performed in comparison to the data \cite{hic1}. Then we study the 
$K^+/\pi^+$ ratio in central Au+Au and Pb+Pb collisions as a function of 
the beam energy and compare our results to the experimental data and to 
other models. Special emphasis is put on the strangeness production and we 
will show in detail the most important production mechanisms at different 
energies.

The structure of the paper is as follows. In Section \ref{model} we describe 
our BUU model. In Section \ref{ssd} we study the influence of the system size 
on the pion and kaon production. In Section  \ref{excit} we show the 
excitation functions of  pions, kaons, $\Lambda$ and $\Sigma$ hyperons in 
central heavy-ion collisions.
In Section \ref{medi} we discuss a medium modification of the FRITIOF string 
model and its influence on pion and kaon production. The summary and outlook
are given in Section  \ref{sum}.
\section{\label{model}The BUU model}

Our calculations are based on the BUU model described in Refs. 
\cite{BUU,MEPhD}. In the high-energy range ($\sqrt s > 2$ GeV) we adopt the treatment of Falter et al \cite{thomas}. Thus we will drop details concentrating only on the main ingredients and modifications. 

The model treats a nucleus-nucleus collision explicitly in time as a 
sequence of baryon-baryon, meson-baryon and meson-meson collisions. If not 
specified explicitly, the calculation is always done in the cascade mode, 
i.e. particles propagate freely between the two-body collisions. The 
baryon-baryon collisions at the invariant energy $\sqrt{s} < 2.6$ GeV are 
treated via a resonance scenario, whereas at $\sqrt{s} > 2.6$ GeV a FRITIOF 
string model \cite{fritiof} is applied. In the case of the meson-baryon 
collisions the resonance (FRITIOF) model is used at $\sqrt{s} < (>) 2$ GeV. 
In the most part of calculations we use an energy dependent strangeness 
suppression factor from Ref. \cite{Gei98}:
\begin{equation}
              \gamma \equiv \frac{P(s)}{P(u)}=
\begin{cases}
0.3 \hspace{1cm} \hbox{for} \hspace{1cm} \sqrt{s}\geq 20\hbox{GeV}\\
0.4 \hspace{1cm} \hbox{for} \hspace{1cm} \sqrt{s}\leq 5\hbox{GeV}\\
0.433-\frac{1}{150}\sqrt{s}\hbox{[GeV]}^{-1} \hspace{1cm} \hbox{for} \hspace{1cm} 5\hbox{GeV}<\sqrt{s}<20\hbox{GeV}\\
\end{cases}
\label{eq1}
\end{equation}
Sometimes we will also apply an energy independent strangeness suppression 
factor $\gamma=0.3$ which will be mentioned explicitly in the text.

The most important modifications are the implementation of the strangeness 
production channels in meson-meson collisions and the possibility for a 
baryon and a meson to annihilate into a resonance with the 
invariant mass more than 2 GeV, whose decay is simulated by the string model.
These two modifications will be explained in detail below. In Section 
\ref{excit} we show how the meson-meson collisions and the effective 
resonance channel influence the $K^+/\pi^+$ ratio (see Fig. \ref{fexcit5}).

\subsection{\label{mesm}Meson-meson reactions}

Our BUU model explicitly propagates $\pi,\eta,\rho,\sigma,\omega,\phi,K$ and 
the $K^\ast$-me\-sons. Ch\-ar\-med mesons are also included, but they are not 
relevant for the energies under consideration and thus we will not mention 
them further. At the beam energy up to 2 A GeV production of 
mesons heavier than pions is negligible and the only relevant meson-meson 
channel of 
strangeness production is  $\pi\pi\leftrightarrow K\overline K$, which
was included in the BUU model earlier \cite{MEPhD}. At higher energies
heavier mesons are produced more abundantly and, therefore, one also has to 
take into account the strangeness production in other 
meson-meson collisions, e.g. $\pi\rho \leftrightarrow K \overline K$.
The problem is, however, that the cross sections of these processes are 
not measured experimentally. In Ref. \cite{mesmes1} the cross sections of
the processes $\pi\pi\rightarrow K\overline K$, 
$\pi\rho\rightarrow K\overline K$ and $\rho\rho\rightarrow K\overline K$ were 
calculated. We use a parametrisation of $\pi\pi\rightarrow K\overline K$ 
from Ref. \cite{mesmes2}, which is based on these calculations:
\begin{equation}
\sigma_{\pi\pi\rightarrow K\overline K} = C\, 6.075\, 
\left(1-\frac{(2m_K)^2}{s}\right)^{0.76}~~~ (\hbox{mb}),
\end{equation}
where the factor $C$ is the combination of  the Clebsch-Gordan coefficients 
for the respective isospin channels:
\begin{equation}
 C=\sum_{I=0,1} \mid\langle i_1i_2m_1m_2\mid i_1i_2IM\rangle\mid^2 \mid\langle
 i_3i_4m_3m_4\mid
 i_3i_4IM\rangle\mid^2~,
\end{equation} 
where $i_k$ and $m_k$ are the total isospin and the third  isospin component 
of the particle $k$. Incoming and outgoing particles are enumerated by the 
pair of indices 1,2 and 3,4 respectively. For simplicity we take the same 
cross section for $\rho\rho\rightarrow K\overline K$ and 
$\pi\rho\rightarrow K\overline K$ since the isospins of the incoming particles
are the same; this  is not exactly the result from Ref. \cite{mesmes1}. 
For all other reactions with two nonstrange mesons in the incoming channel 
we assume a constant value of 2 mb for the cross section. The back reactions
are included and their cross sections are calculated according to detailed 
balance.

By setting the cross sections constant we did not take into account the p-wave suppression of e.g. $\pi \rho \rightarrow K\overline K$ near threshold. Due to the spin of the $\rho$-meson the strangeness production in that reaction is suppressed up to the threshold of $\pi \rho \rightarrow \overline K K^\ast$ and $\pi \rho \rightarrow \overline K^\ast K$. Including this suppression, however, has only a small effect on our results (see discussions in section 4 and Fig.\ref{fexcit6}). 

Since the elementary reactions are not accessible experimentally, we test our 
choice of the cross sections by looking on the yields of kaons in heavy-ion collisions. Although this will give us only a rough estimate of our cross sections due to presence of the baryon-baryon and the meson-baryon channels of the kaon production, there is no other 
way to get more reliable cross sections for meson-meson reactions.

\subsection{\label{annproc}Annihilation processes}

\begin{figure}
\begin{center}
\begin{tabular}{c}
\includegraphics[height=7cm]{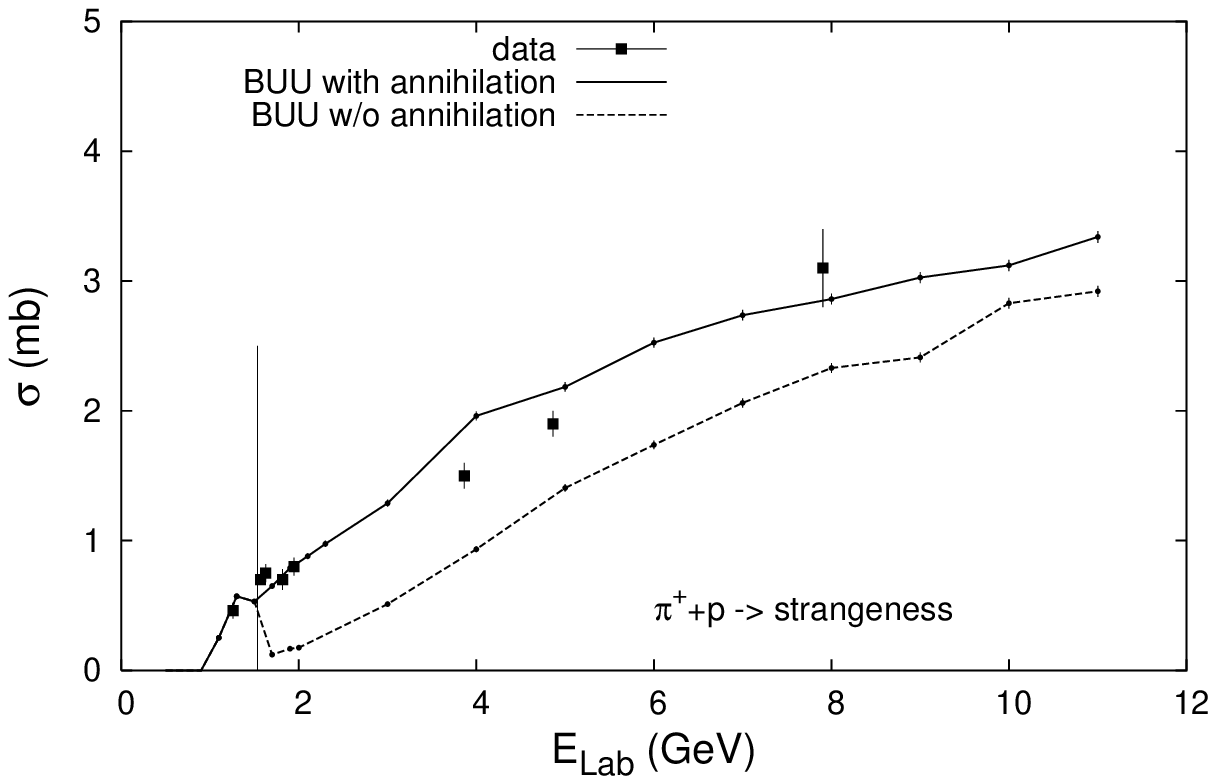}\\ 
\includegraphics[height=7cm]{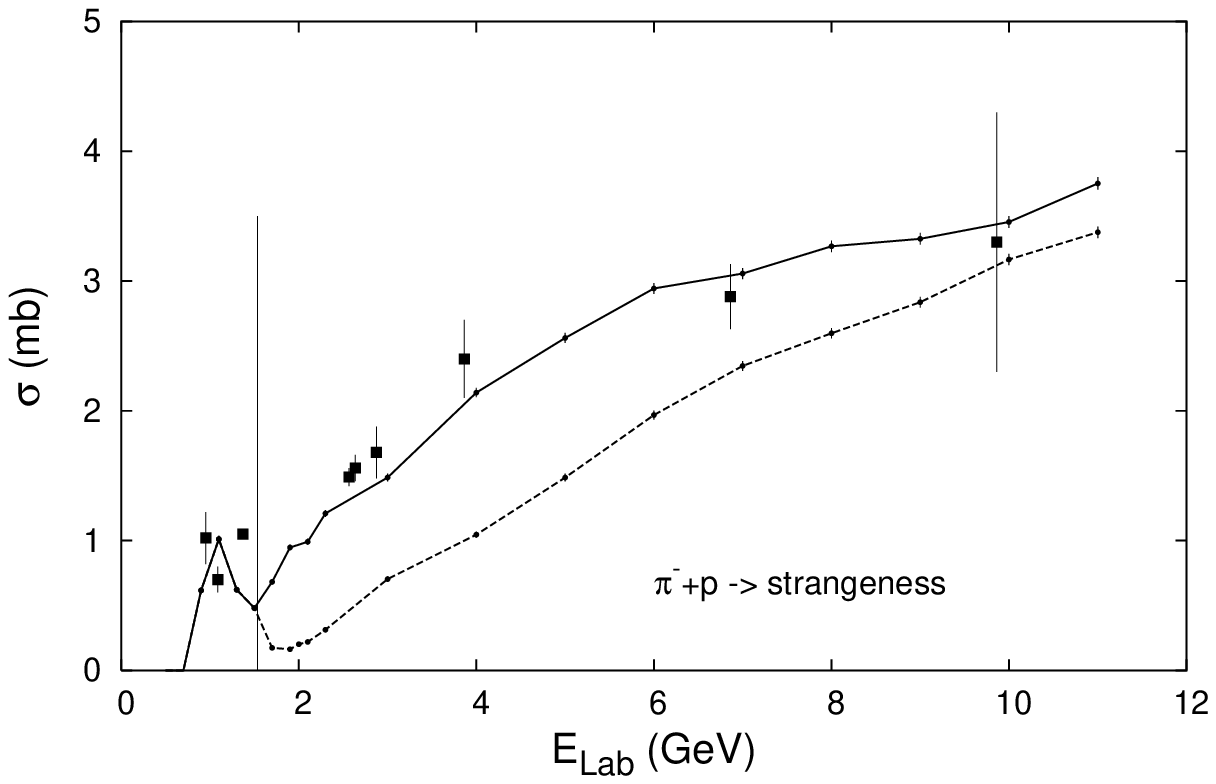}\\ 
\end{tabular}
\caption{The cross section of the strangeness production in $\pi^+ p$ collisions (upper panel) and $\pi^-p$ collisions (lower panel) as a function of beam energy in comparison to data from \cite{land}. The vertical line corresponds to the threshold for the string model ($\sqrt s =2$ GeV). Solid and dashed lines show results without and with $q\overline q$ annihilation channels (see subsection \ref{annproc}). Errorbars on calculations are statistical.} 
\label{anni1}
\end{center}
\end{figure}

Fig. \ref{anni1} (see dashed lines) shows
\footnote{ In Figs.~\ref{anni1},\ref{pberap},\ref{paurap},\ref{siaupic1}
the statistical errorbars are shown for the theoretical curves.
They are calculated assuming a Poisson distribution, i.e. by dividing
the plotted physical value by $\sqrt{N}$, where $N$ is the total
accumulated number of events which is used to construct the value.
In all other figures the statistical errors of theoretical results are either
negligibly small or visible from the histogram representation.} 
that we underestimate the strangeness production in the region just above 
the FRITIOF threshold ($\sqrt{s} > 2$ GeV). We see that directly above the 
threshold the cross section without the annihilation descends to almost zero.

\begin{figure}
\begin{center}
\epsfig{file=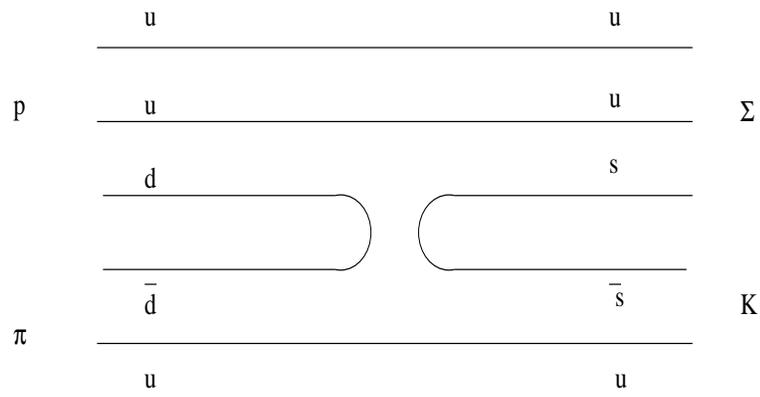,width=10cm,height=5cm}\caption{Quark Diagram for 
the process $\pi^+p \rightarrow \Sigma^+ K^+$.}
\label{quarkdia}
\end{center}
\end{figure}

This is due to the fact that the FRITIOF model is only capable to produce two excited hadrons, which fragment separately. Thus it is not possible to describe Drell-Yan like processes in which, e.g., a quark from an incoming baryon and an antiquark from an incoming meson annihilate (c.f. Fig. \ref{quarkdia}). An example for such a process is 
$\pi N \rightarrow Y K$, where $Y$ stands for a hyperon.

For that reason we have included the annihilation channel phenomenologically. In the case 
of a reaction of a baryon with a meson we check if an annihilation between a 
quark and an antiquark is possible; we split each of the interacting hadrons into 
their constituents and check whether a quark and antiquark with the same flavour
exist. If they exist we annihilate the quark and antiquark with probability (\ref{anniprob}) and (\ref{anniprobabel}), neglecting any particles that might be created in this process. In order to make up for this neglect we put all the energy and momentum of both incoming hadrons into the remaining quark content of the baryon and the meson. The fragmentation of this hadron is then done according to the Lund Model. UrQMD \cite{HWPhD} interprets meson-baryon reactions in a similar way. RQMD \cite{Matt89} also includes baryon resonances with mass $> 2$ GeV whose decay is described by the Lund model. 

The probability for the annihilation is chosen such that we agree with experiment for the 
strangeness production in $\pi p \rightarrow$ strange particles (see solid
lines in Fig. \ref{anni1}):
\begin{equation}\label{anniprob}
\hbox{Prob(annihilation)} = 
              \hbox{max}(0.85-0.17\cdot\frac{\sqrt{s}}{\text{GeV}},0).
\end{equation}
For the constant strangeness suppression factor $\gamma=0.3$ the probability 
for the annihilation processes is readjusted:  
\begin{equation}\label{anniprobabel}
\hbox{Prob(annihilation)}_{\gamma=0.3} = 
              \hbox{max}(1.2-0.2\cdot\frac{\sqrt{s}}{\text{GeV}},0).
\end{equation}
There are two main reasons for the increase of the strange particle production
by including the annihilation. First, we include new channels, as 
discussed above. Another point is that the invariant energy per string decay 
is higher. If we have two strings instead of one, the two strings decay 
separately and therefore it will often occur that every string alone is below 
the threshold for strangeness production. By 
putting all the energy into one string, the invariant energy becomes higher 
and the production of strangeness more probable.

\section{\label{ssd}System-size dependence}

In order to clarify the reaction mechanisms in heavy-ion collisions, it is 
instructive, first, to understand the proton and light ion induced reactions.
For larger mass numbers of colliding nuclei, the effect of secondary 
hadron-hadron collisions becomes more and more important, which drives the 
system towards thermal equilibrium and enhances the maximum baryon density
reached in the collision process. Thus by increasing the system size  we can 
also see how particle spectra evolve with increasing density and if our model 
within the standard parameters is able to reproduce the experimental 
measurements. In the discussion of numerical results it is assumed, if the
opposite is not stated explicitly, that the meson-meson cross sections and
the $q \bar q$ annihilation are included as described in the previous section.

\begin{figure}
\begin{center}
\begin{tabular}{cc}
\includegraphics[height=6.5cm]{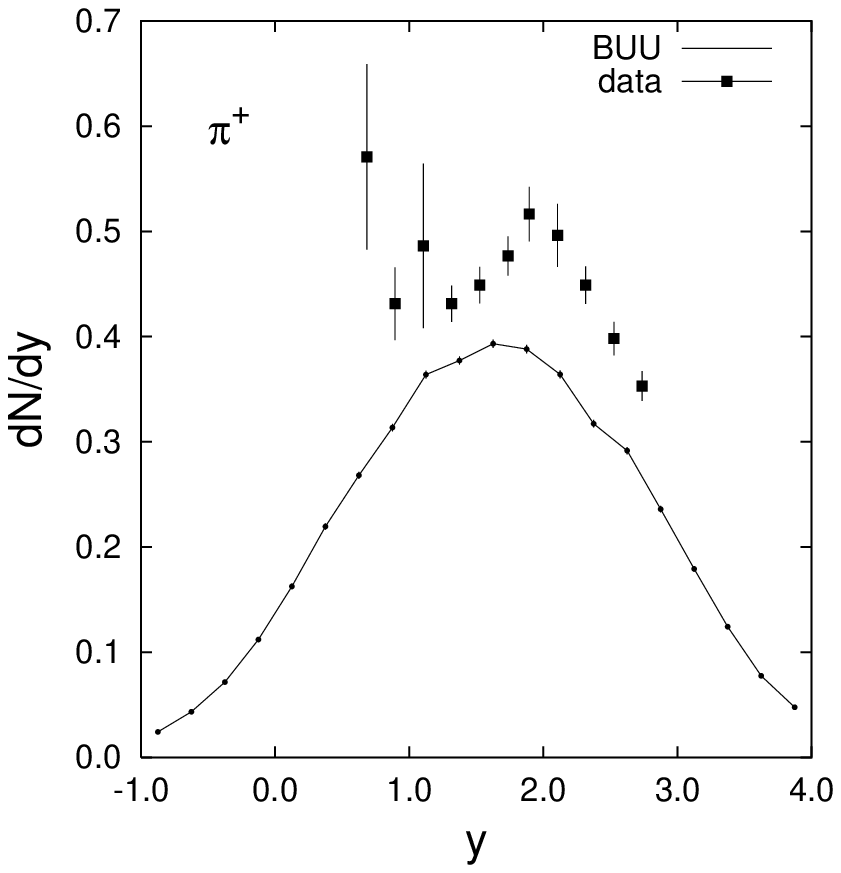} & 
\includegraphics[height=6.5cm]{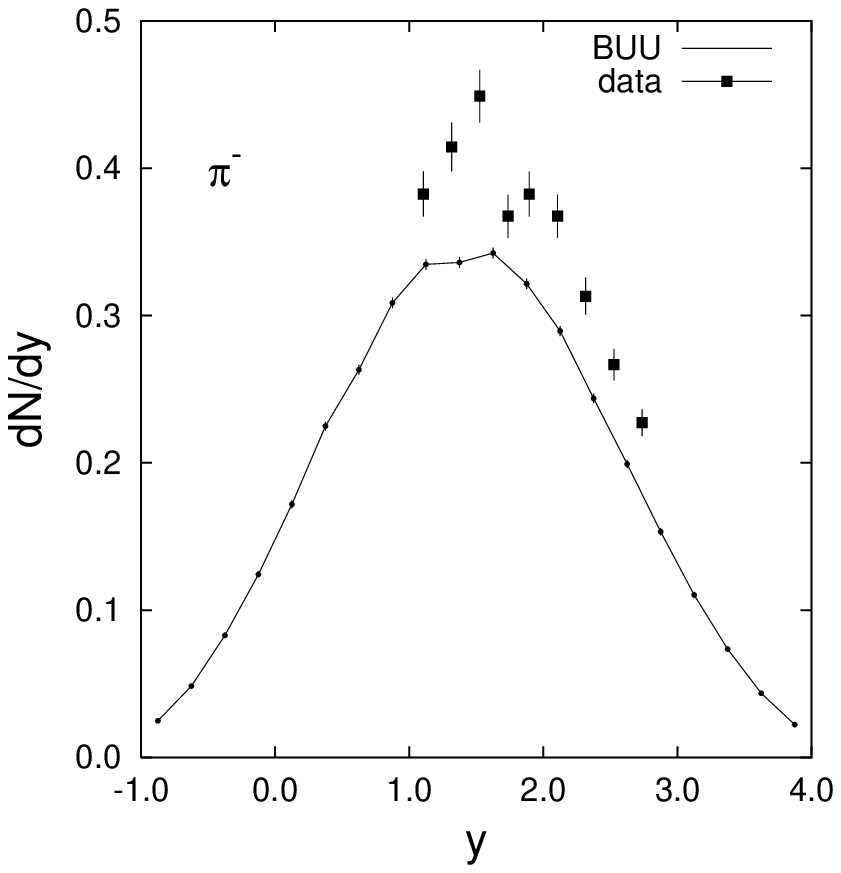} \\ 
\includegraphics[height=6.5cm]{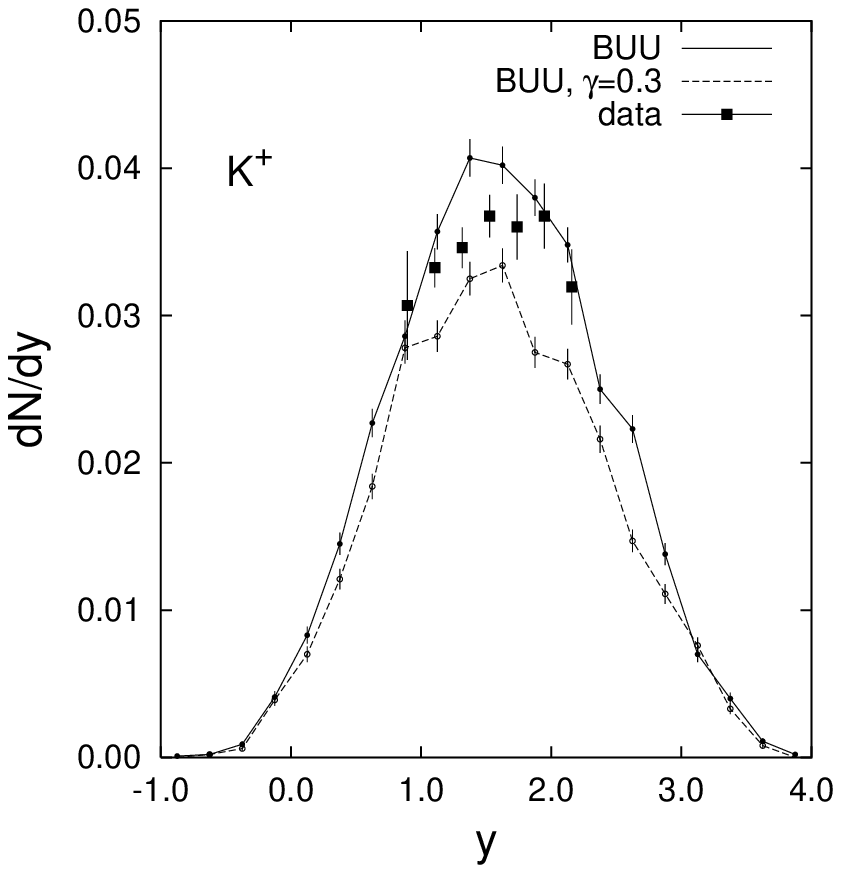} & 
\includegraphics[height=6.5cm]{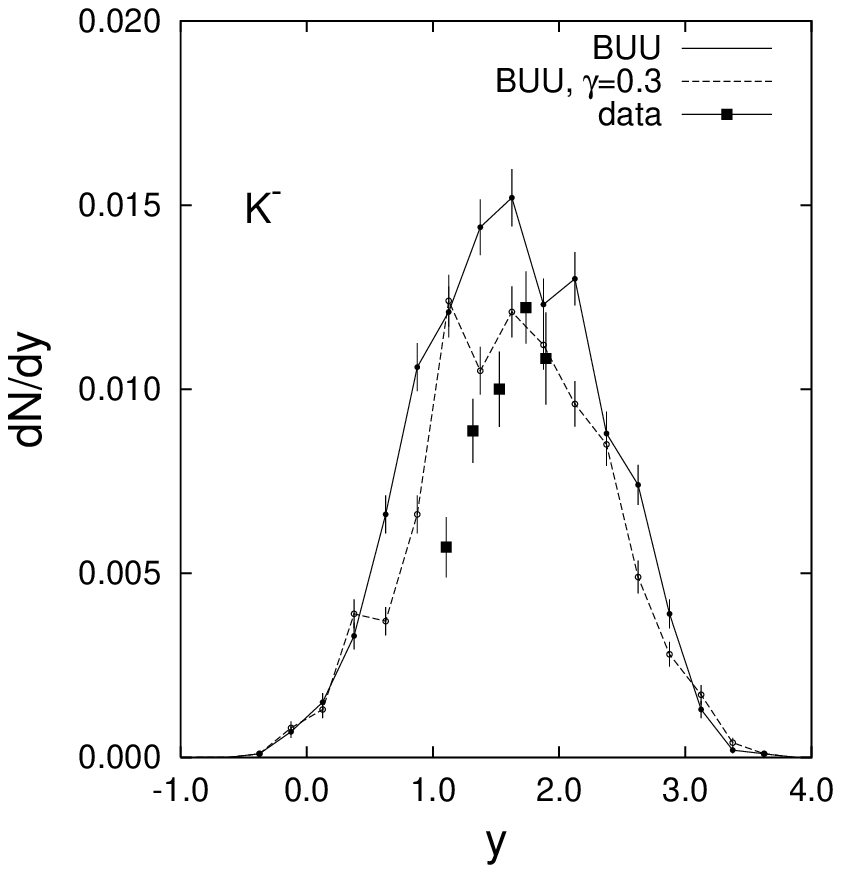} \\ 
\end{tabular} 
\caption{Rapidity distributions of $\pi^\pm$ and $K^\pm$ for $p$+Be at 14.6 GeV/c in comparison to data from \cite{pa1}. Solid and dashed lines show results with energy  dependent (see Eq.(\ref{eq1})) and constant ($\gamma=0.3$) strangeness suppression factor, respectively. Errorbars on calculations are statistical.}\label{pberap}
\end{center}
\end{figure}

First, we study the proton-induced reactions p+Be and p+Au at the beam
momentum of 14.6 GeV/c measured at BNL-AGS \cite{pa1}. Fig. \ref{pberap}
shows rapidity distributions of produced $\pi^\pm$ and $K^\pm$ for the
p+Be system. In this case of light target, the pions and kaons are produced 
mostly in the first-chance nucleon-nucleon (NN) collisions and have only a small
probability to rescatter afterwards. Therefore, their rapidity distributions
are centered near the NN center-of-mass (c.m.) rapidity $y_{NN}=y_{beam}/2 = 
1.72$. The pion yield is underestimated by $\sim 20\%$,
whereas the kaon and antikaon yields are well described by BUU for
the p+Be system. 

\begin{figure}
\begin{center}
\begin{tabular}{cc}
\includegraphics[height=5cm]{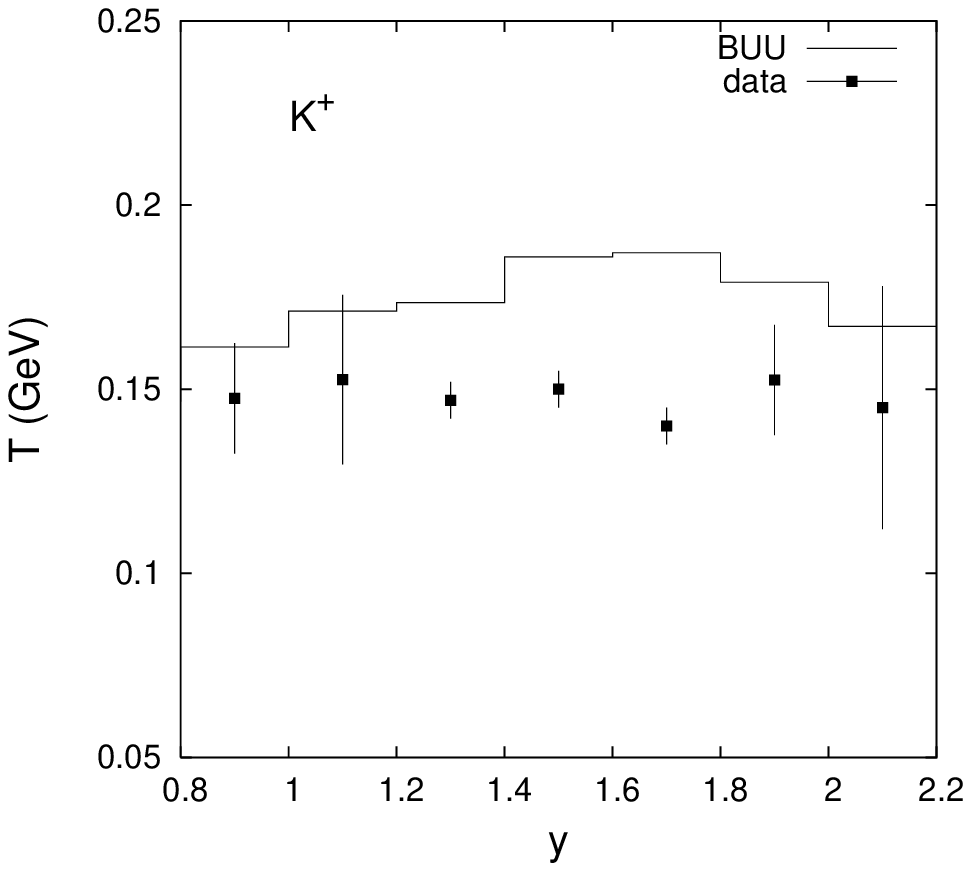} & 
\includegraphics[height=5cm]{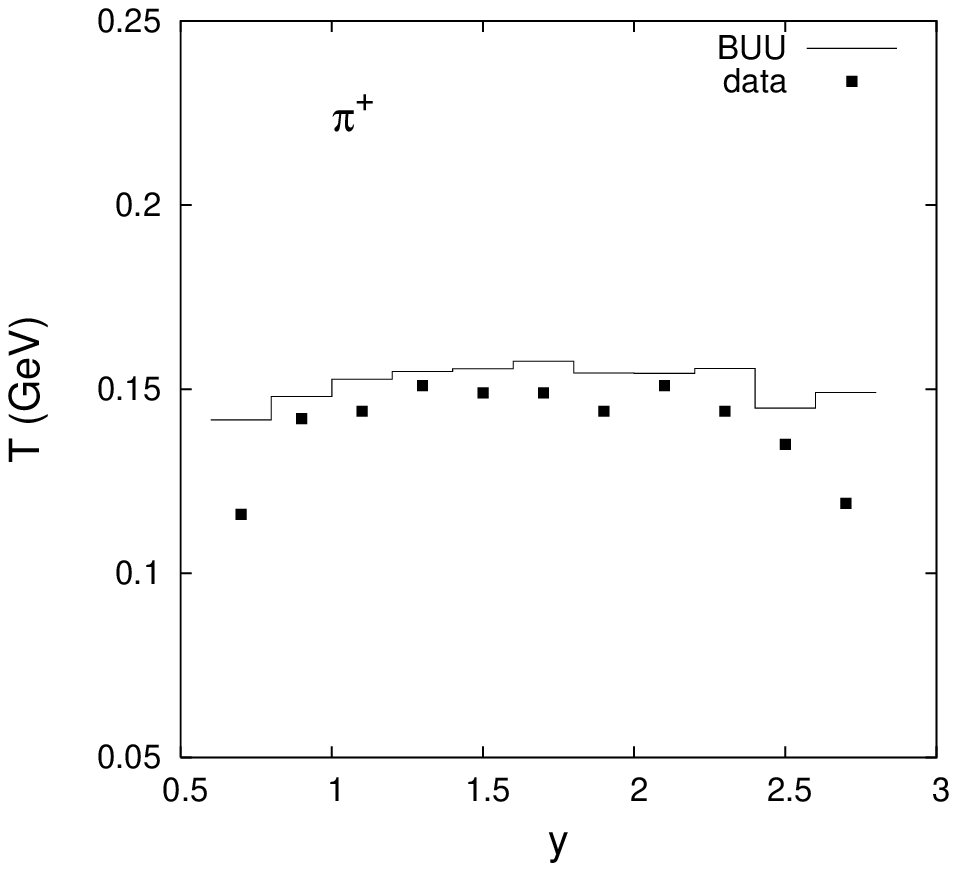} \\ 
\end{tabular} 
\caption{Rapidity dependence of the inverse slope parameter $T$ of the 
transverse mass spectra of $\pi^+$ and 
$K^+$ for $p+$Be at 14.6 GeV/c in comparison to data from \cite{pa1}.}
\label{slopespbe}
\end{center}
\end{figure}

According to Ref. \cite{pa1}, we have fitted the calculated
transverse mass spectra with an exponential function
\begin{equation}
   {d^2 \sigma \over 2 \pi m_\perp d m_\perp d y} 
 = a \exp\{-m_\perp/T\}                                 \label{expfit}
\end{equation}
at various rapidities $y$. Fig. \ref{slopespbe} presents the inverse slope
parameter $T$ as a function of rapidity for $\pi^+$ and $K^+$ in the case
of p+Be collisions. The pion inverse slope parameter is well reproduced by 
BUU except for the very forward and the very backward rapidities in the NN 
c.m. system. The calculated kaon inverse slope parameter overestimates 
the data by $\sim 20\%$ at midrapidity.   

\begin{figure}
\begin{center}
\begin{tabular}{cc}
\includegraphics[height=5cm]{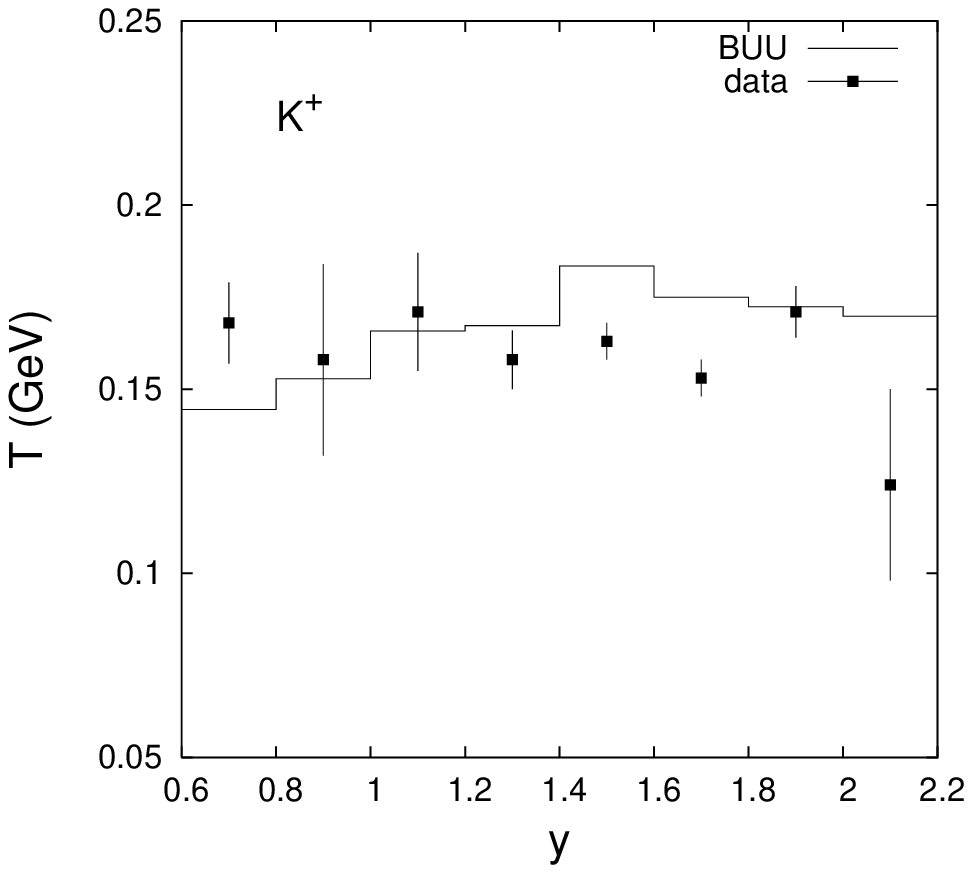} & 
\includegraphics[height=5cm]{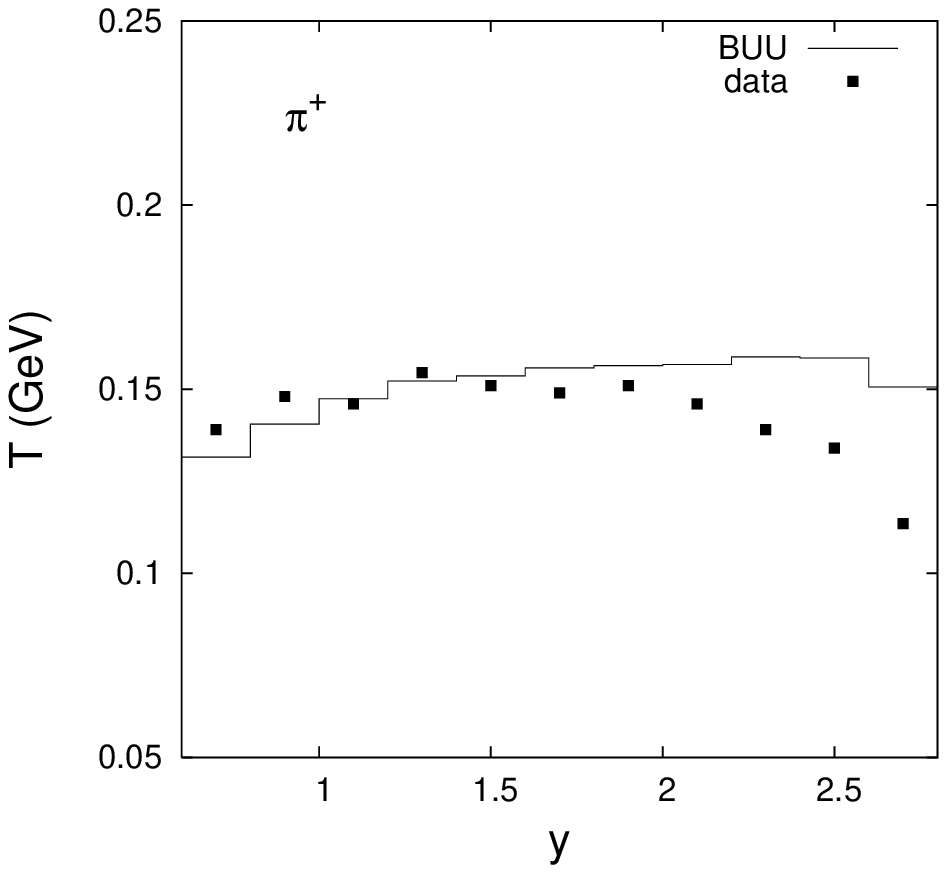}\\ 
\end{tabular} 
\caption{Rapidity dependence of the inverse slope parameter $T$ of the 
transverse mass spectra of $\pi^+$ and 
$K^+$ for $p+$Au at 14.6 GeV/c in comparison to data from \cite{pa1}.}
\label{slopespau}
\end{center}
\end{figure}

Fig. \ref{paurap} shows the rapidity distributions of $\pi^\pm$ and
$K^\pm$ for the p+Au collisions. These distributions are shifted 
to smaller rapidities $y < y_{NN}$ with respect to the case of p+Be
reaction (Fig. \ref{pberap}) due to contribution of the secondary 
NN and $\pi$N collisions to the meson production and rescattering of the 
produced mesons on the target nucleons. The $K^-$ rapidity distribution 
is narrower and is shifted somewhat less than the $K^+$ distribution, 
since an antikaon is always produced 
together with a kaon, while a kaon can be also produced in association with 
a hyperon which requires less c.m. energy \cite{pa1}. Thus,  
the secondary NN and $\pi$N collisions contribute more to the $K^+$ 
than to the $K^-$ production. BUU describes the experimental pion 
and kaon rapidity distributions within $\sim 20\%$. In Fig. \ref{slopespau}
we present the rapidity dependence of the inverse slope parameter $T$ of 
the $K^+$ and $\pi^+$ transverse mass spectra for the p+Au reaction. 
There is a good agreement between BUU and the data except for the very
forward rapidity in the $\pi^+$ case where we overpredict the experiment 
by $\sim 25 \%$.

In agreement with the data, we observe little change in the 
value ($\sim 150$ MeV) of the inverse slope parameter for $\pi^+$ and $K^+$ 
with increasing target mass (c.f. Figs. \ref{slopespbe} and \ref{slopespau}).
The $K^+$ yield at $y=y_{NN}$ shows a factor of two enhancement, while the 
$K^-$ yield at $y=y_{NN}$ stays practically unchanged both in BUU and in the 
data (c.f. Figs. \ref{pberap} and \ref{paurap}). We attribute this behaviour to the stronger absorption of $K^-$ in the heavier target.
The experimental pion yield at $y=y_{NN}$ is the same for both systems,
whereas in BUU we observe a slight enhancement of the pion yield at $y=y_{NN}$
with increasing target mass.

Next, we present results for Si+Au collisions at the beam momentum 
of 14.6 A GeV/c, which were studied experimentally in Ref. \cite{siau1}.
This reaction  has earlier been studied theoretically in 
Refs. \cite{Matt89,Pang92}. In Ref. \cite{Matt89} the RQMD model has been
employed, which includes known nonstrange baryon resonances with 
$m < 2$ GeV. RQMD also describes the production of baryon resonances with 
$m > 2$ GeV in high-energy collisions; their decay is simulated by a string 
model. Shapes of
the $\pi^\pm$ and $K^\pm$ transverse momentum spectra are quite well
described by RQMD. No conclusions on the agreement of the absolute yields
of the produced particles with data have been drawn in \cite{Matt89} due
to the absolute normalization on the experimental $\pi^+$ spectra.
However, ratios $K^+/\pi^+$ and $K^-/\pi^-$ computed within RQMD 
agree well with data. In Ref. \cite{Pang92} a relativistic hadronic cascade
(ARC) model has been used; a pure hadronic scenario without string excitation 
was assumed. A good agreement between ARC and E-802 data
on proton $m_\perp$ spectra, proton, $\pi^+$ and $K^+$ rapidity distributions
was reached within the resonance model, i.e. when, e.g. a three-pion
production channel in a NN collision is simulated as
$N N \to \Delta \Delta \pi$ rather than directly as $N N \to N N \pi \pi \pi$.
Inverse slope parameters of the $m_\perp$ spectra for protons and pions
are well described by ARC, but underestimated by $\sim 20 \%$ for kaons.

\begin{figure}
\begin{center}
\begin{tabular}{cc}
\includegraphics[height=6.5cm]{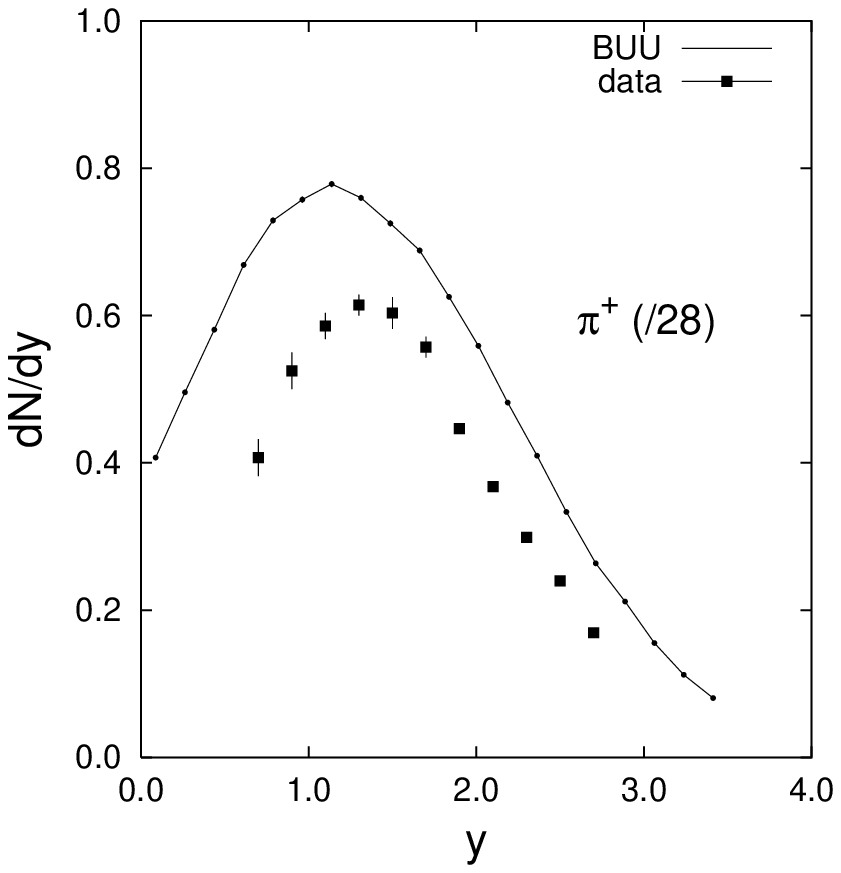} & 
\includegraphics[height=6.5cm]{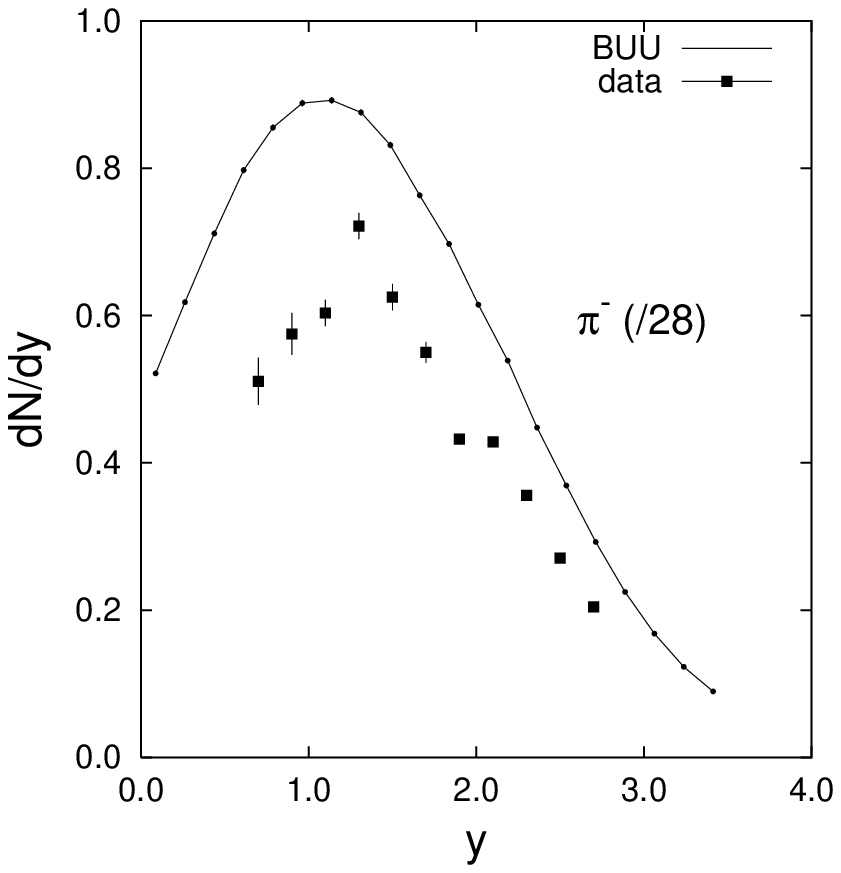} \\ 
\includegraphics[height=6.5cm]{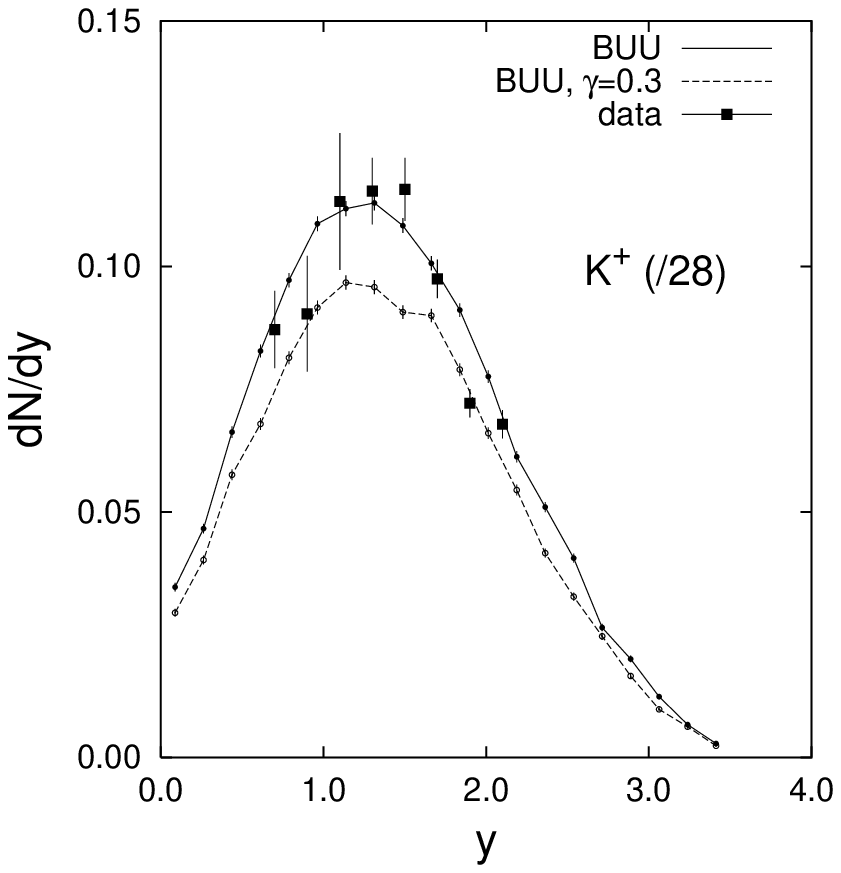} & 
\includegraphics[height=6.5cm]{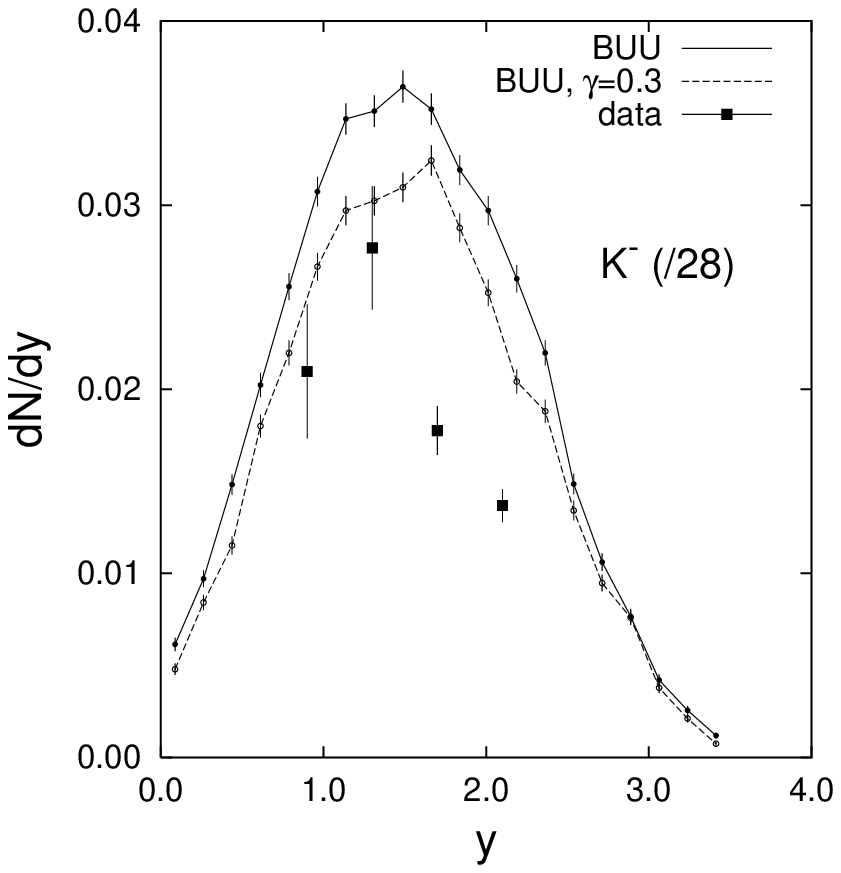} \\ 
\end{tabular} 
\caption{Rapidity distributions of $\pi^\pm$ and $K^\pm$ for central collisions Si+Au at 14.6 A GeV/c in comparison to data from \cite{siau1}. Solid and dashed lines show results with energy  dependent (see Eq.(\ref{eq1})) and constant ($\gamma=0.3$) strangeness suppression factor, respectively. The spectra are divided by 28 in order to be able to compare to proton induced reactions. Errorbars on calculations are statistical.}
\label{siaupic1}
\end{center}
\end{figure}

We have considered only central collisions Si+Au corresponding to 7$\%$
of the inelastic cross section selected on multiplicity of charged
particles \cite{siau1}. In the theoretical calculations we selected
the central collisions in the same way.
Fig. \ref{siaupic1} shows the calculated $\pi^\pm$
and $K^\pm$ rapidity distributions which were divided by the projectile mass (28) in order to be able to directly compare them with the rapidity distributions from
the proton induced reactions.

By comparing the data points in Figs. \ref{siaupic1} and \ref{paurap}
we see that the pion yields at $y=y_{NN}$ are, practically, the same
in p+Au and Si+Au systems. This feature is not reproduced by BUU:
there is an enhancement of the pion yield per projectile nucleon
in the system Si+Au with respect to the p+Au system in our calculations. This may indicate a problem with pion production (or reabsorption) in the heavy system.
The experimental $K^\pm$ yields per nucleon are higher in the Si+Au case 
than in the p+Au case, which is well reproduced by BUU.

\begin{figure}
\begin{center}
\begin{tabular}{cc}
\includegraphics[height=5cm]{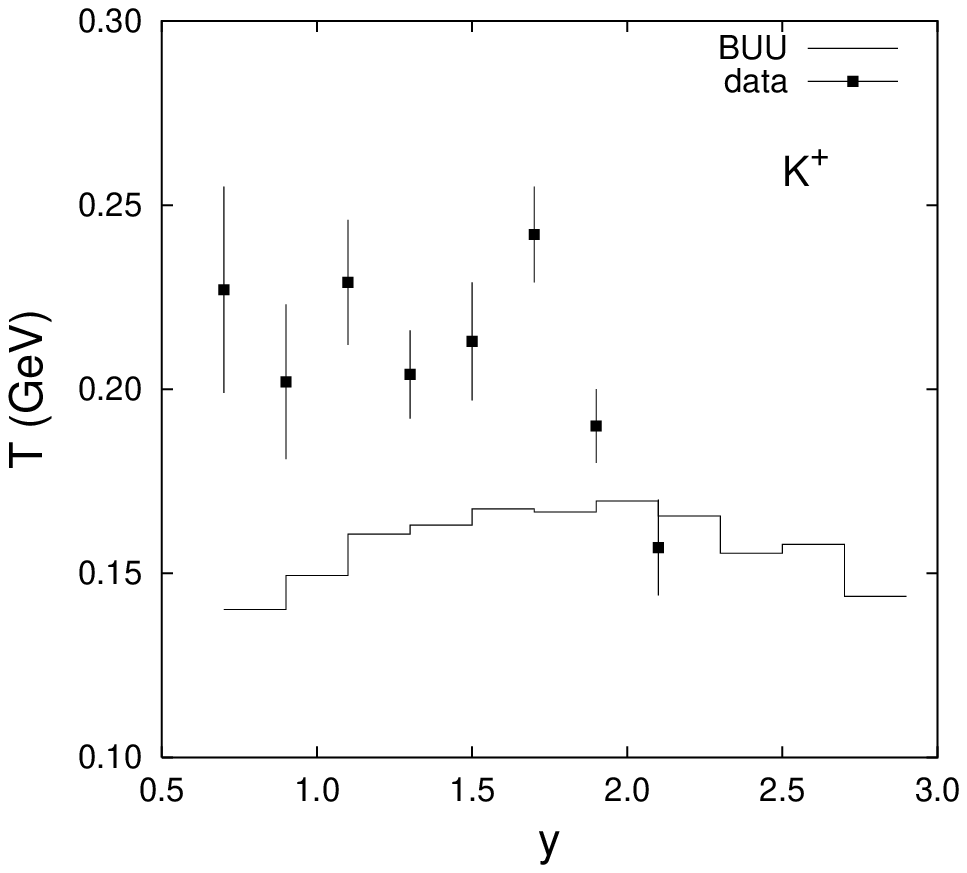} & 
\includegraphics[height=5cm]{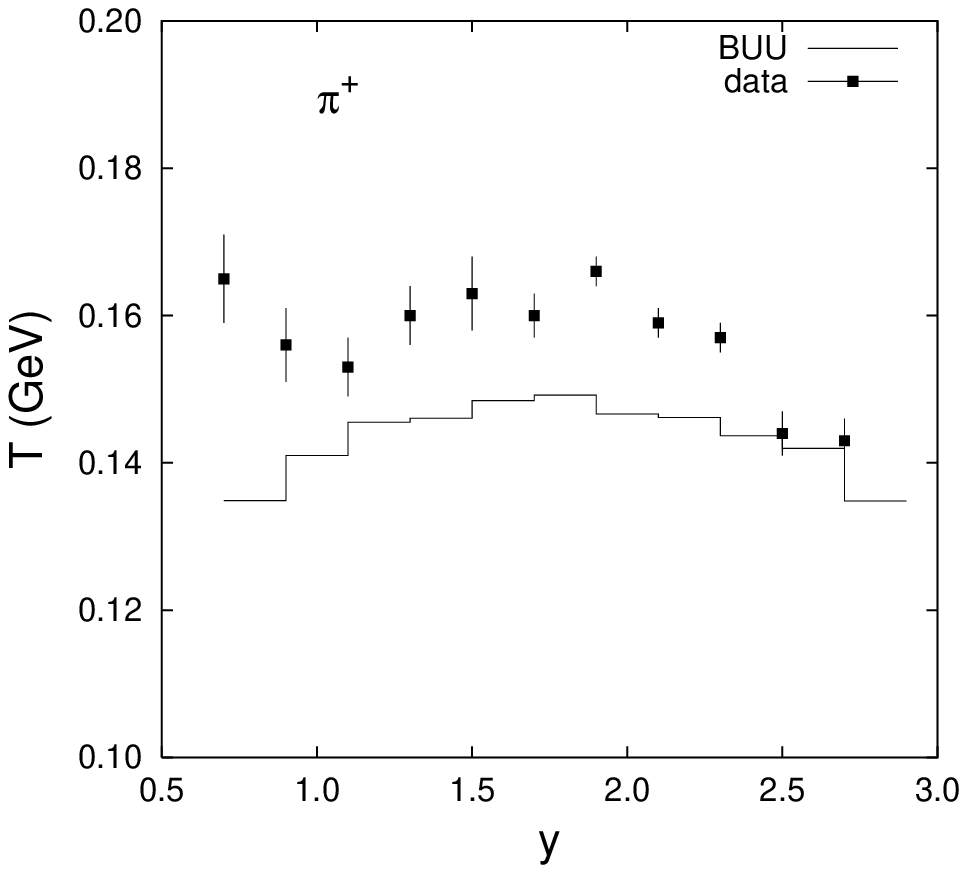} \\ 
\end{tabular} 
\caption{Rapidity dependence of the inverse slope parameter $T$ of the 
transverse mass spectra of $\pi^+$ and 
$K^+$ for Si+Au at 14.6 A GeV/c in comparison to data from \cite{siau1}.}\label{slopessiau}
\end{center}
\end{figure}

In Fig. \ref{slopessiau} we present the inverse slope parameters of
the $K^+$ and $\pi^+$ transverse mass spectra. Despite of the big
errorbars plus systematic errors of $\pm 10$ \% which are not included
into the errorbars of the experimental data \cite{siau1}, we see that
BUU underpredicts the inverse slope parameter for $K^+$'s by about 25 \%
and for $\pi^+$'s by 15 \%. The calculated inverse slope parameter stays,
practically, constant $T \simeq 150-160$ MeV for all three systems
p+Be, p+Au and Si+Au both for $\pi^+$'s and $K^+$'s, whereas the
experimental data show higher $T \simeq 200$ MeV for $K^+$'s in the
Si+Au system.

Studying strangeness production in more detail, we have also performed
the BUU calculations using the constant energy-independent strangeness
suppression factor $\gamma=0.3$ (dashed lines in Figs. \ref{pberap},
\ref{paurap}, \ref{siaupic1}). The $K^+$ rapidity distributions favour
the energy-dependent strangeness suppression factor, while the $K^-$ spectra
are better described with $\gamma=0.3$.

In order to demonstrate an effect of the $q \bar q$ annihilation on
the $K^+$ production (see discussion in the previous section) we also show
in Fig. \ref{paurap} the results without the annihilation. In the p+Au
system the secondary $\pi N$ and $\rho N$ collisions play already an 
important role. Thus, including the annihilation improves
an agreement with the data (c.f. solid and dotted lines in
the lower left panel of Fig. \ref{paurap}).

The heaviest colliding system measured at AGS is Au+Au at the beam 
energies of 2-10.7 A GeV \cite{hic1,ags1,ags2}.
Before discussing the beam energy dependence (see next section), we
will consider the centrality dependence of the pion and kaon production
for Au+Au collisions at the top AGS energy of 10.7 A GeV \cite{hic1}.

In Ref. \cite{hic1} the collision centrality was determined by using
two criteria: (i) The energy deposited in the zero-degree calorimeter
$E_{ZCAL}$, which gives an estimate of the projectile participant number
$N_{pp}$ as
\begin{equation}
    N_{pp} = 197 \times 
    \left( 1 - {E_{ZCAL} \over E_{beam}^{kin}} \right)~,     \label{Npp}
\end{equation}
where $E_{beam}^{kin} = 2123$ GeV is the kinetic energy of the beam.
The smaller $E_{ZCAL}$ is, the larger is the size of the participant
zone, which selects geometrically more central events.
(ii) The multiplicity of particles with velocity $\beta > 0.8$ in 
the New Multiplicity Array (NMA) $mult_{NMA}$. The velocity cut
filters out the slow protons, whereas the produced mesons (mostly pions)
are accepted. Thus, the larger $mult_{NMA}$ corresponds to the larger energy
transfer from the longitudinal motion of colliding nuclei to the
meson production. In average, events with smaller impact parameter $b$ have
larger $mult_{NMA}$. However, at fixed $b$ the multiplicity $mult_{NMA}$
fluctuates stochastically event-by-event depending on the amount of stopping 
of the counterstreaming nuclear matter in the interaction zone.

According to Ref. \cite{hic1} we, first, divided BUU events using $E_{ZCAL}$
into eight classes with increasing  $E_{ZCAL}$ (decreasing centrality) from 
the first to the eighth class (see Table II in \cite{hic1}). Second,
by modelling the NMA acceptance, we subdivided each of the first three
$E_{ZCAL}$ event classes to the three $mult_{NMA}$ classes with decreasing
$mult_{NMA}$ (decreasing centrality) from the first to the third 
multiplicity class (see Table III in \cite{hic1}). 

Fig. \ref{auauprot} shows the proton rapidity distributions for 
various combinations of $E_{ZCAL}$- and $mult_{NMA}$- event classes.
We see that in each case BUU overestimates stopping. Agreement with experiment
can be improved by taking into account the in-medium corrections
to the FRITIOF model (see Sect V).

\begin{figure}
\begin{center}
\begin{tabular}{cc}
\includegraphics[height=4.5cm]{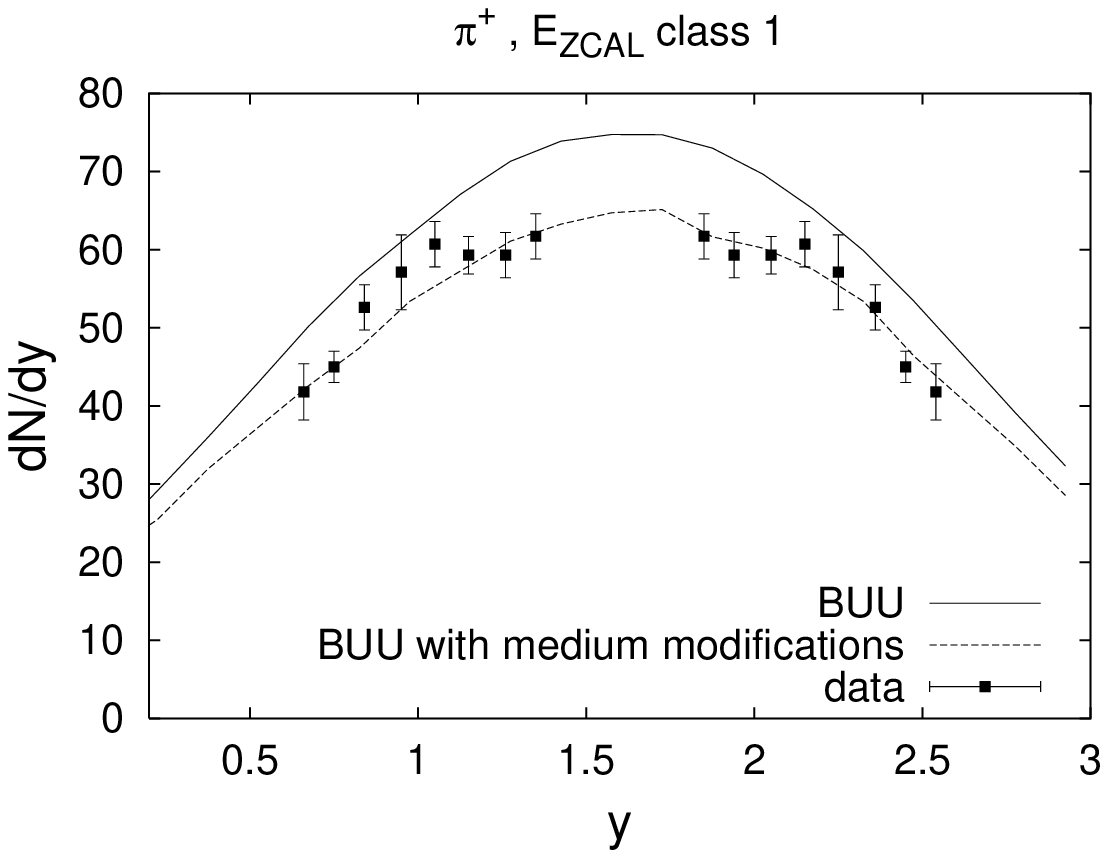} & 
\includegraphics[height=4.5cm]{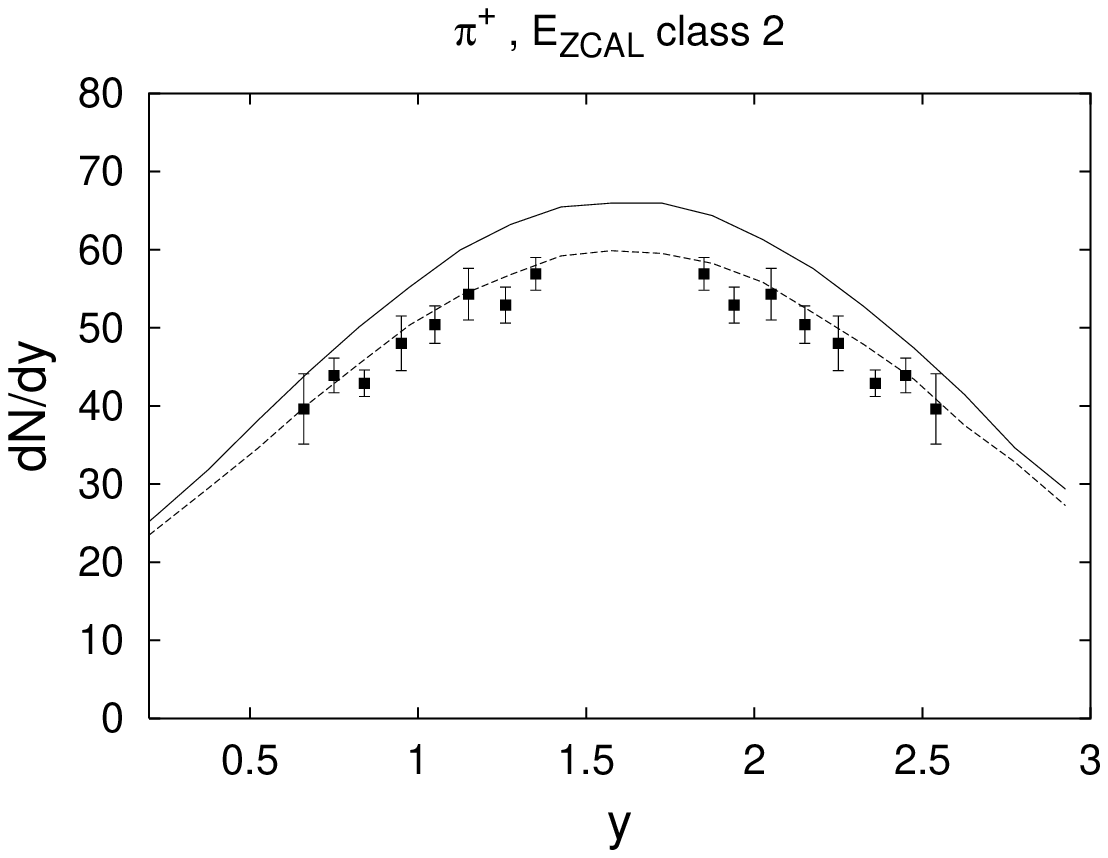} \\ 
\includegraphics[height=4.5cm]{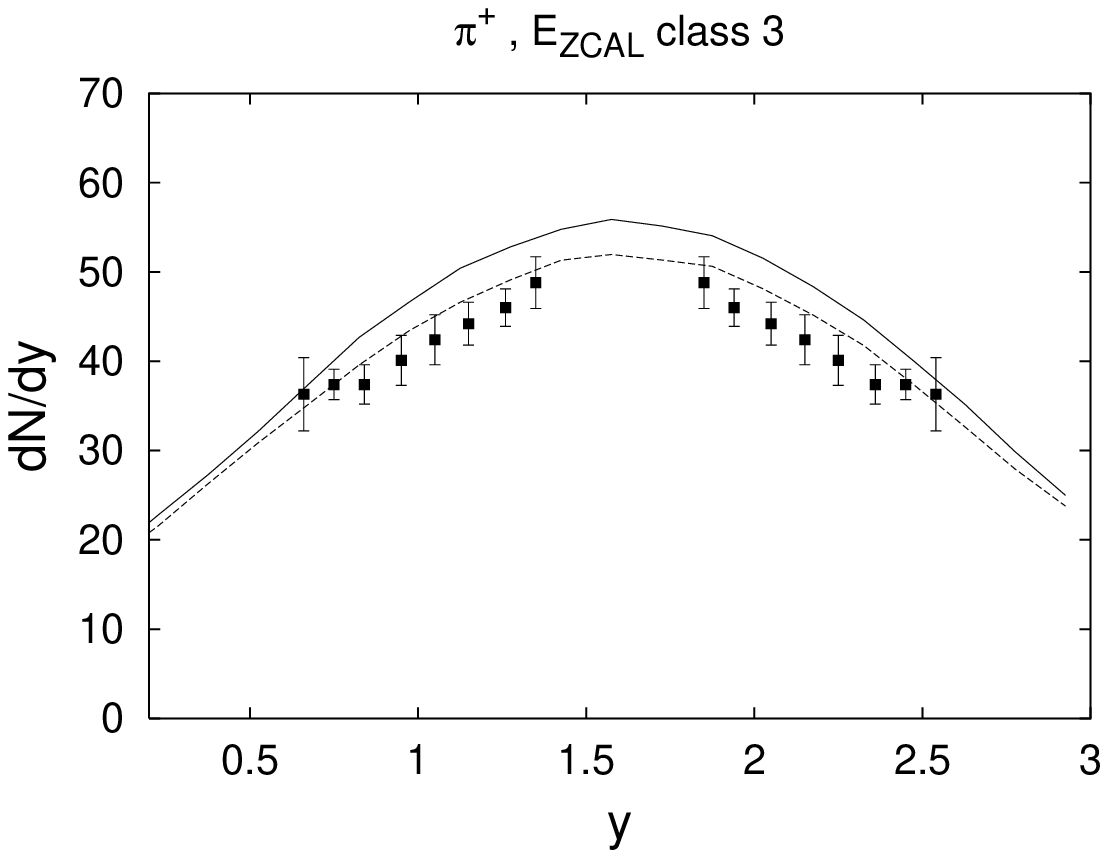} & 
\includegraphics[height=4.5cm]{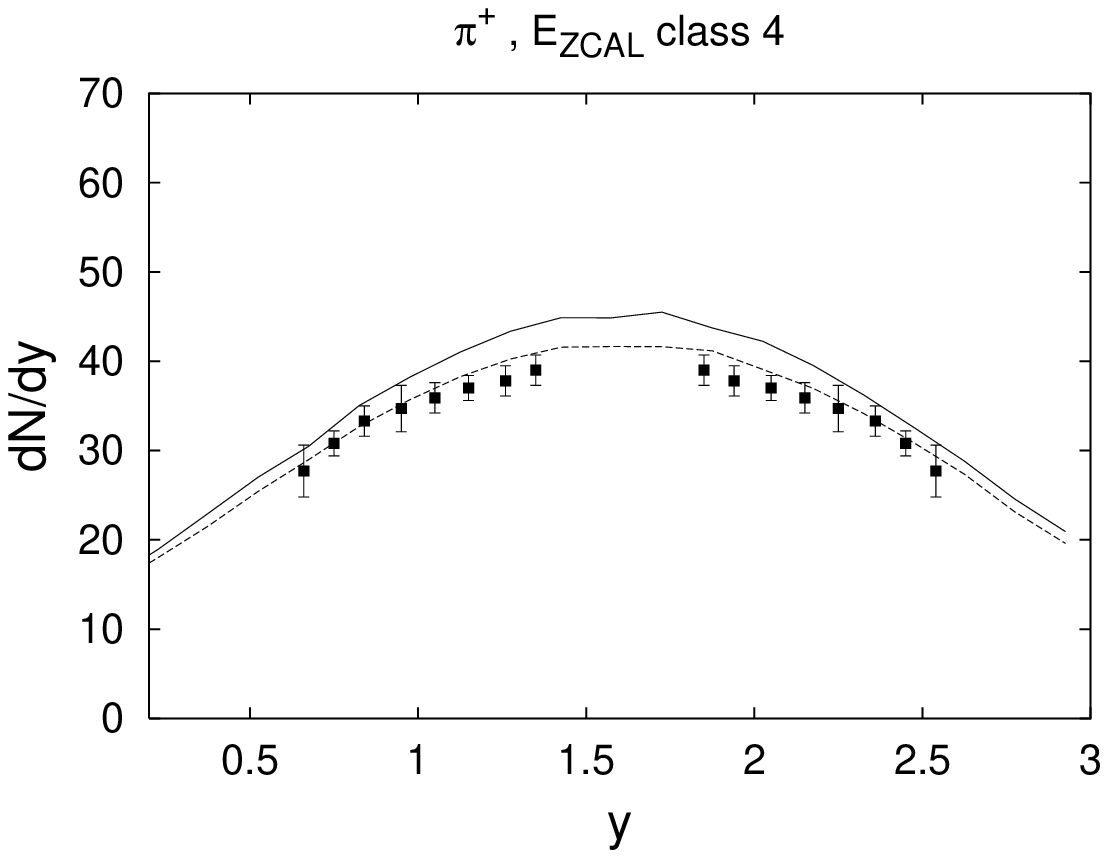} \\ 
\includegraphics[height=4.5cm]{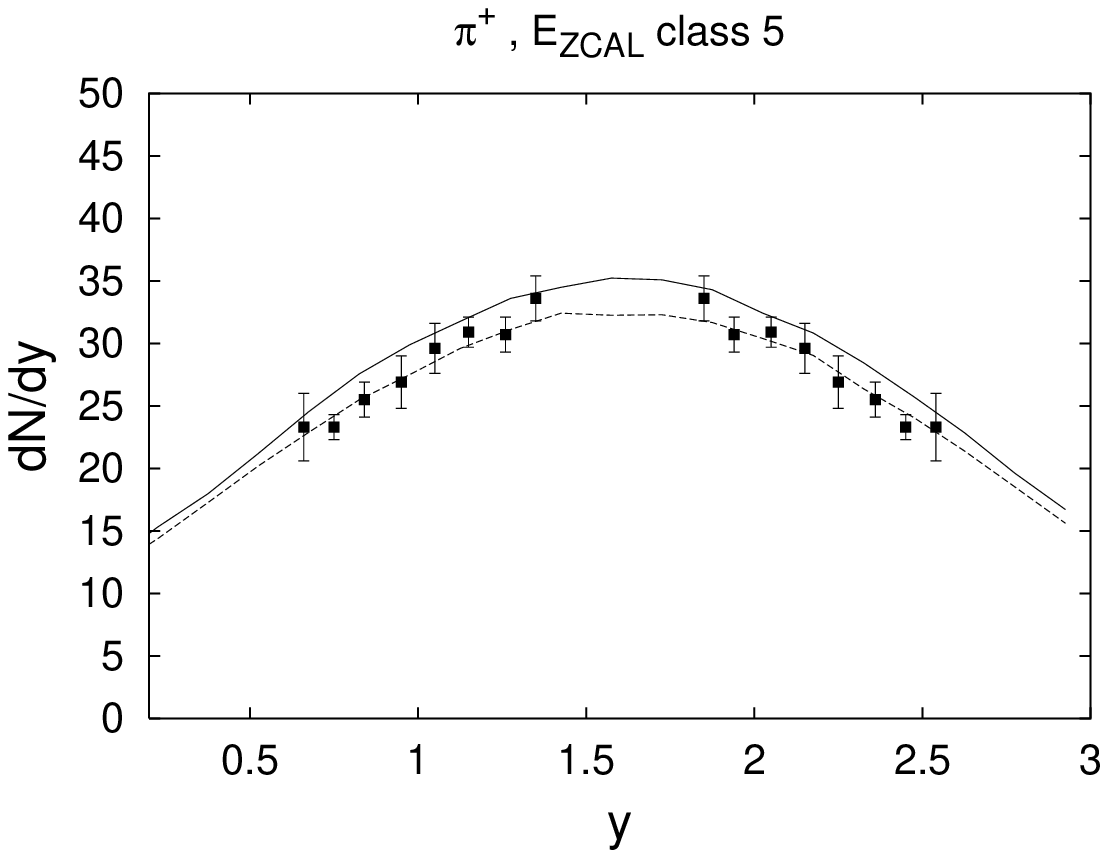} & 
\includegraphics[height=4.5cm]{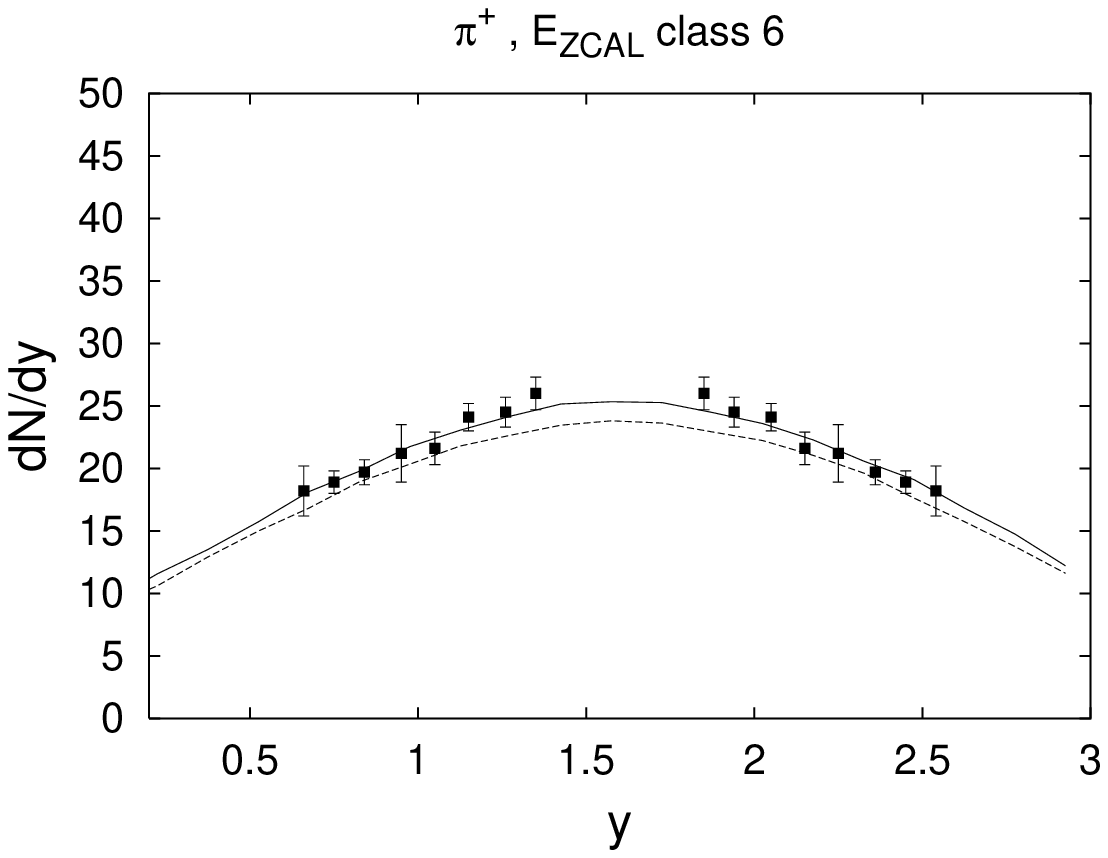} \\ 
\includegraphics[height=4.5cm]{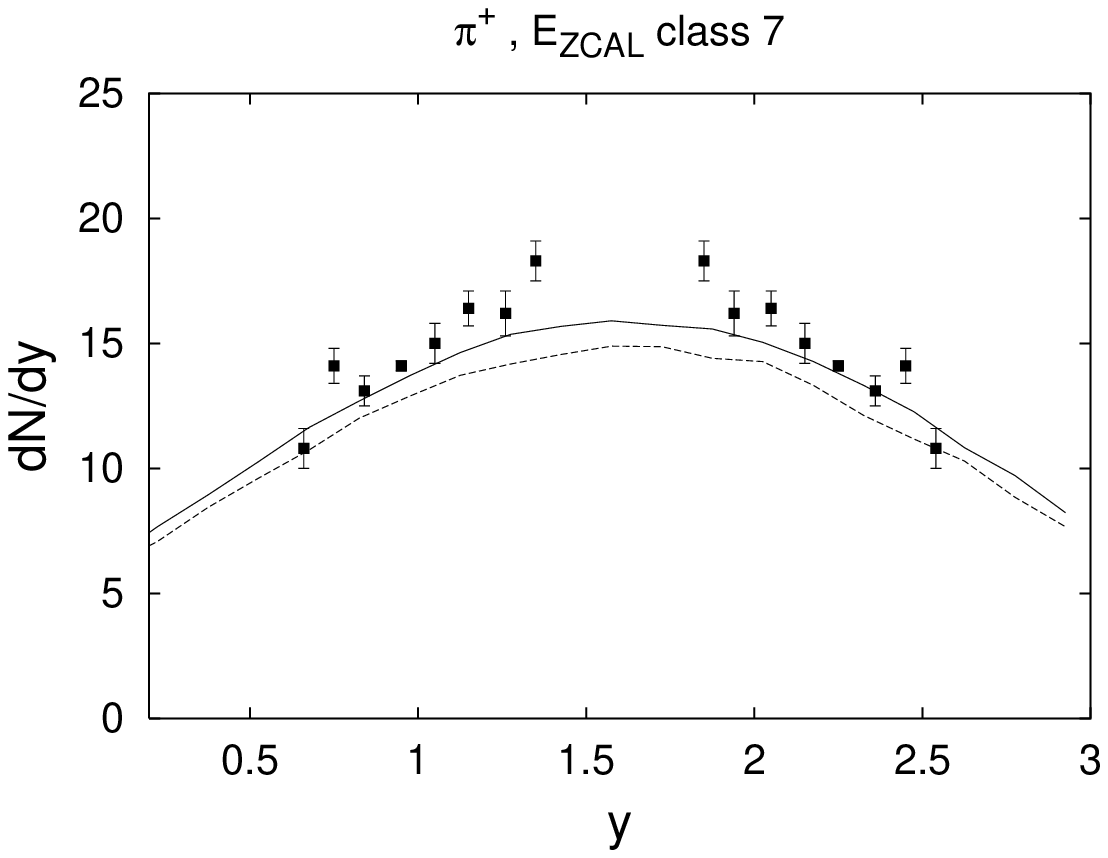} & 
\includegraphics[height=4.5cm]{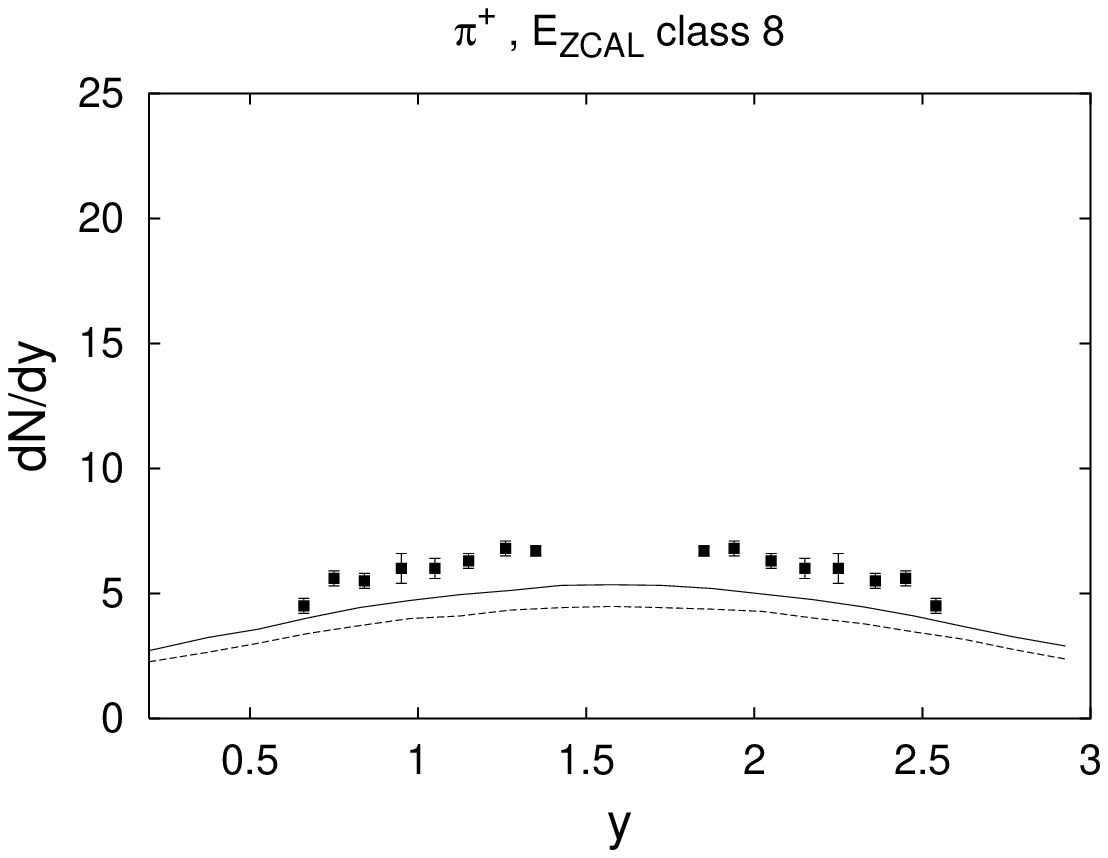} \\ 
\end{tabular} 
\caption{Rapidity spectra of $\pi^+$ from BUU events Au+Au at 10.7 A GeV selected by the zero degree energy in comparison to data from \cite{hic1}. The centrality decreases from the upper left corner to the lower right. The solid line shows the standard BUU calculation, whereas the dashed line is calculated with the medium modification described in section \ref{medi}.}\label{auaupi}
\end{center}
\end{figure}

\begin{figure}
\begin{center}
\begin{tabular}{cc}
\includegraphics[height=4.5cm]{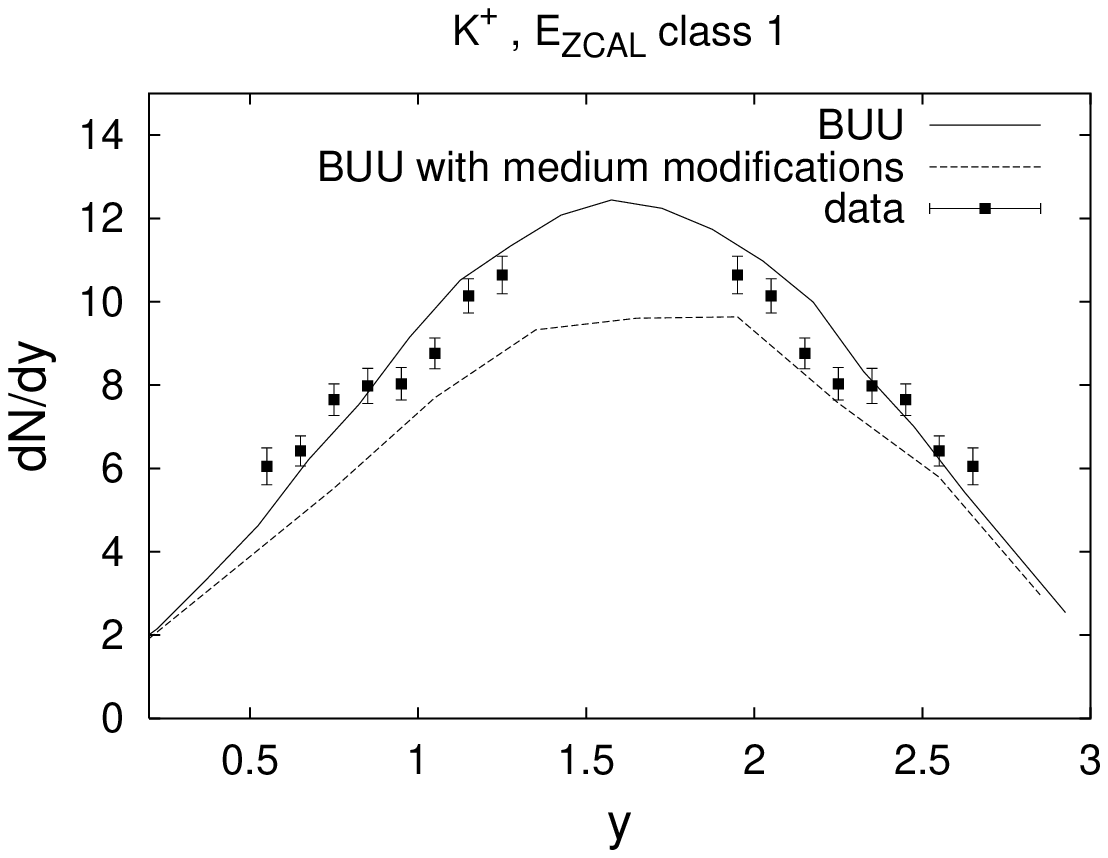} & 
\includegraphics[height=4.5cm]{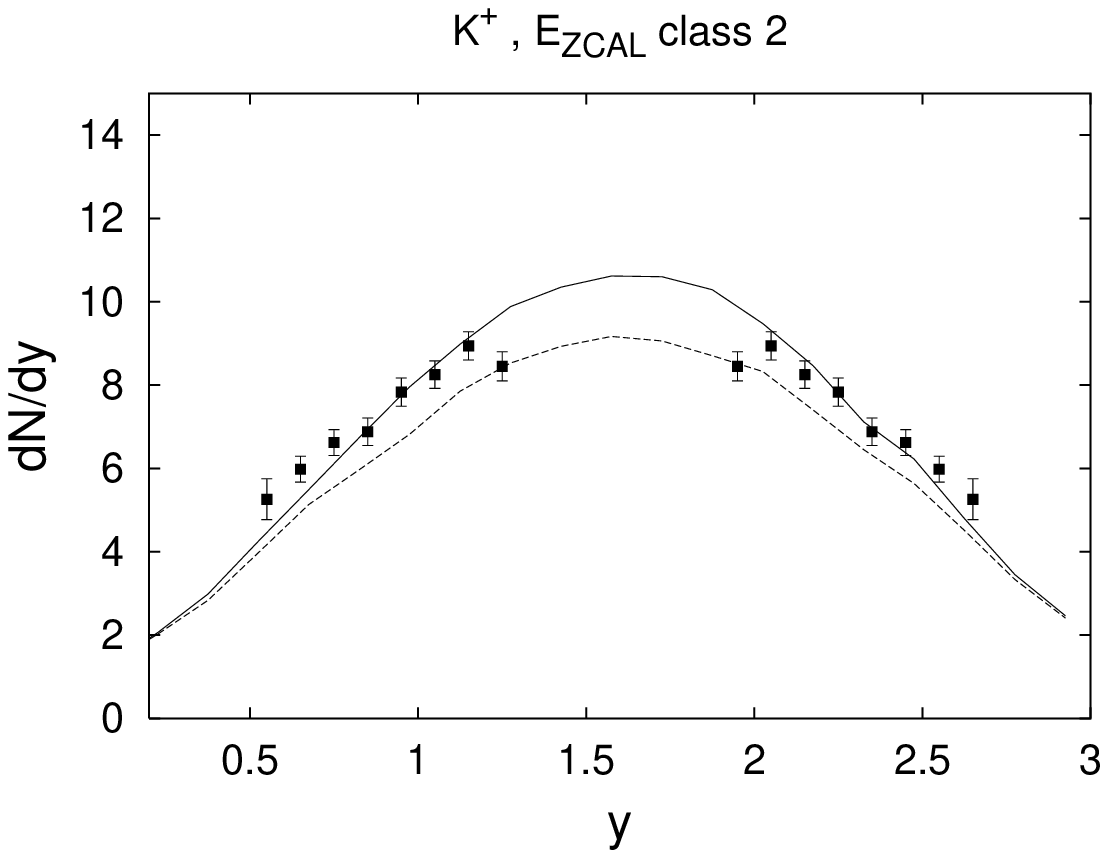} \\ 
\includegraphics[height=4.5cm]{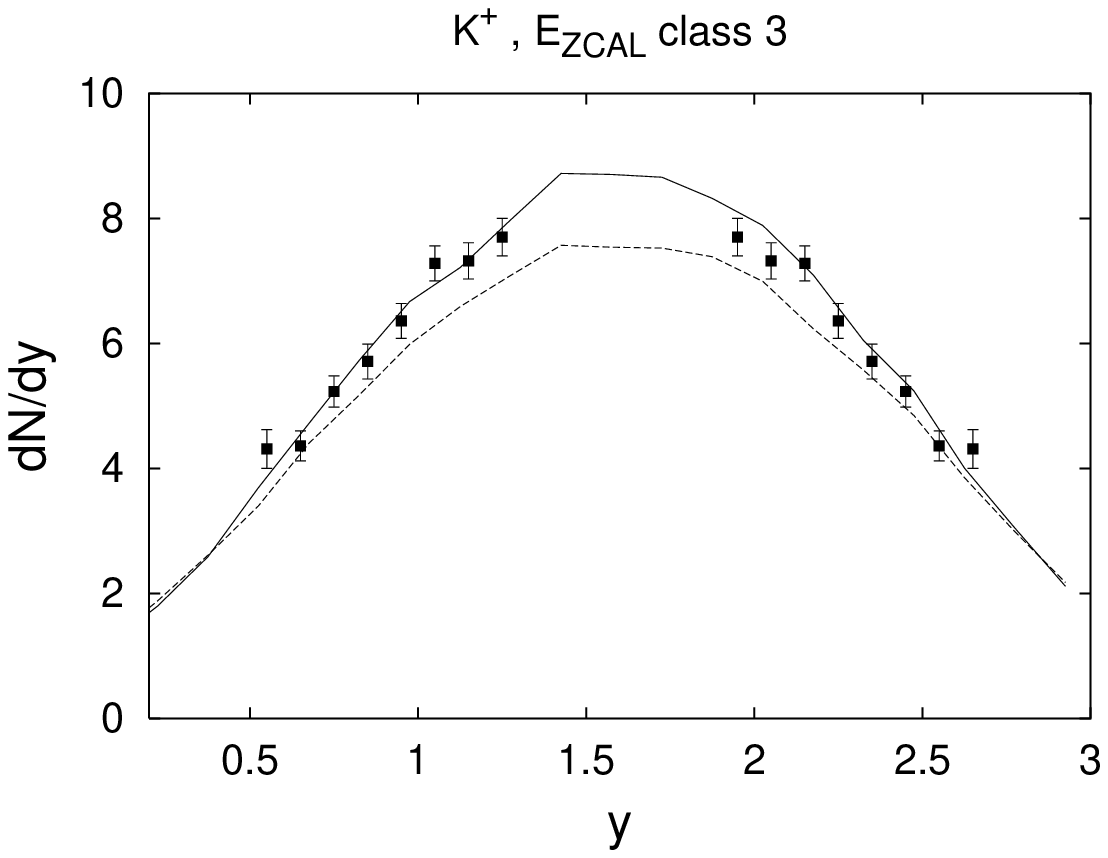} & 
\includegraphics[height=4.5cm]{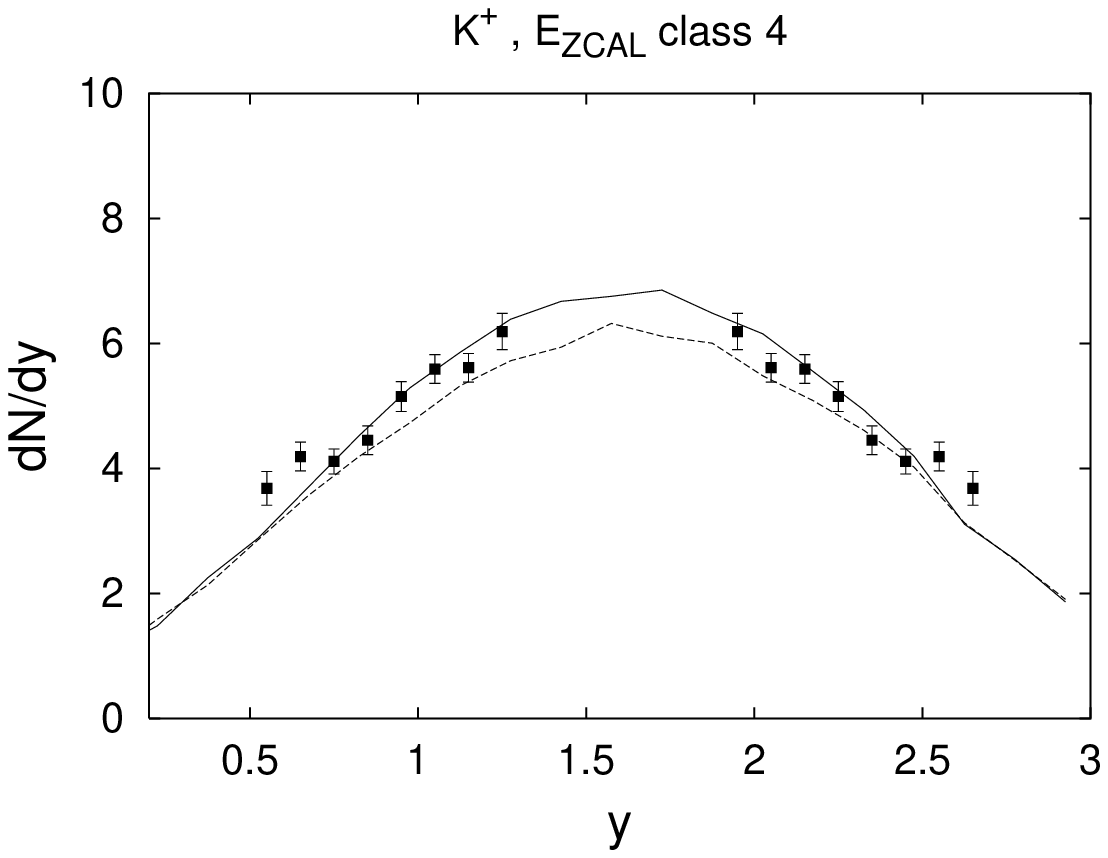} \\ 
\includegraphics[height=4.5cm]{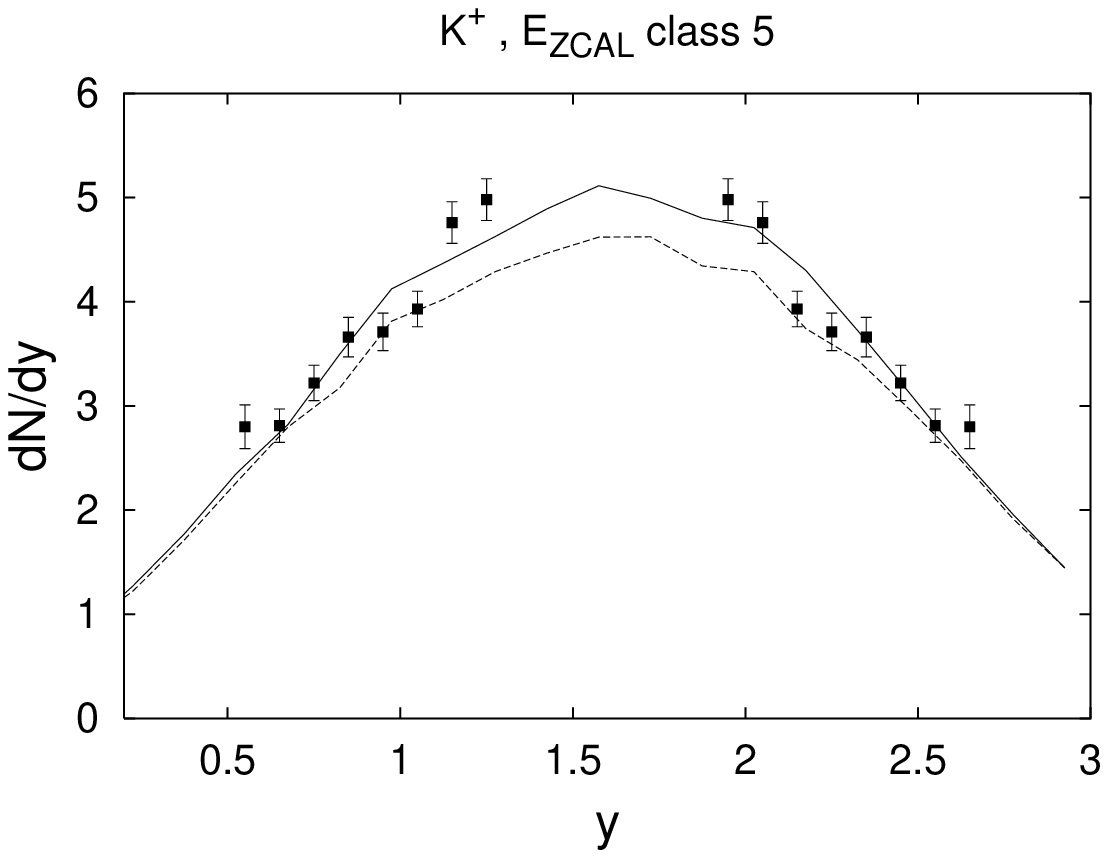} & 
\includegraphics[height=4.5cm]{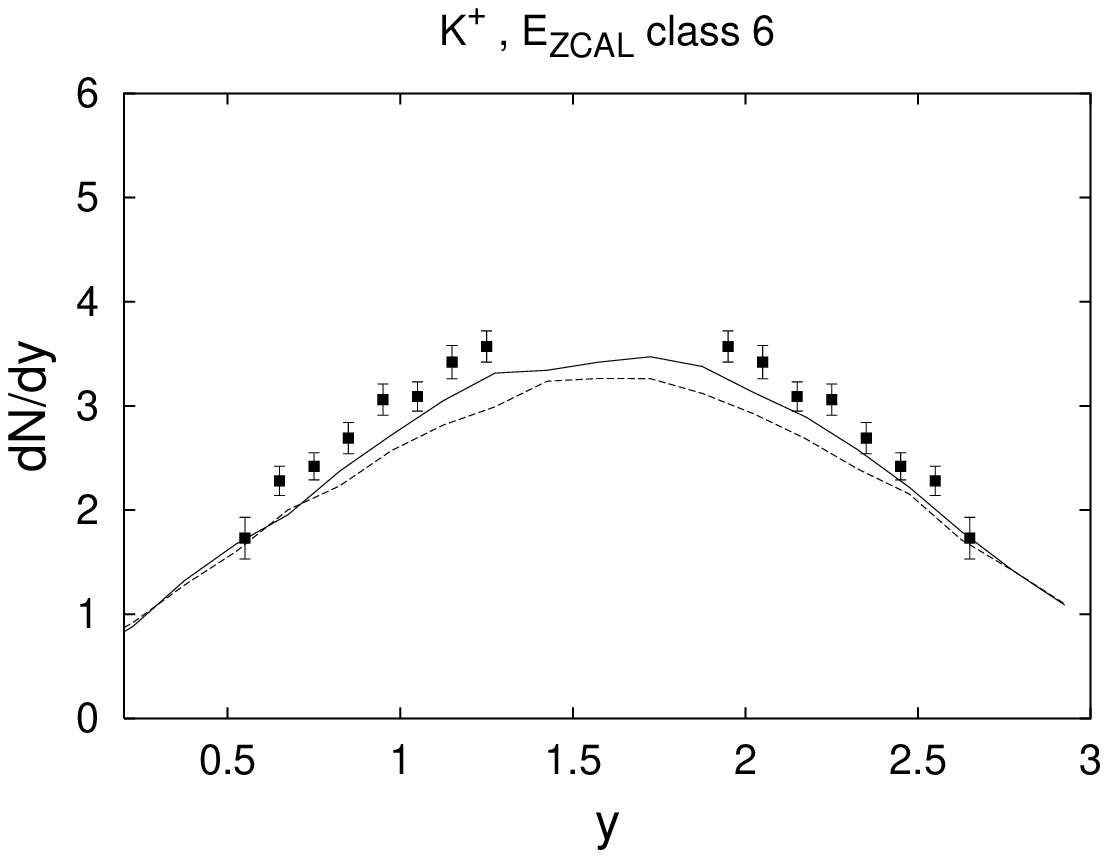} \\ 
\includegraphics[height=4.5cm]{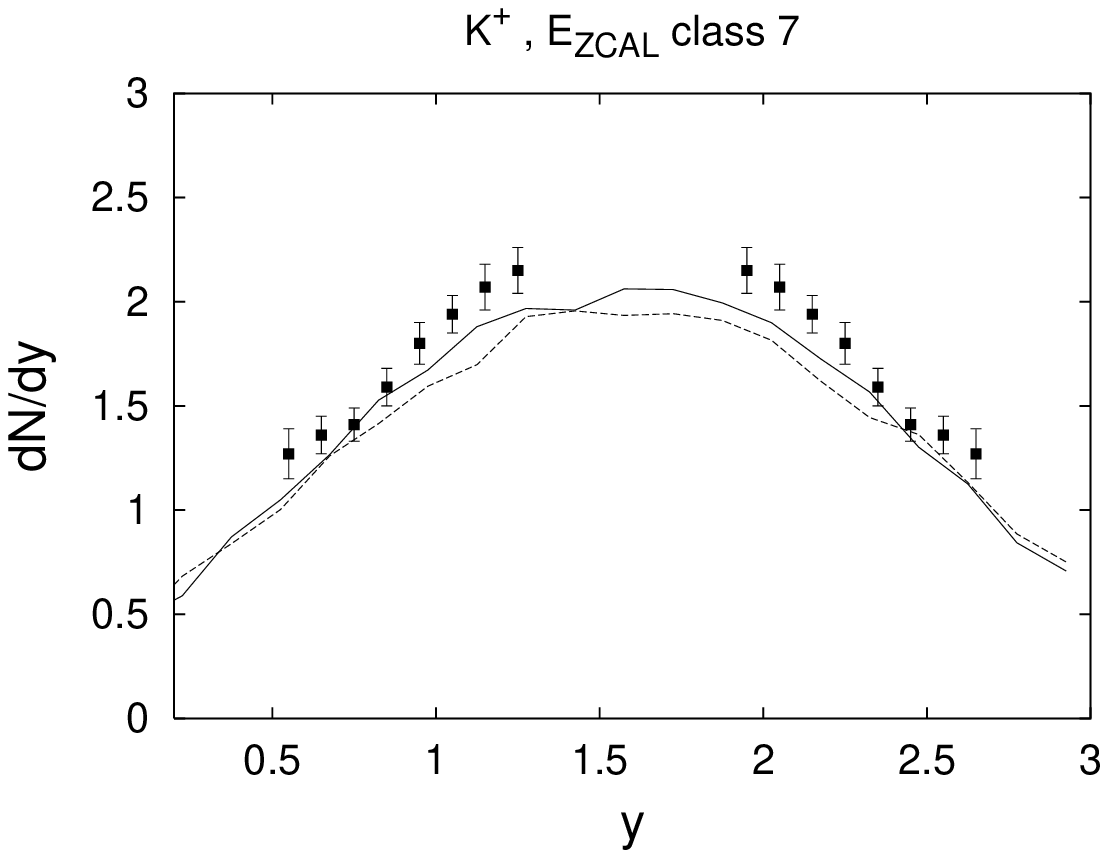} & 
\includegraphics[height=4.5cm]{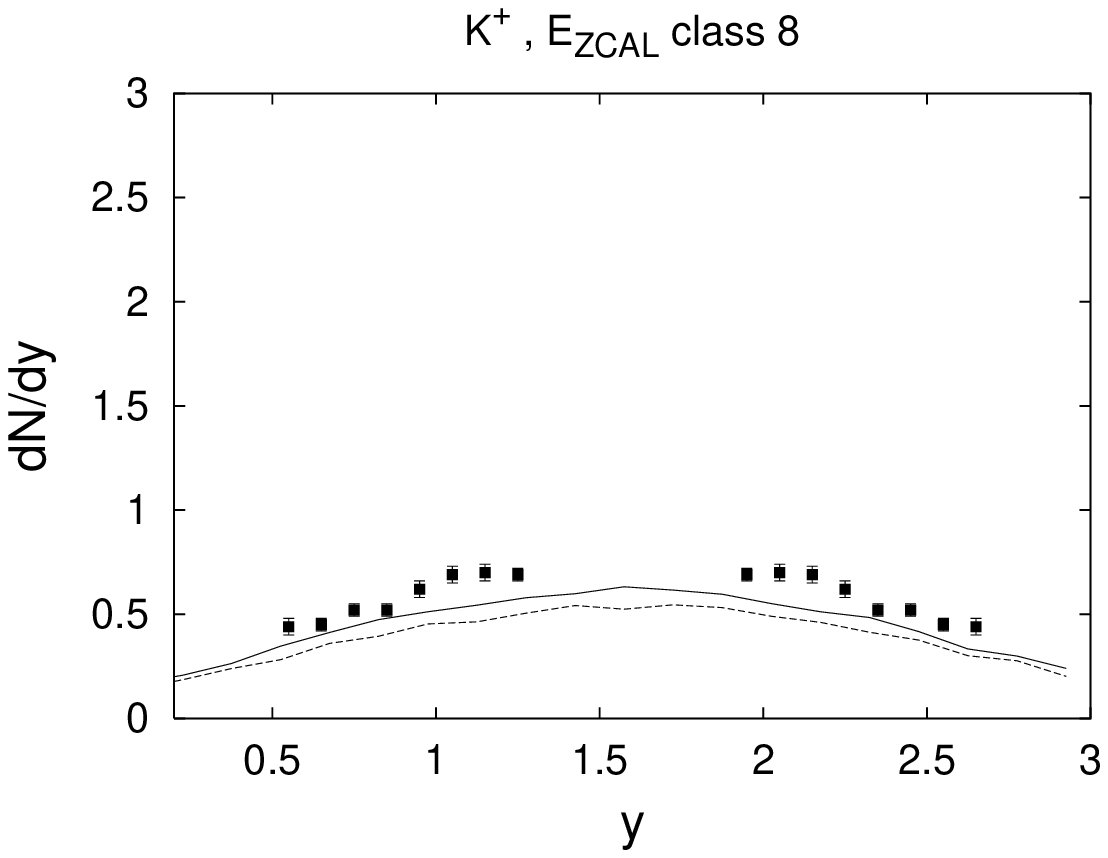} \\ 
\end{tabular}
\caption{Rapidity spectra of $K^+$ from BUU events Au+Au at 10.7 A GeV selected by the zero degree energy in comparison to data from \cite{hic1}. The centrality decreases from the upper left corner to the lower right. The solid line shows the standard BUU calculation, whereas the dashed line is calculated with the medium modification described in section \ref{medi}.}\label{auauka}
\end{center}
\end{figure}

Figs. \ref{auaupi} and \ref{auauka} show the rapidity spectra of
pions and kaons for different event classes selected by applying
the $E_{ZCAL}$ cut only. The $K^+$ rapidity spectra are very well 
described for all $E_{ZCAL}$ classes. For the $\pi^+$ rapidity 
spectra we see deviations from the data: 
In the most central collisions there are excessive pions in BUU 
produced mainly at midrapidity. With decreasing centrality the deviation
from the data disappears gradually, and in the most peripheral collisions
there is even an underprediction of the pion multiplicity by BUU.
These results are consistent with Figs. \ref{pberap}, \ref{paurap}, 
\ref{siaupic1}, where one can also observe a tendency to overpredict 
pion production with increasing size of the participant zone.

Figs. \ref{auaupitr} and \ref{auauktr} show transverse mass spectra of
$\pi^+$'s and $K^+$'s for the central collisions of Au+Au at 10.7 A GeV.
The spectra are shown for various rapidities starting from the backward
rapidity in the c.m. frame (upper line) to the midrapidity (lower line).
The shapes of the $\pi^+$ spectra are well described by standard BUU,
however, the $\pi^+$ yields are slightly overpredicted at small $m_\perp$
(see also upper left panel in Fig. \ref{auaupi}). The agreement of BUU
with the $K^+$ spectra is much poorer. The low $m_\perp$ part of
the $K^+$ spectra is, typically, overestimated by BUU, whereas the
high $m_\perp$ part of the spectra is underestimated by our calculations.
Thus, BUU underestimates the inverse slope parameter of the $ K^+$
transverse mass spectra, while the $K^+$ yield is well described
(c.f. upper left panel in Fig. \ref{auauka}). This problem has been
pointed out earlier in Ref. \cite{slopeel}.
  
Fig. \ref{hicfid} shows a fiducial yield of $K^+$ and $\pi^+$
divided by the projectile participant number $N_{pp}$ 
as a function of $N_{pp}$. 
The fiducial yield is defined as follows \cite{hic1}:
\begin{equation}
\text{fiducial yield}=\sum_{0.6 < y < 1.3} \frac{dN}{dy}dy~,
\end{equation}
where the $dN/dy$ are the rapidity distributions selected by the zero degree 
energy. The $K^+$ fiducial yield, as expected, agrees quite well with
the data except for a slight underprediction at peripheral collisions.
The $\pi^+$ fiducial yield increases with $N_{pp}$ faster
than the data do. In the absence of secondary NN collisions the fiducial yields divided by $N_{pp}$ would be constant.

\section{\label{excit}Excitation functions}

In this section we show the excitation functions of pions, kaons, $\Lambda$-
and $\Sigma$-hyperons from central Au+Au and Pb+Pb collisions in comparison 
to data, two other transport models HSD and UrQMD \cite{strange} 
and the statistical model \cite{thermoref1} . The calculations performed with 
the transport models have all been done in the cascade mode, which makes the 
comparison easier. We selected the data sets for the Au+Au system at
1.96, 4.00, 5.93, 7.94 and 10.7 A GeV \cite{ags1,ags2} with 5 \% of the most 
central events and for the Pb+Pb system at 30 and 40 A GeV \cite{na491,na492}
with 7 \% of the most central events. In the theoretical calculations we used 
a sharp impact parameter cut off at 3.5 fm for AGS energies and 4 fm for SPS 
energies. The influence of the centrality selection was tested at 10.7 A GeV 
by comparing calculations with a sharp cutoff to calculations which were done 
by employing the centrality criteria described in section \ref {ssd}. No 
deviations were seen in the observables which will be presented in the 
following.

\begin{figure}
\begin{center}
\epsfig{file=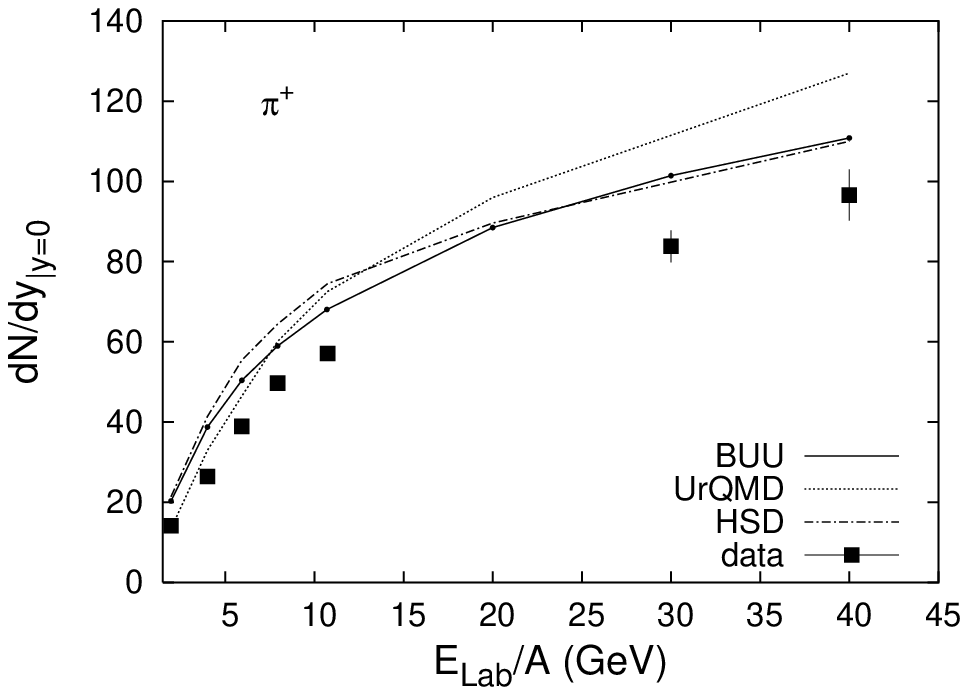,height=7cm}\caption{Midrapidity yields for $\pi^+$ as a function of energy in comparison to results of HSD, results of UrQMD and data from \cite{ags1,na491,na492}.}
\label{fexcit1}
\end{center}
\end{figure}

Fig. \ref{fexcit1} shows the midrapidity yield of positive pions as a function
of the beam energy. We see that all three  models overpredict the pion
yield in the considered beam energy range (2-40 A GeV). Our model (solid
line) overpredicts the $\pi^+$ midrapidity yield by $\sim 10$ \%
at 40 A GeV to $\sim 50$ \% at 2 A GeV. However, the shape of the experimental
excitation function  $dN/dy$ vs $E_{\text{Lab}}$ for $\pi^+$ is remarkably well
described by BUU. The HSD model (dot-dashed line) produces the $\pi^+$ 
yields close to the BUU results excepting the points at 6, 8 and 10.7 A GeV,
where HSD has $\sim 10$ \% more pions than BUU. The UrQMD model (dotted
line) agrees well with the pion data at the smallest energy of 2 A GeV,
but the pion yield grows too fast with energy within UrQMD producing
the discrepancy $\sim 30$ \% with data at the highest considered energy
of 40 A GeV.

\begin{figure}
\begin{center}
\epsfig{file=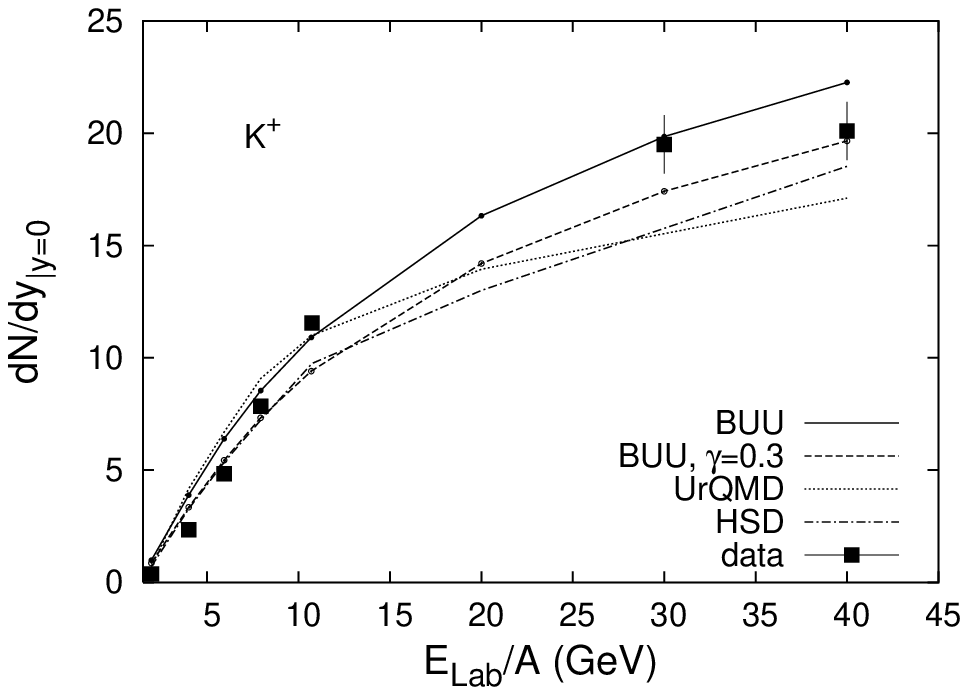,height=7cm}\caption{Midrapidity yields for $K^+$ as a function of energy in comparison to results of HSD, results of UrQMD and data from \cite{ags1,na491,na492}.}
\label{fexcit2}
\end{center}
\end{figure}

\begin{figure}
\begin{center}
\epsfig{file=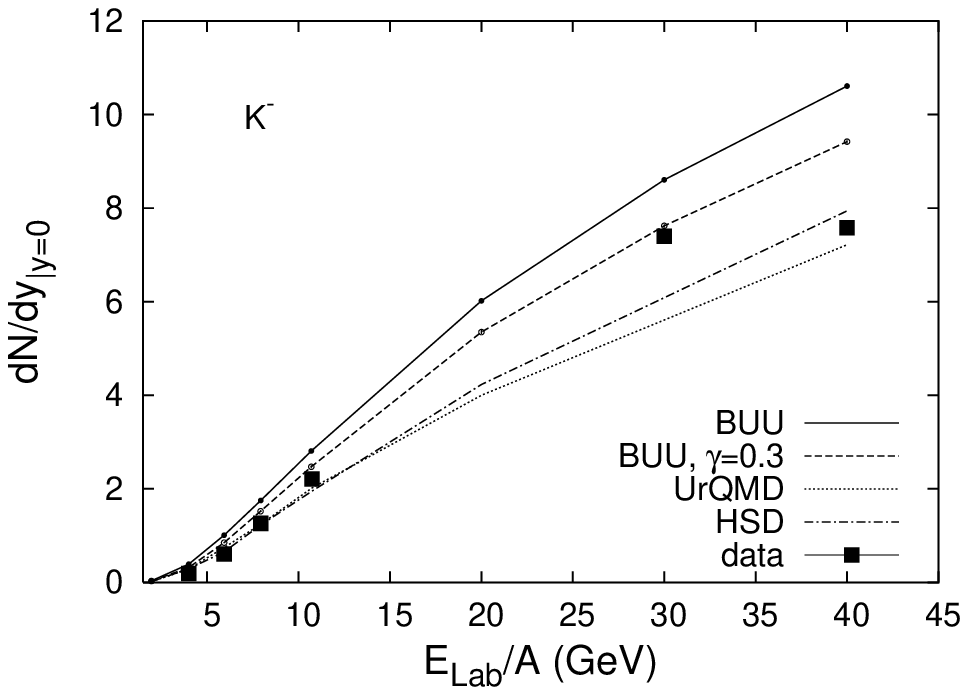,height=7cm}\caption{Midrapidity yields for $K^-$ as a function of energy in comparison to results of HSD, results of UrQMD and data from \cite{ags2,na491,na492}.}
\label{fexcit3}
\end{center}
\end{figure}

\begin{figure}
\begin{center}
\epsfig{file=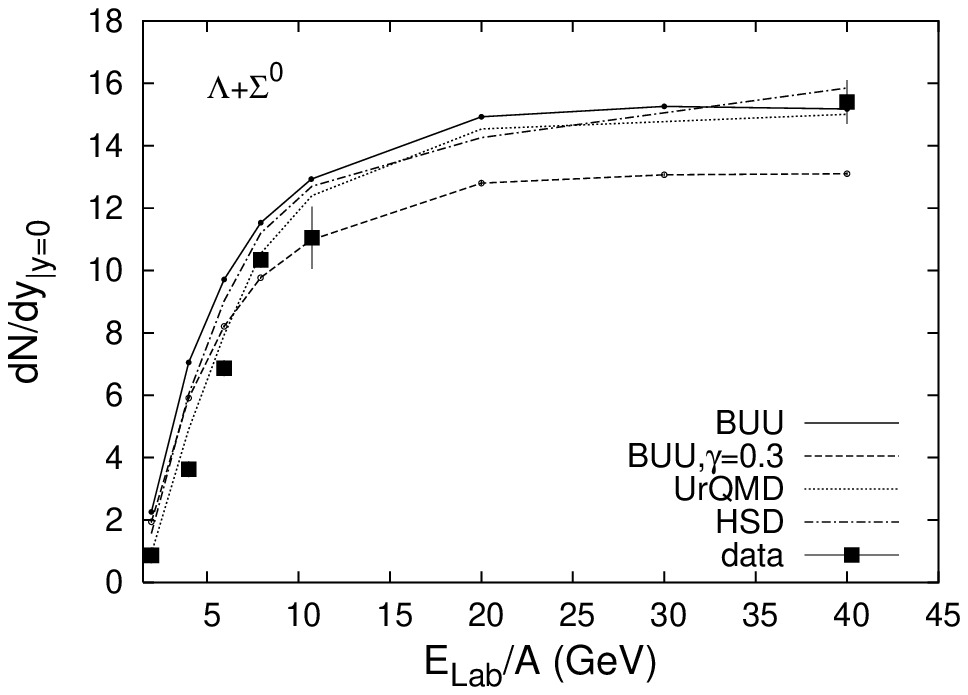,height=7cm}\caption{Midrapidity yields for $\Lambda + \Sigma_0$ as a function of energy in comparison to results of HSD, results of UrQMD and data from \cite{lambda1,lambda2,lambda3,lambda4,lambda5}.}
\label{fexcit4}
\end{center}
\end{figure}

Figs. \ref{fexcit2}, \ref{fexcit3} and \ref{fexcit4} show the midrapidity 
yields of $K^+, K^-$ and ($\Lambda + \Sigma^0$), respectively, as functions 
of the beam energy. BUU quite well describes the $K^+$ midrapidity yield 
excepting the points at 4 and 6 A GeV, where BUU overestimates the data
by 30-50 \%.  

The $K^-$ midrapidity yield and ($\Lambda + \Sigma^0$) midrapidity yield 
at $E_{\text{Lab}} < 40$ A GeV are overestimated by BUU. Using the constant strangeness suppression factor $\gamma=0.3$ (dashed lines) reduces the yields of $K^+, K^-$ and hyperons. 
This leads to a better description of the $K^-$ yields, while in the cases of $K^+$ and 
($\Lambda + \Sigma^0$) it is hard to judge which strangeness suppression
factor works better. In Fig.\ref{fexcit2} and Fig.\ref{fexcit4} we see that up to about 10 A GeV the $K^+$ and hyperon yields are better described with $\gamma=0.3$. Above that energy, however, the energy dependent suppression factor works better. For the antikaons the strangeness exchange processes $\overline K N \leftrightarrow \pi Y$ are important due to the strong in-medium modifications \cite{tolos}, which are not taken into account in our study.  

The HSD results on the $K^+$ production are close to our calculation
with $\gamma=0.3$. The UrQMD model gives the $K^+$ yield at lower energies
similar to our standard calculation, whereas at higher energies UrQMD 
produces substantially less kaons. The $K^-$ yield is rather well described
by both models, HSD and UrQMD, except for the point at 30 A GeV. At lower energies the ($\Lambda + \Sigma^0$) 
yields calculated within HSD and UrQMD are somewhat closer to the data than 
our standard calculation. At higher energies the HSD, UrQMD and our
standard calculation give very close results for the ($\Lambda + \Sigma^0$)
yield.

Fig. \ref{fexcit5} shows the ratio of midrapidity yields of $K^+$ and
$\pi^+$ as a function of the beam energy. In the upper left panel of Fig. \ref{fexcit5} we see that neither BUU nor HSD and UrQMD describe the ratio $K^+/\pi^+$ in the whole beam energy region. At the lowest beam energy of 2 A GeV BUU and UrQMD overpredict the 
ratio by a factor of two, whereas HSD agrees with data. Between 4
and 8 A GeV BUU is quite close to the data. However, the $K^+/\pi^+$ ratio excitation function levels off too early in BUU and, 
as a consequence, we underestimate the ratio by $\sim 25$ \% between
10 and 30 A GeV. The HSD results on the ratio $K^+/\pi^+$ have a similar
beam energy dependence, but the value of the ratio is smaller by
$\sim 20$ \%, which is close to our calculation with $\gamma=0.3$.
UrQMD produces a larger slope at lower energies overestimating the ratio
at $E_{\text{Lab}} < 8$ A GeV, but at higher energies the slope gets negative
which causes a strong discrepancy with data in SPS energy regime.
Overall, we observe that BUU has the best agreement with data on the
$K^+/\pi^+$ ratio in the considered energy regime. At the beam energies
of 4-6 A GeV this, however, comes about due to cancellation of the 
overestimation of the pion and kaon yields.

\begin{figure}
\begin{center}
\epsfig{file=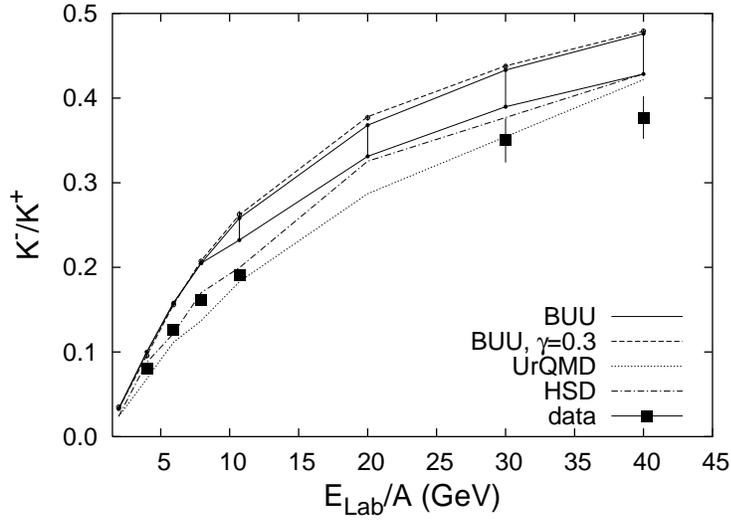,height=7cm}\caption{The $K^-/K^+$ ratio at midrapidity as a function of energy in comparison to results of HSD, results of UrQMD and data from \cite{ags2,na491,na492}. The errorband on the BUU results indicate the uncertainty in the meson-meson channels (see discussions in sections \ref{mesm} and \ref{excit}).}
\label{fexcit6}
\end{center}
\end{figure}

Fig. \ref{fexcit6} shows the ratio of the midrapidity yields $K^-/K^+$
vs the beam energy. The BUU calculations with the energy dependent strangeness suppression factor and our "standard" meson-meson cross sections is shown as the upper boundary of the errorband. In order to estimate an effect of the p-wave suppression on the $\pi \rho \rightarrow K\overline K$ and $\pi \omega \rightarrow K \overline K$ cross section (see section \ref{mesm}), we also performed a calculation by putting the cross section equal to zero below the $K^\ast\overline K$ production threshhold, which is shown by the lower boundary of the errorband in Fig.\ref{fexcit6}. Thus the p-wave suppression could reduce the $K^-$ multiplicity by about 10$\%$. The $K^+$ multiplicity is reduced less than 5$\%$ by this effect, since the fraction of $K^+$ coming from meson-meson collisions is less than the fraction of $K^-$ coming from meson-meson reactions. (For this reason our results on the $K^+/\pi^+$ ratio Fig.\ref{fexcit5} are, practically, untouched by the p-wave suppression effect.) 

Overall, we see that BUU overestimates the $K^-/K^+$ ratio independent on the strangeness suppression factor. This result is expected
from the previous Figs. \ref{fexcit2} and \ref{fexcit3}, where we see 
that the $K^+$ yield is rather well described by BUU, while the $K^-$ yield 
is overestimated substantially. Since the choice $\gamma=0.3$ reduces both
$K^+$ and $K^-$ yields, the ratio $K^-/K^+$ is practically independent
on the strangeness suppression factor. The HSD and UrQMD models describe
the experimental $K^-/K^+$ ratio quite well. This can be also traced back
to Figs. \ref{fexcit2} and \ref{fexcit3}.

\begin{figure}
\begin{center}
\epsfig{file=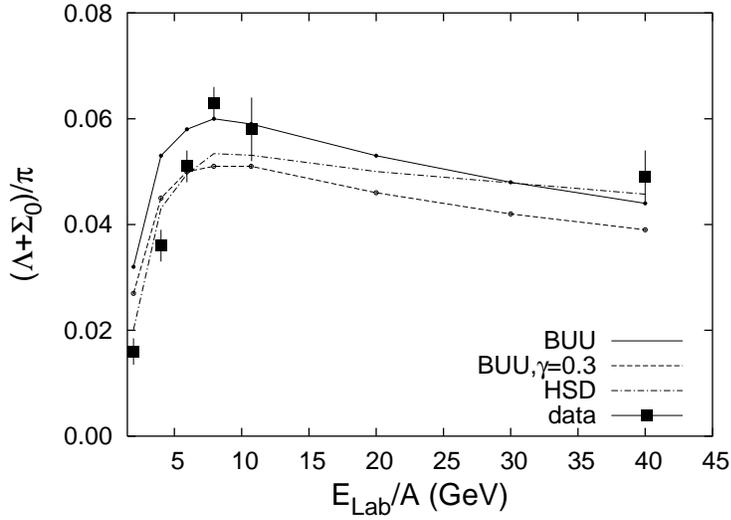,height=7cm}\caption{The $(\Lambda+\Sigma^0)/\pi$ ratio at midrapidity , where $\pi=1.5(\pi^++\pi^-)$, as a function of energy in comparison to data from \cite{lambda1,lambda2,lambda3,lambda4,lambda5}.}
\label{fexcit7}
\end{center}
\end{figure}

Fig. \ref{fexcit7} shows the ratio of the midrapidity yields 
$(\Lambda + \Sigma^0)/\pi$ vs the beam energy. This ratio has a peak
near $E_{\text{Lab}}=8$ A GeV which is reproduced by BUU and HSD models. 
As far as the absolute values of this ratio are concerned at small
energies $E_{\text{Lab}} < 10$ A GeV, we would like to remind that both
pion and hyperon yields are overestimated by standard BUU (see Figs. 
\ref{fexcit1} and \ref{fexcit4}). Thus, the agreement of standard BUU
with data on $(\Lambda + \Sigma^0)/\pi$ at 8 and 10.7 A GeV is again
a result of a mutual cancellation of the $(\Lambda + \Sigma^0)$
and $\pi$ excesses. The choice of $\gamma=0.3$ which describes the 
$(\Lambda + \Sigma^0)$ midrapidity yield at small energies better 
(c.f. Fig. \ref{fexcit4}) leads to the underestimation of the
$(\Lambda + \Sigma^0)/\pi$ ratio at the peak due to the overestimation
of the pion yield. 

All calculations discussed above were performed in the cascade mode.
There is an option in our BUU model, which switches on a nuclear mean 
field potential. The nuclear mean field potential is necessary, 
in particular, for a description of the experimental data on collective
in-plane and out-of-plane proton and neutron flows \cite{pot1} 
at 0.15-2 A GeV. At higher energies, however, the parametrisation of
the momentum-dependent interaction used in \cite{pot1} leads to too 
strong repulsive in-plane flow (see also Ref. \cite{pot2}). 
Nevertheless, in order to estimate the mean field effect on pion and
kaon production, we have also done the calculation with the mean field
potential (incompressibility $K=215$ MeV, soft momentum-dependent mean
field SM). The lower left panel of Fig. \ref{fexcit5} shows the results of this calculation 
(dashed line) in comparison with our standard calculation in the cascade
mode (solid line) and with experimental data. We see that at $E_{\text{Lab}} < 40$
A GeV the ratio $K^+/\pi^+$ is reduced due to the mean field potential,
since the pion yield is relatively insensitive to the nuclear mean field,
whereas the kaon yield is reduced. Indeed, a part of the kinetic energy
of the counterstreaming nucleon flows transforms now to the potential
energy. This reduces, generally, particle production. However, the kaon
production is closer to its threshold than the pion production. Therefore,
kaons are more strongly influenced by the mean field potential than pions.  

As we described in Section \ref{model}, the meson-meson collisions and
the $q \bar q$ annihilation channel for the meson-baryon collisions are
implemented in our BUU model. The lower right panel of Fig. \ref{fexcit5} shows an effect of
these implementations on the $K^+/\pi^+$ midrapidity ratio. The result
of our standard calculation including both the meson-meson collisions and 
the $q \bar q$ annihilation is shown by the solid line in Fig. 
\ref{fexcit5}. The dotted and dashed lines represent the calculations 
without the meson-meson collisions but with annihilation and without the 
annihilation but with the meson-meson collisions, respectively.  
The meson-meson collisions strongly enhance the $K^+/\pi^+$ ratio
above 6 A GeV due to the increased $K \bar K$ production.
An effect of the $q \bar q$ annihilation channel is less
pronounced: only a slight enhancement of the $K^+/\pi^+$ ratio is visible
at 5-10 A GeV.  

The upper right panel of Fig. \ref{fexcit5} compares the BUU and the statistical model
\cite{thermoref1} calculations for the $K^+/\pi^+$ ratio at midrapidity.
Since we use a string model, which produces a multiparticle final state
for the two colliding particles, the thermal equilibrium would be only
reached if we also included the corresponding back reactions
(c.f. Ref. \cite{cassin}). However, this was out of scope of the present 
work. Nevertheless, there is a surprisingly good
agreement between BUU and the statistical model. The statistical model
is closer to the data, but it is also unable to describe the data points 
at 10 and 30 A GeV.

Finally, in Fig. \ref{fexcit8} we show the inverse slope parameter $T$
of the $K^+$ transverse mass spectra vs the laboratory energy. At the
energies 2, 4, 6 and 8 A GeV the inverse slope parameter was obtained
by fitting the exponential function (\ref{expfit}) to the $m_\perp$-spectrum
of kaons in the rapidity range $|(y-y_{NN})/y_{NN}| < 0.25$ \cite{ags2}.
At 10.7 A GeV the rapidity range was $|(y-y_{NN})/y_{NN}| < 0.125$ 
\cite{ags2}. At the SPS energies of 30 and 40 A GeV the rapidity range
was taken as $|y-y_{NN}| < 0.1$ \cite {na491,na492}. Our model underestimates
the inverse slope parameter by 30-40 \% (see also Fig. \ref{auauktr}). 
A similar problem was reported before in Ref. \cite{slopeel,cass2} and ascribed to the lack of pressure due to missed nonhadronic degrees of freedom in the transport models. We speculate here that also the inclusion of multi-baryon collisions would tend to make the spectrum harder. The probability for such processes naturally increases with a high power of baryon density.

\begin{figure}
\begin{center}
\begin{tabular}{c}
\includegraphics[height=6cm]{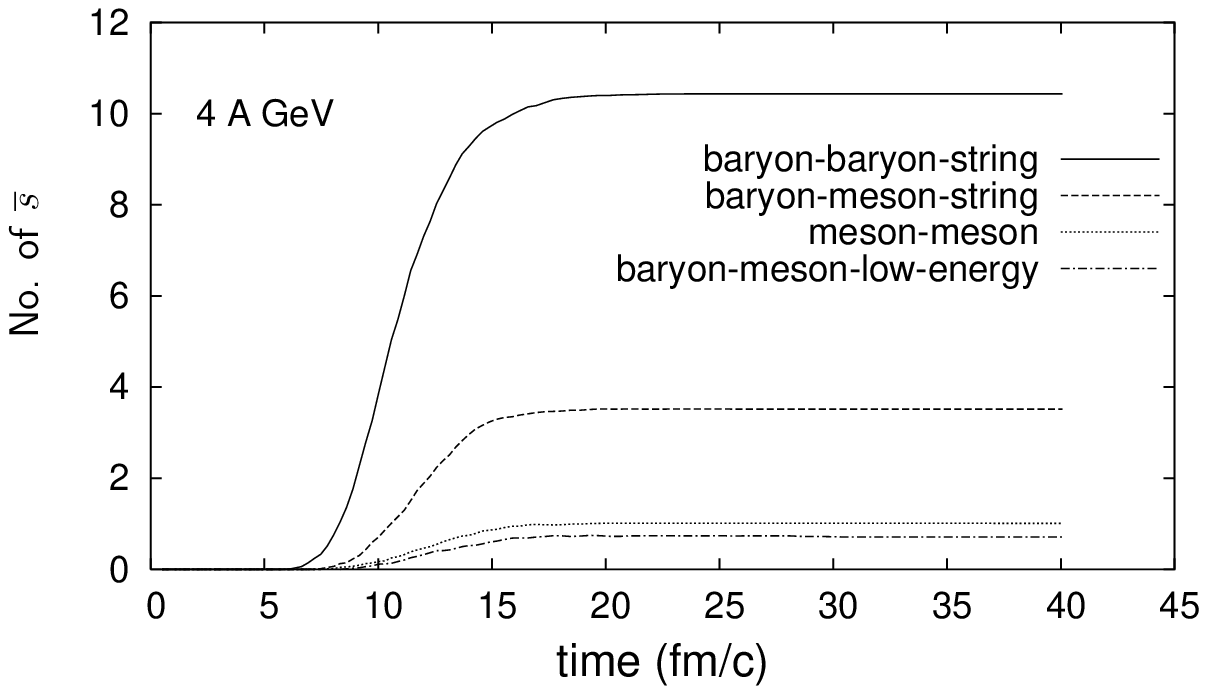} \\
\includegraphics[height=6cm]{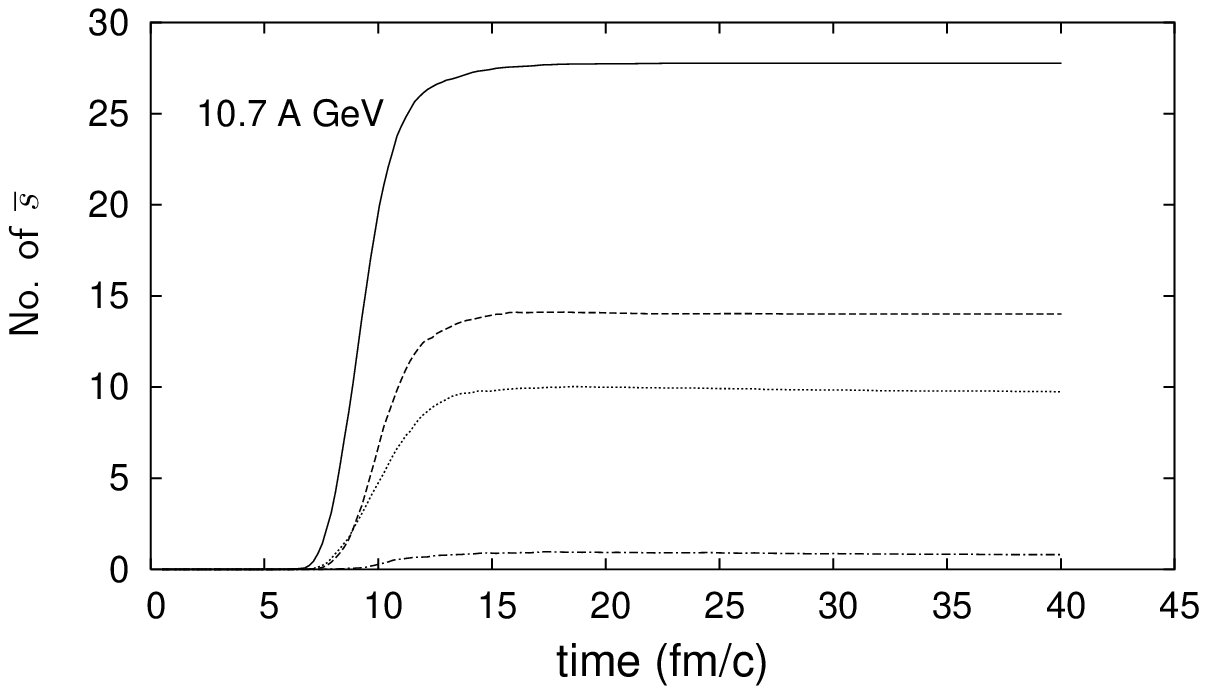} \\
\includegraphics[height=6cm]{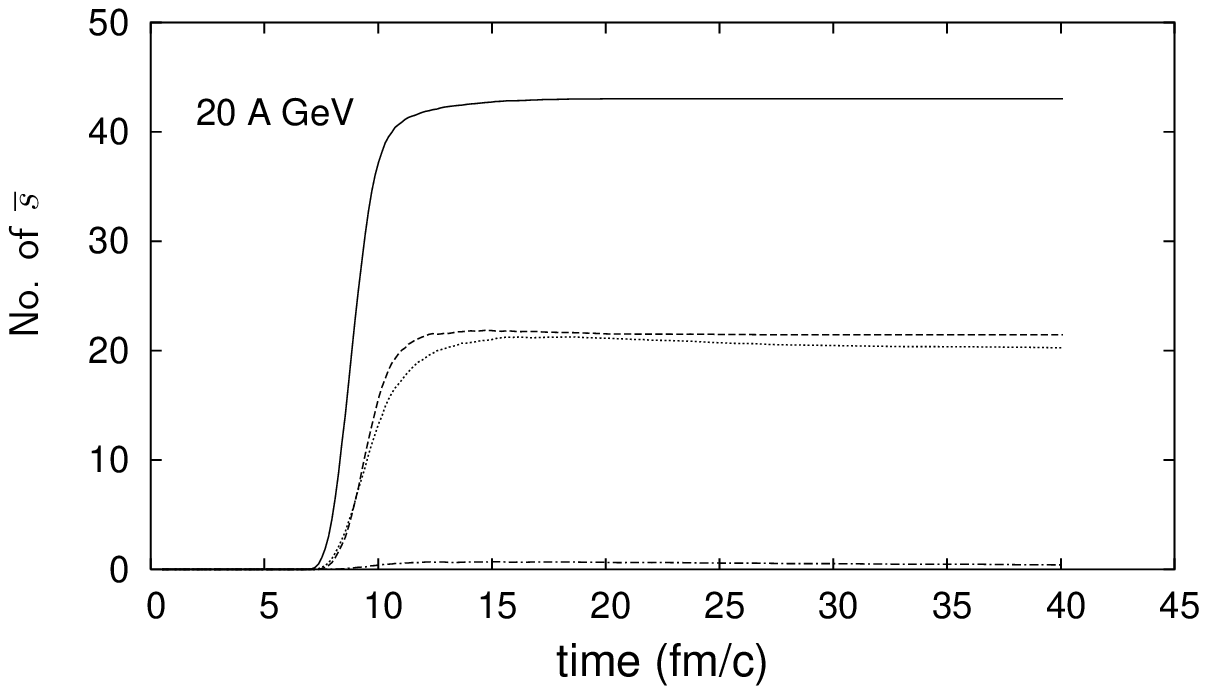} \\
\end{tabular} 
\caption{Contribution of different channels to strangeness production in 
Au+Au collisions with $b=1$ fm at 4 A GeV (upper panel), 
10.7 A GeV (middle panel) and 20 A GeV (lower panel).}
\label{chann}
\end{center}
\end{figure}

In order to see an origin of the produced kaons we performed a channel
decomposition of the $\bar s$-quark production for the central Au+Au
collisions at 4, 10.7 and 20 A GeV. Fig. \ref{chann} shows the number 
of the produced $\bar s$-quarks vs time for four different channels:
(i) baryon-baryon channel at high energies ($\sqrt{s} > 2.6$ GeV) or,
in other words, the baryon-baryon reactions simulated by the FRITIOF
string model (solid line), (ii) baryon-meson collisions at
$\sqrt{s} > 2$ GeV simulated by the string model (dashed line),
(iii) baryon-meson collisions at $\sqrt{s} < 2$ GeV, i.e. below the
string model threshold (dash-dotted line), (iv) meson-meson collisions
(dotted line). The low energy ($\sqrt{s} < 2.6$ GeV) baryon-baryon
collisions do not contribute to the $\bar s$-production significantly
at the considered beam energies. Thus, this channel is not shown in
Fig. \ref{chann}. We counted only the creation of the $\bar s$-quark 
and we did not consider reactions or decays, as e.g. 
$K^\ast \rightarrow K \pi$ where the $\bar s$-quark is only shifted from 
a $K^\ast$ to a $K$. 

The baryon-baryon-string channel plays the dominant role in the whole
beam energy region. This channel includes mainly the first-chance $NN$ 
collisions between the projectile and the target nucleons.
The meson-meson channel is not important at 4 A GeV, but its role grows
quickly with energy and at 20 A GeV it includes already 25 \% of
the produced $\bar s$-quarks. A relative contribution of the 
baryon-meson-string channel also increases with energy, while the low
energy baryon-meson collision relative contribution stays always very small
and decreases with energy.

\begin{figure}
\begin{center}
\epsfig{file=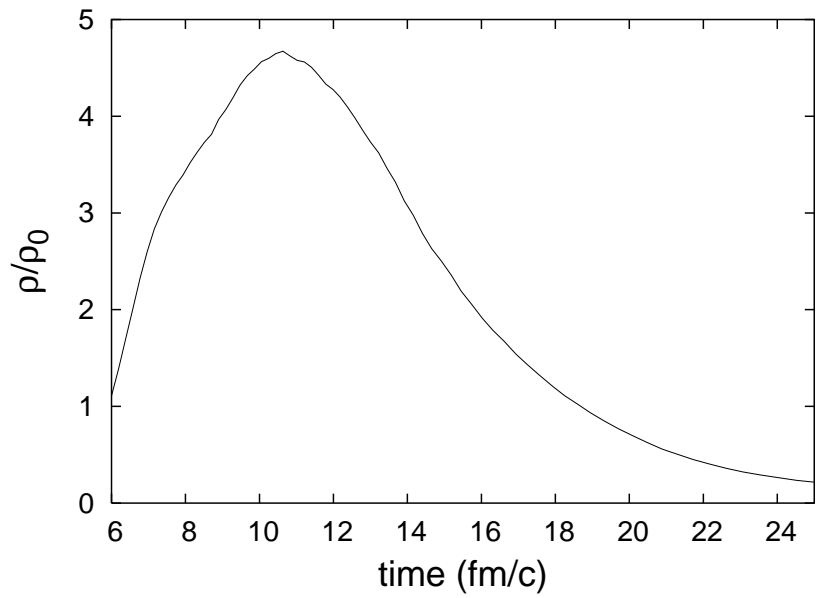,height=8cm}
\caption{Time evolution of the central baryon density in Au+Au collisions at 10.7 A GeV, b=0 fm.}
\label{cdens}
\end{center}
\end{figure}

The time evolution of the $\bar s$-quark production can be better
understood if one looks also on the central density time evolution
shown in Fig. \ref{cdens} for the central Au+Au collision at 10.7 A GeV.
The central density reaches its maximum value $\sim 4.5 \rho_0$ at 
$t\simeq 10.5$ fm/c and stays above $3\rho_0$ in the time interval
$t=7.5-14$ fm/c, where the $\bar s$-quark production just takes place.
Here $\rho_0 = 0.16$ fm$^{-3}$ is the nuclear saturation density.
Thus, strangeness is produced during the high-density stage of a
heavy-ion collision. It is evident also from Fig. \ref{chann}, that
the $\bar s$-quark production from the meson-meson and the baryon-meson
channels, which contain the secondary collisions, starts later than from
the baryon-baryon channel.
\section{\label{medi}In-medium modification of the FRITIOF\\ model }

In the course of a heavy-ion collision the elementary hadron-hadron
collisions happen at a finite baryon density. Therefore, the wave
functions of incoming and outgoing particles are the in-medium plane
waves rather than the vacuum ones.\footnote{Exchange particles expressed by
the propagators get also in-medium modified. However, this last
effect would be strongly dependent on the model used for the description
of an elementary collision. Thus we will, for simplicity, neglect it by
using the vacuum matrix elements on the place of the in-medium ones.}

In order to take into account the in-medium modifications of the incoming
and outgoing particles in the FRITIOF events, we follow here the
approach of \cite{quench} generalized to the processes with many-meson
final states. Only the events with two colliding nonstrange baryons
will be modified. The in-medium modifications of the meson-baryon
and meson-meson collisions are neglected, since as we expect, they are
small with respect to the baryon-baryon case (see below).

Let us consider the process
\begin{equation}
    B_1 B_2 \to B_3 B_4 M_5 M_6 ... M_N~,                 \label{proc}
\end{equation}
where $B_1$,$B_2$ and $B_3$,$B_4$ are the incoming and the outgoing baryons,
respectively; $M_5$,$M_6$,...,$M_N$ are the produced mesons.
The in-medium differential cross section of this process is given
by the following expression
\begin{equation}
    d\sigma^{med} = (2\pi)^4
                    {(2m_1^*)(2m_2^*)(2m_3^*)(2m_4^*) \over 4 I^*} 
                    \overline{|T|^2}
d\Phi_{N-2}(p_1^*+p_2^*;p_3^*,p_4^*,k_5^*,k_6^*,...,k_N^*)~,  \label{dsig}
\end{equation}
where $\overline{|T|^2}$ is the matrix element squared in the normalization
of Ref. \cite{BD} averaged over spins of initial particles and summed
over spins of final particles,
\begin{equation}
\begin{split}
 &d\Phi_{N-2}(p_1^*+p_2^*;p_3^*,p_4^*,k_5^*,k_6^*,...,k_N^*) 
 = \delta^{(4)}(p_1^*+p_2^*-p_3^*-p_4^*-k_5^*-k_6^*-...-k_N^*) \\
 &\times {d^3p_3^* \over (2\pi)^3 2(p_3^*)^0}
          {d^3p_4^* \over (2\pi)^3 2(p_4^*)^0}
          {d^3k_5^* \over (2\pi)^3 2(k_5^*)^0} \cdots 
          {d^3k_N^* \over (2\pi)^3 2(k_N^*)^0}                \label{dphi}
\end{split}
\end{equation}
is the $(N-2)$-body phase space \cite{PDG02} with
$(p_i^*)^0 = (({\bf p}_i^*)^2+(m_i^*)^2)^{1/2}~,~~i=1,2,3,4$ and
$(k_i^*)^0 = (({\bf k}_i^*)^2+(m_i^*)^2)^{1/2}~,~~i=5,6,...,N$ being
the zeroth components of the kinetic four-momenta, and the $m^\ast_i$ being the effective masses of the particles involved. 
In Eq. (\ref{dsig}) the flux factor is
\begin{equation}
    I^* = q(\sqrt{s^*},m_1^*,m_2^*)\sqrt{s^*}~,                 \label{Ist}
\end{equation}
where $s^* \equiv (p_1^*+p_2^*)^2$ and
\begin{equation}
    q(\sqrt{s^*},m_1^*,m_2^*)
= [(s^*+(m_1^*)^2-(m_2^*)^2)^2/(4s^*)-(m_1^*)^2]^{1/2}          \label{q}
\end{equation}
is the c.m. momentum of incoming baryons. 

The matrix element $\overline{|T|^2}$ entering into Eq. (\ref{dsig})
can be extracted from the vacuum cross section by dividing out the
vacuum  phase space and multiplying by the vacuum flux factor.
Thus, our final result for the in-medium total cross section of
the process (\ref{proc}) is
\begin{equation}
    \sigma^{med}(\sqrt{s^*}) = F \sigma^{vac}(\sqrt{s})~.       \label{sig}
\end{equation}
The modification factor $F$ is
\begin{equation}
    F \equiv {m_1^* m_2^* m_3^* m_4^* \over m_1 m_2 m_3 m_4}
             {I \over I^*}
{\Phi_{N-2}(\sqrt{s^*};m_3^*,m_4^*,...,m_N^*)  \over
 \Phi_{N-2}(\sqrt{s};m_3,m_4,...,m_N)}~,                         \label{fact}
\end{equation}
where $I=q(\sqrt{s},m_1,m_2)\sqrt{s}$. In Eq. (\ref{sig}) $\sqrt{s}$
is the c.m. energy of the colliding baryons in vacuum, which is directly 
provided by the BUU calculations in the cascade modus performed in
the present work. The in-medium c.m. energy $\sqrt{s^*}$ is then
determined from the condition that the energy excess above threshold
is the same as in vacuum, i.e.
\begin{equation}
    \sqrt{s^*} = \sqrt{s} - m_3 - m_4 - \cdots - m_N +
                 m_3^* + m_4^* + \cdots + m_N^*~.           \label{srtsst}
\end{equation}
Since the modification factor $F$ is proportional to the product of the ratios
of the Dirac mass to the bare mass for incoming and outgoing fermions, we
expect that the meson-baryon and meson-meson channels will be modified
relatively weaker.

In Eq.(\ref{dsig}) we replaced the canonical four-momenta by the kinetic ones in the $\delta$-function entering the phase-space volume element (\ref{dphi}). This is possible only if the vector fields cancel each other, which is valid in the case of the Walecka model (c.f. ref.\cite{Fang94}), but would be violated in a more sophisticated relativistic mean field model with momentum-dependent scalar and vector fields \cite{weberk}. Taking into account the momentum dependence, in particular, for the vector field, which drops with momentum, is important for the description of the baryon flow in heavy ion collisions above 1 A GeV \cite{sahup}. However, in the present exploratory work we will neglect the momentum dependence of the $\sigma$ and $\omega$ fields, which would strongly complicate the calculation of the phase space volume.

We evaluate the in-medium masses using a nonlinear version NL2 \cite{Lee86} of the relativistic mean field model and assuming
that the nucleons and all nonstrange baryonic resonances
are coupled to the scalar mean field $\sigma$ and to the vector mean field
$\omega$ by the same universal coupling constants $g_\sigma$ and
$g_\omega$ \cite{Wehr89}. This gives the Dirac effective masses  
\begin{equation}
    m_B^* = m_B + g_\sigma \sigma                         \label{mstar}
\end{equation}
and the kinetic four-momenta
\begin{equation}
    p_B^* = p_B - g_\omega \omega                         \label{pstar}
\end{equation}
of the nonstrange baryons. The hyperon coupling constants are
(c.f. Ref. \cite{Fang94}): 
\begin{equation}
    g_\sigma^Y = {2 \over 3} g_\sigma~,~~~
    g_\omega^Y = {2 \over 3} g_\omega,~                   \label{gY}
\end{equation}
where $Y = \Lambda$ or $\Sigma$. The baryon single-particle energy is
\begin{equation}
    \varepsilon({\bf p}_B) = g_\omega \omega^0 + 
    \sqrt{({\bf p}_B^*)^2 + (m_B^*)^2}~.                  \label{eB}
\end{equation}
For the mesons $\pi$, $\rho$ and $\omega$ we neglect any in-medium
modifications, while for the $K$ and $\bar{K}$ single-particle energies
we use the model of Ref. \cite{Zheng02} with parameters of
Ref. \cite{BR96}:
\begin{equation}
    \omega({\bf k},\rho) = \sqrt{ ({\bf k}^*)^2 + (m_K^*)^2 }
                           \pm V^0~,                       \label{ompm}
\end{equation}
where the upper (lower) sign corresponds to the $K$ ($\bar{K}$) case,
\begin{equation}
    {\bf k}^* = {\bf k} \mp {\bf V}                        \label{kkmom}
\end{equation}
is the kaon kinetic momentum,
\begin{equation}
    m_K^* = \sqrt{m_K^2 -
            {\Sigma_{KN} \over f_\pi^2} \rho_s + V^2}      \label{mkst}
\end{equation}
is the kaon effective (Dirac) mass, and
\begin{equation}
    V^\mu = {3 \over 8 (f_\pi^*)^2} j^\mu                  \label{Vmu}
\end{equation}
is the kaon vector field. $\rho_s$ and $j^\mu$ are the scalar density
and the baryonic four-current, respectively. The parameters which appear 
in Eqs. (\ref{mkst}),(\ref{Vmu}) are $\Sigma_{KN} = 450$ MeV,
$f_\pi = 93$ MeV and $(f_\pi^*)^2 = 0.6 (f_\pi)^2$ \cite{BR96}.
Within these parameters the following relation is expected to hold \cite{BR96}
\begin{equation}
    V^\mu \simeq {1 \over 3} g_\omega \omega^\mu~.         \label{rel}
\end{equation}
Taking into account relations (\ref{gY}) and (\ref{rel}) one can see
that the vector field is, indeed, completely excluded
from the energy-momentum conservation conditions for the strange particle
production processes like $B_1 B_2 \to B_3 Y_4 K$ or
$B_1 B_2 \to B_3 B_4 K \bar{K}$, which gives a possibility to simplify
the in-medium calculations by just replacing the bare masses of particles
by the Dirac masses and canonical four-momenta by the kinetic four-momenta.

\begin{figure}
\begin{center}
\epsfig{file=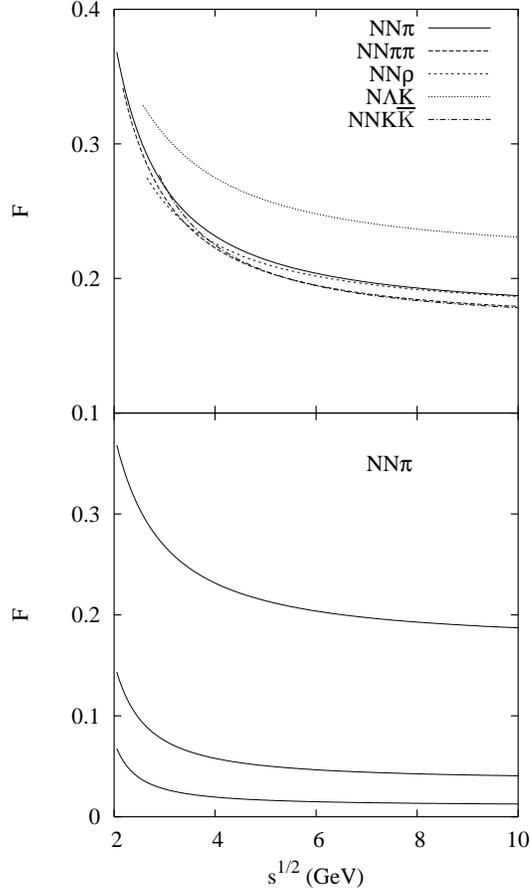,height=12cm}
\caption{In-medium modification factor $F$ of the
$NN$ cross section vs c.m. energy. Upper panel shows $F$ at $\rho=\rho_0$
for various outgoing channels: $NN\pi$ -- solid line, $NN\pi\pi$ --
long-dashed line, $NN\rho$ -- short-dashed line, $N\Lambda K$ --
dotted line and $NNK\bar{K}$ -- dash-dotted line. 
On the lower panel the factor $F$ is presented for the $NN\pi$ channel
at $\rho = \rho_0$, $2\rho_0$ and $3\rho_0$ in the order from the upper 
to the lower line.}
\label{ratio}
\end{center}
\end{figure}

We have calculated the modification factor $F(\sqrt{s},\rho)$ as
a function of the c.m. energy $\sqrt{s}$ and the baryon density $\rho$
for various outgoing channels with no more than four mesons in the final 
state. We assume that the incoming baryons are nucleons, but
an outgoing baryon can be either nucleon or $\Lambda$-hyperon. For
outgoing mesons we have considered all possible combinations of
pions, $\rho$-mesons, kaons and antikaons with no more than one kaon
and one antikaon in the final state. The upper panel of Fig. \ref{ratio}
shows the medium modification factor $F(\sqrt{s},\rho_0)$ for some selected 
processes: $NN \to NN\pi$, $NN\pi\pi$, $NN\rho$, $N\Lambda K$, $NNK\bar{K}$.
We see that the modification factor depends on the outgoing channel 
rather weakly. In particular, the addition of a pion does not change the 
factor. On the lower panel of Fig. \ref{ratio} we 
demonstrate the density dependence of the modification factor for the 
one-pion production channel. One can observe a strong decrease of the 
factor with the baryon density.

For the application to the FRITIOF model built in the BUU code,
we have stored the factors $F$ on a two-dimensional grid $(\sqrt{s},\rho)$.
Once some final state is generated by FRITIOF, it is accepted with
the probability $F$. In the case where at least one of incoming baryons is
a resonance, we use the modification factor for incoming nucleons at
the same $\sqrt{s}$. If an outgoing baryon is the $\Delta$-resonance,
the factor for an ougoing nucleon is applied, shifted by the production
threshold, i.e. $F(\sqrt{s}-m_\Delta+m_N)$. An analogous threshold
correction is performed if $\Sigma$-hyperon and/or $K^*$ are produced.

\begin{figure}
\begin{center}
\begin{tabular}{ccc}
\includegraphics[height=4cm]{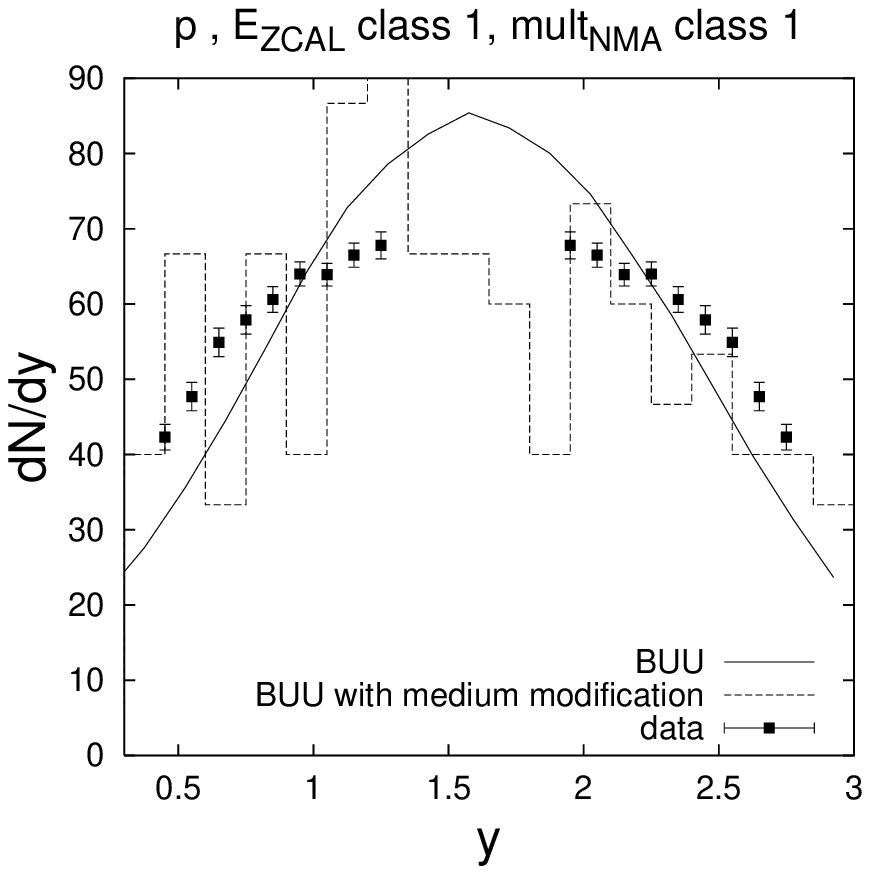} & 
\includegraphics[height=4cm]{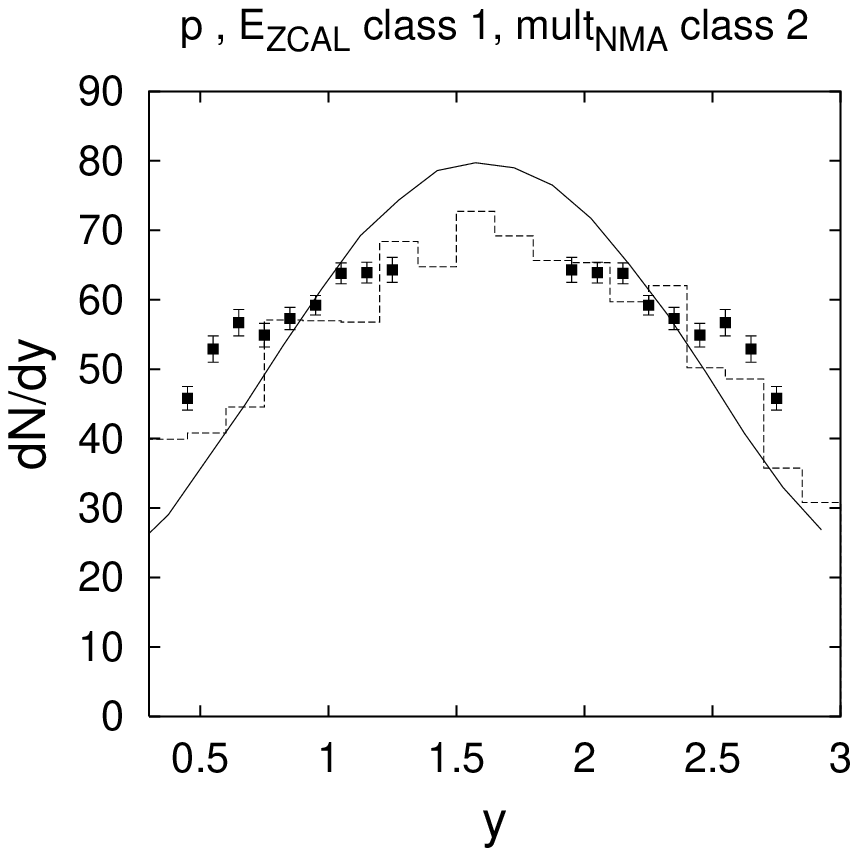} &
\includegraphics[height=4cm]{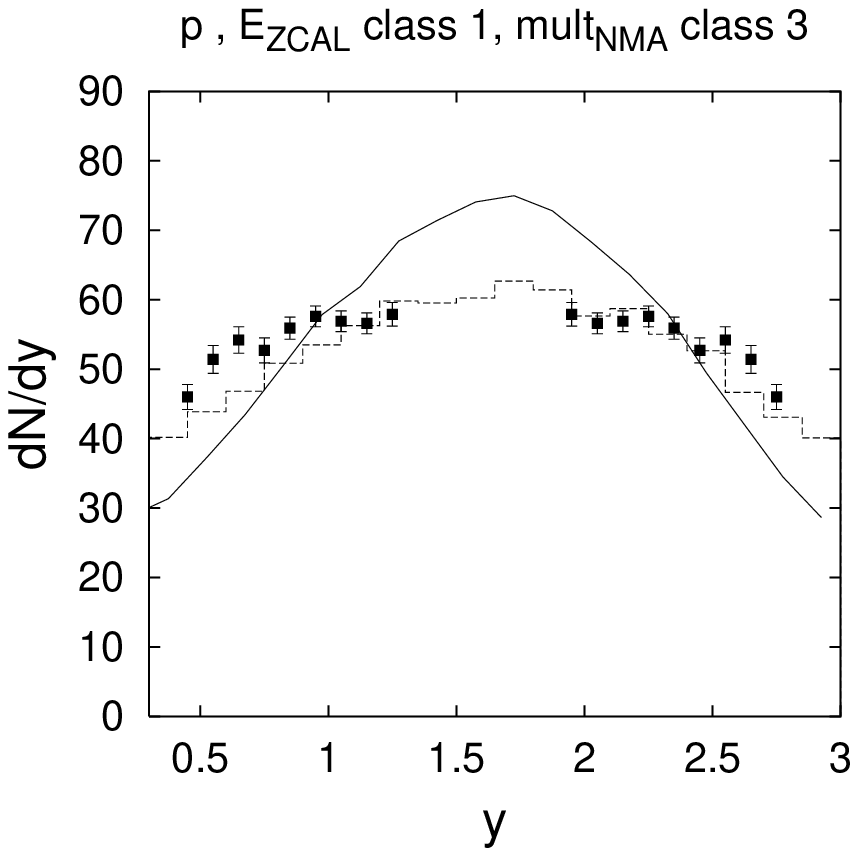} \\ 
\includegraphics[height=4cm]{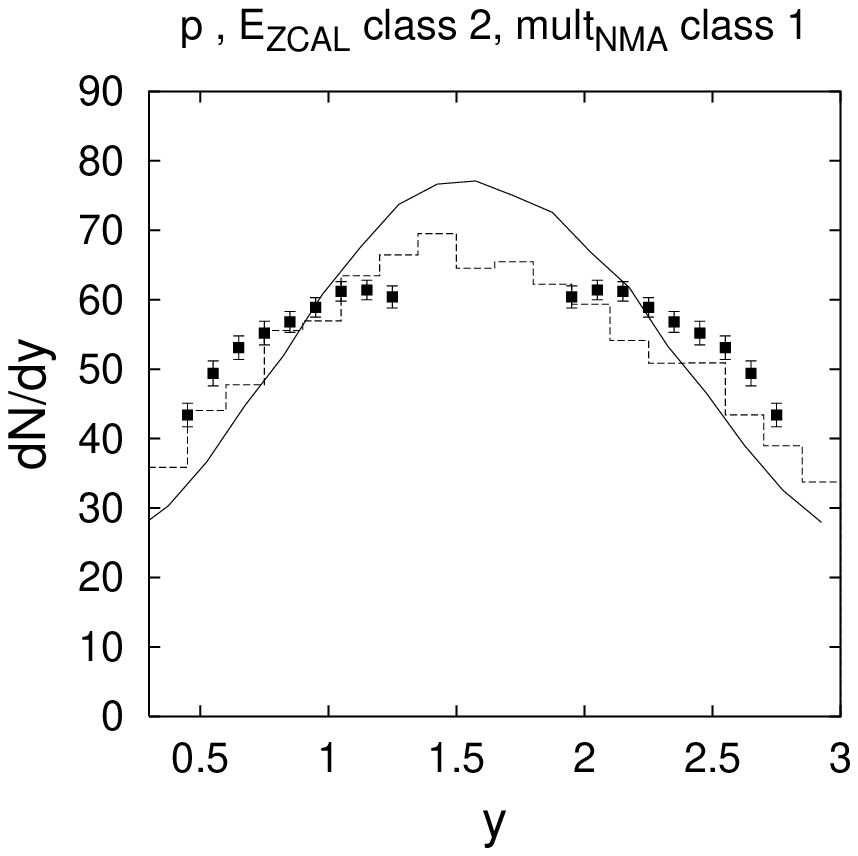} & 
\includegraphics[height=4cm]{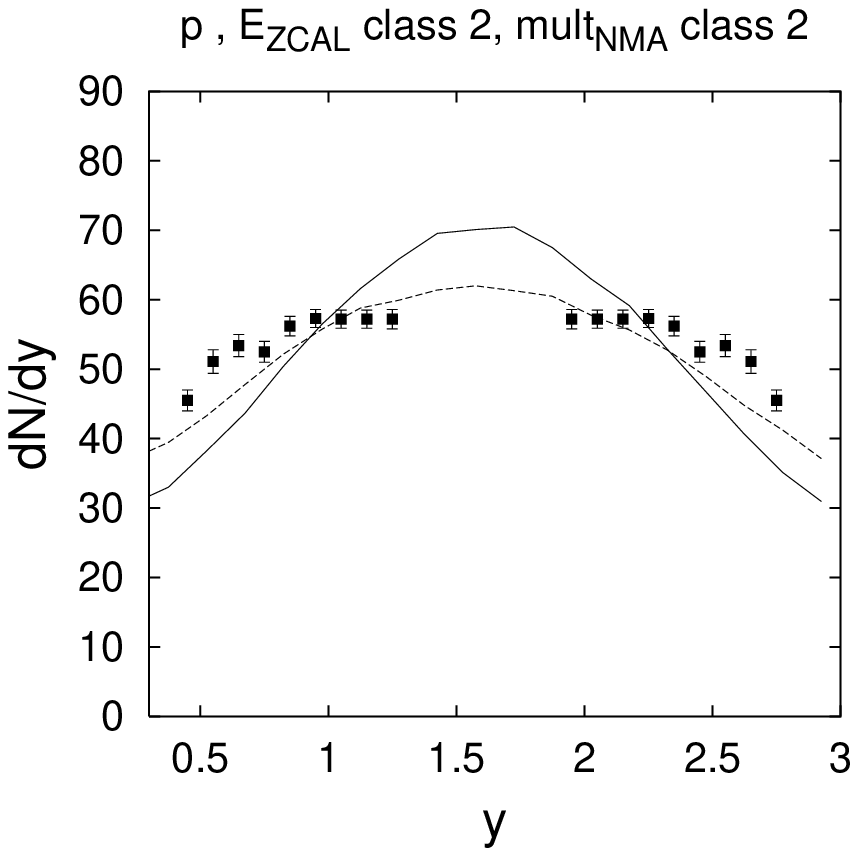} & 
\includegraphics[height=4cm]{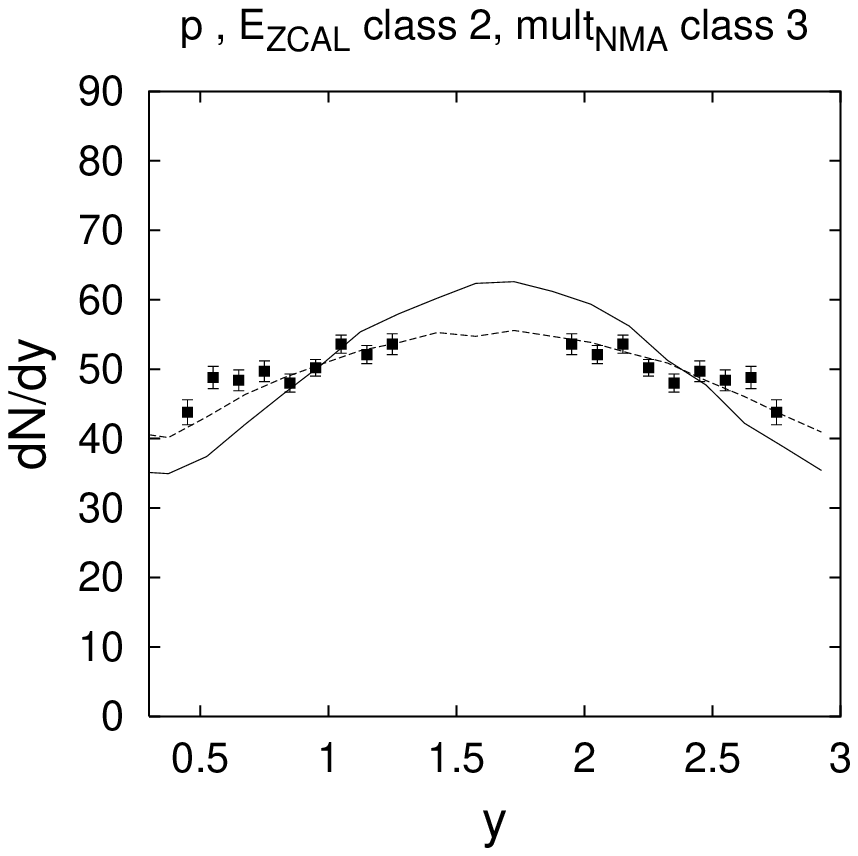} \\
\includegraphics[height=4cm]{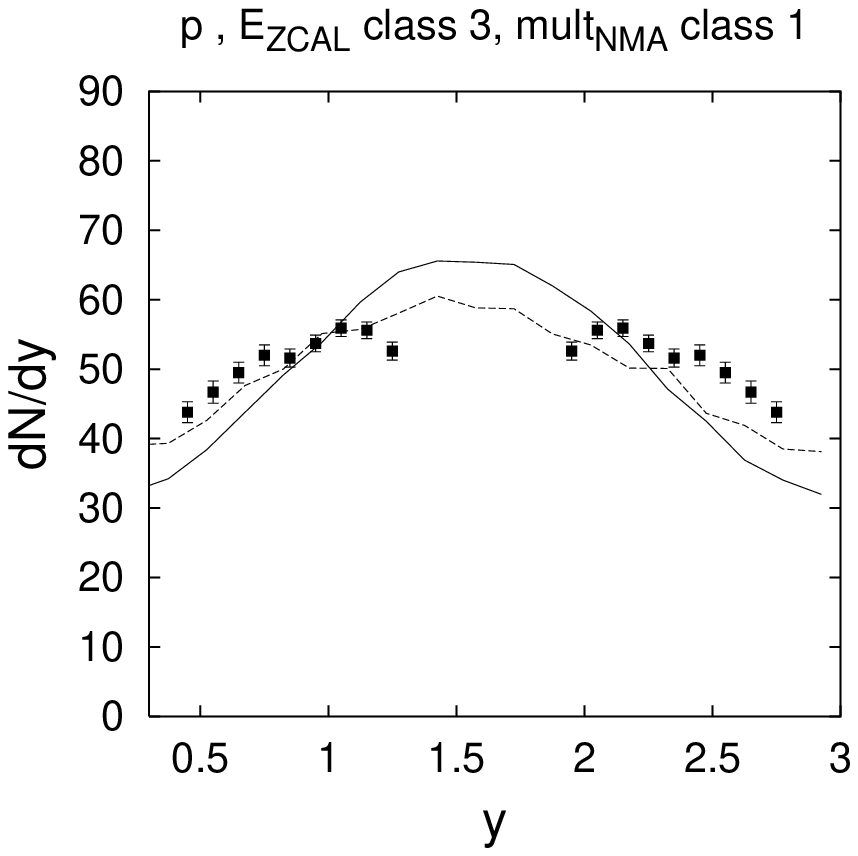} &
\includegraphics[height=4cm]{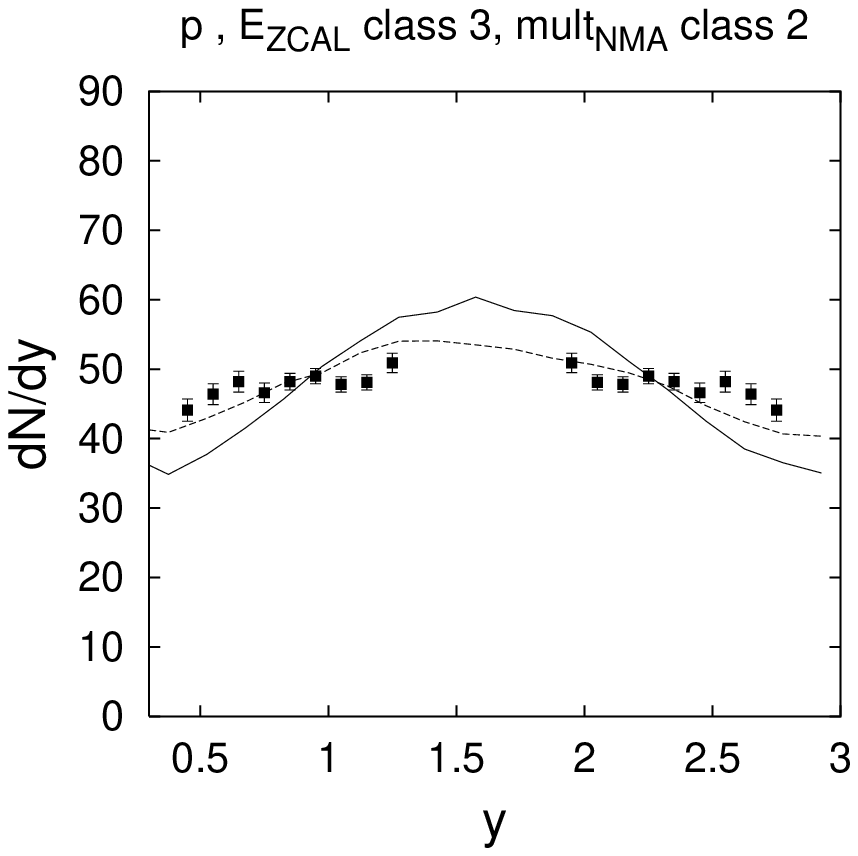} &
\includegraphics[height=4cm]{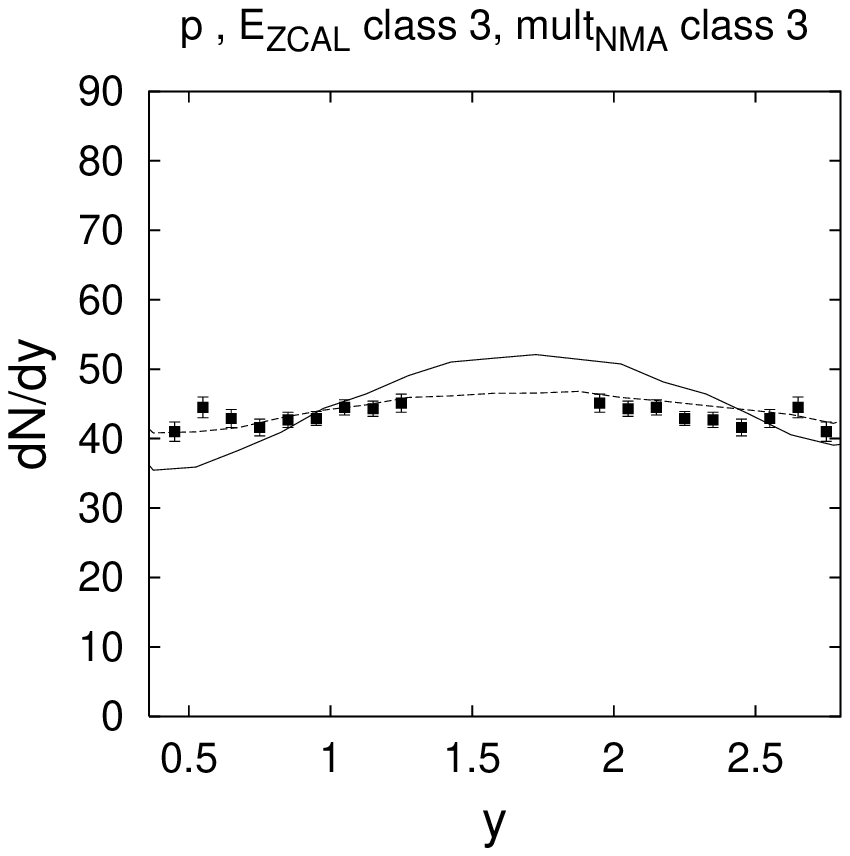} \\
\end{tabular} 
\end{center}
\caption{Proton rapidity spectra from BUU events Au+Au at 10.7 A GeV double selected by the total multiplicity and the zero degree energy in comparison to data from \cite{hic1}. The centrality decreases from the upper left corner to the lower right. The solid line shows the standard BUU calculation, whereas the dashed line is calculated with the medium modification described in section \ref{medi}.}\label{auauprot}
\end{figure}

\begin{figure}
\begin{center}
\begin{tabular}{cc}
\includegraphics[height=6cm]{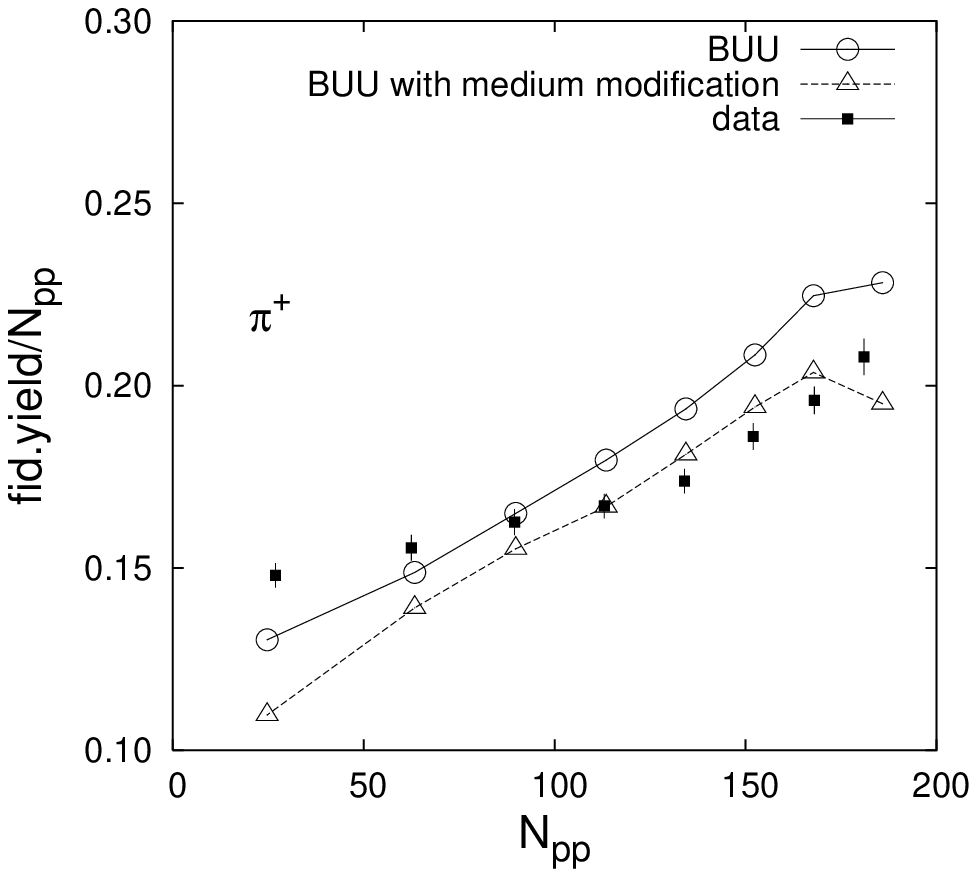} & 
\includegraphics[height=6cm]{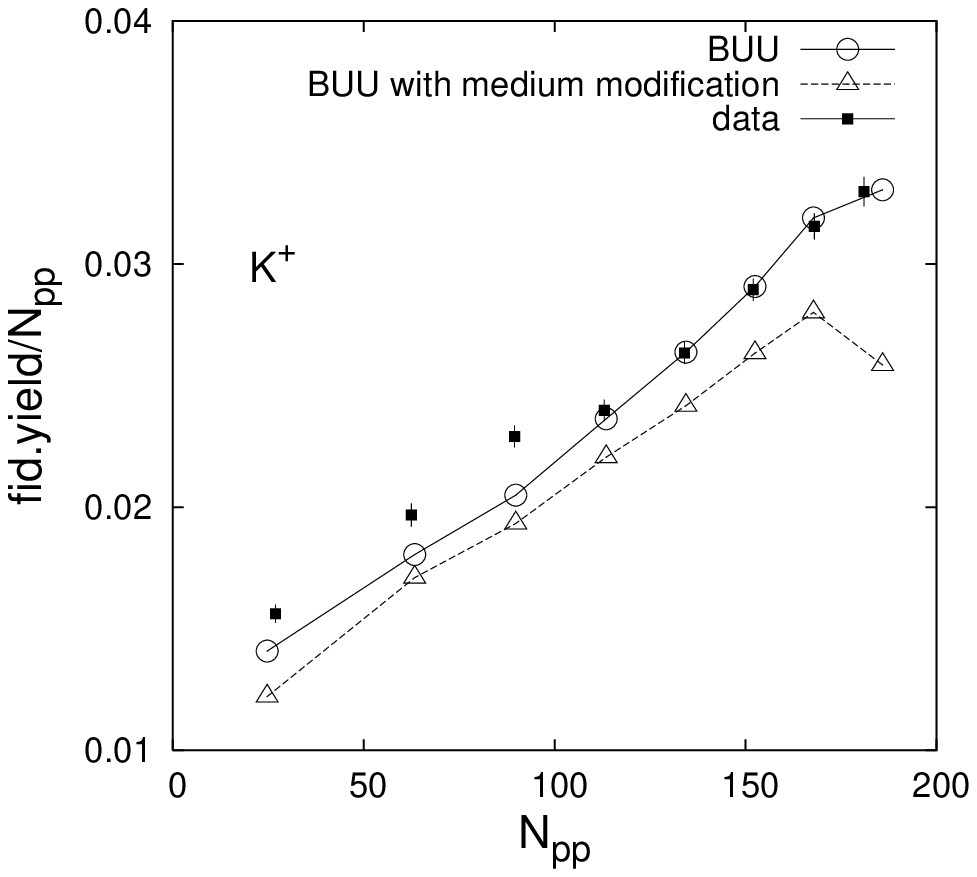} \\ 
\end{tabular} 
\caption{Fiducial yields of $\pi^+$ and $K^+$ divided by the number of projectile participants $N_{pp}$ as functions of $N_{pp}$ from Au+Au collisions at 10.7 A GeV in comparison to data from \cite{hic1}.}\label{hicfid}
\end{center}
\end{figure}

In order to see an effect of the in-medium string model modifications
in heavy-ion collisions, we have performed the calculation for the
Au+Au system at 10.7 A GeV. The results of this calculation are shown
by the dashed lines in Figs. \ref{auauprot}-\ref{hicfid}. 

The proton rapidity spectra (Fig. \ref{auauprot}) get now wider, in agreement
with the experiment. This is expected since the inelastic NN cross section is 
reduced by the in-medium effects. The results with in-medium modifications for the most central events (upper left panel in Fig. \ref{auauprot}) have a big statistical error due to a small number of events in this centrality class.

The $\pi^+$ rapidity spectra 
(Fig. \ref{auaupi}) are reduced in a closer agreement with the data,
except for the very peripheral collisions. The $K^+$ rapidity spectra
(Fig. \ref{auauka}) are also reduced. This, however, makes the agreement
with the $K^+$ data worse.

The transverse mass spectra of $\pi^+$'s (Fig. \ref{auaupitr}) and
$K^+$'s (Fig. \ref{auauktr}) are reduced more at high transverse masses,
since the high-$m_\perp$ particles are emitted from hard collisions
which happen at the high baryon density where the in-medium modifications
are stronger. Thus, the $m_\perp$-spectra get steeper, which, again,
leads to a worse description of the $K^+$ data. The  $\pi^+$' transverse
mass spectra at small $m_\perp$ are now better described, whereas at high
$m_\perp$ the in-medium modifications result in a slight underestimation
of the experiment.

Fig. \ref{hicfid} summarizes our findings on the in-medium modifications.
The $\pi^+$ fiducial yield is described better with the in-medium 
modifications, except for the very peripheral collisions. The $K^+$
fiducial yield is underestimated at all collision centralities 
with the in-medium modifications.
 
\section{\label{sum} Summary and outlook}

In the present work, the BUU model developed earlier in Refs.
\cite{BUU,MEPhD} is further improved by including the heavy
($m > 2$ GeV) resonance or $q \bar q$-annihilation channel in the
meson-baryon collisions and by the new meson-meson channels
($\pi \rho \to K \bar K,~\rho \rho \to  K \bar K$) for the strangeness
production. Moreover, an in-medium modification of the FRITIOF
model by taking into account the effective (Dirac) masses
of the incoming and outgoing particles is formulated and implemented 
in BUU. The BUU model is applied to the nucleus-nucleus collisions 
at 2-40 A GeV.

\begin{figure}
\begin{center}
\begin{tabular}{cc}
\includegraphics[height=6.5cm]{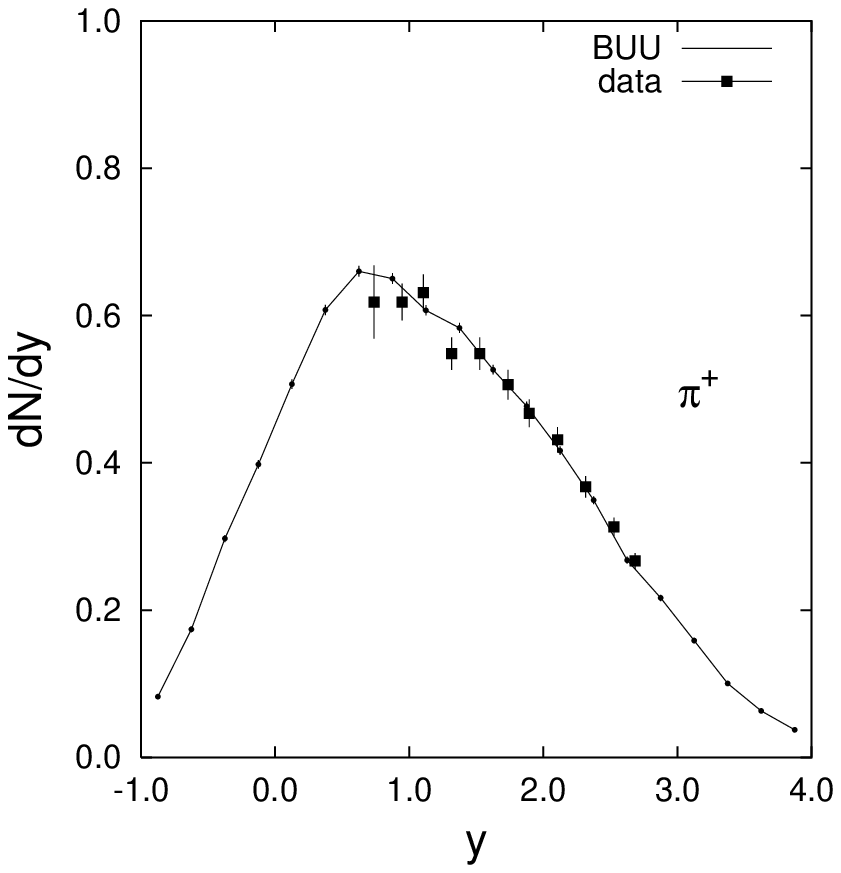} & 
\includegraphics[height=6.5cm]{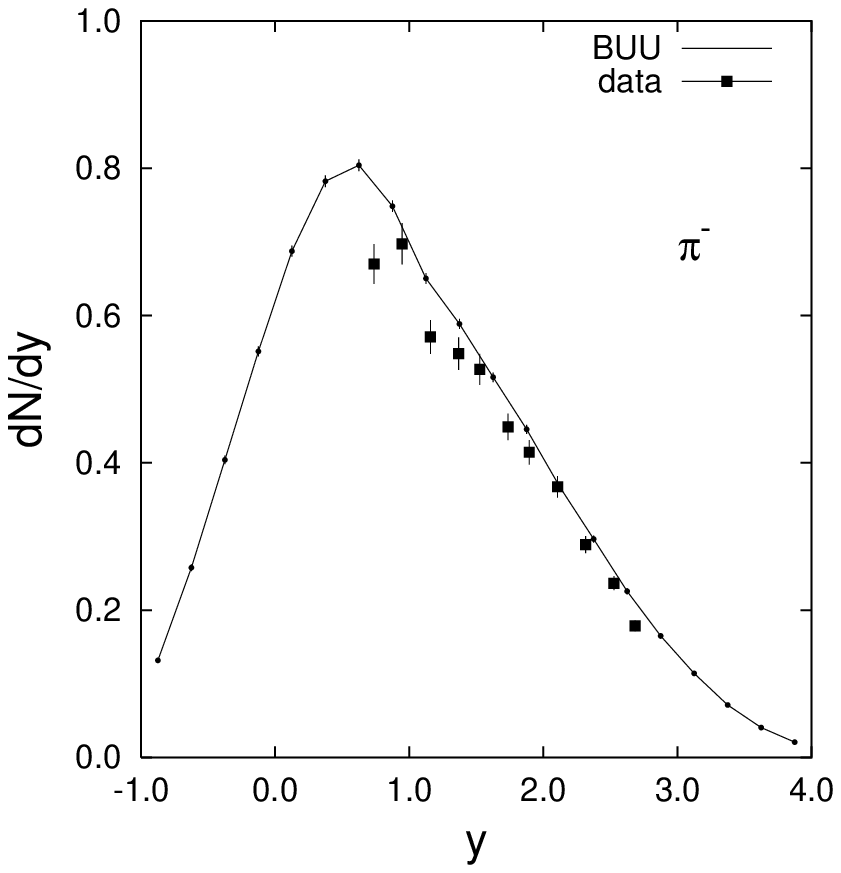} \\ 
\includegraphics[height=6.5cm]{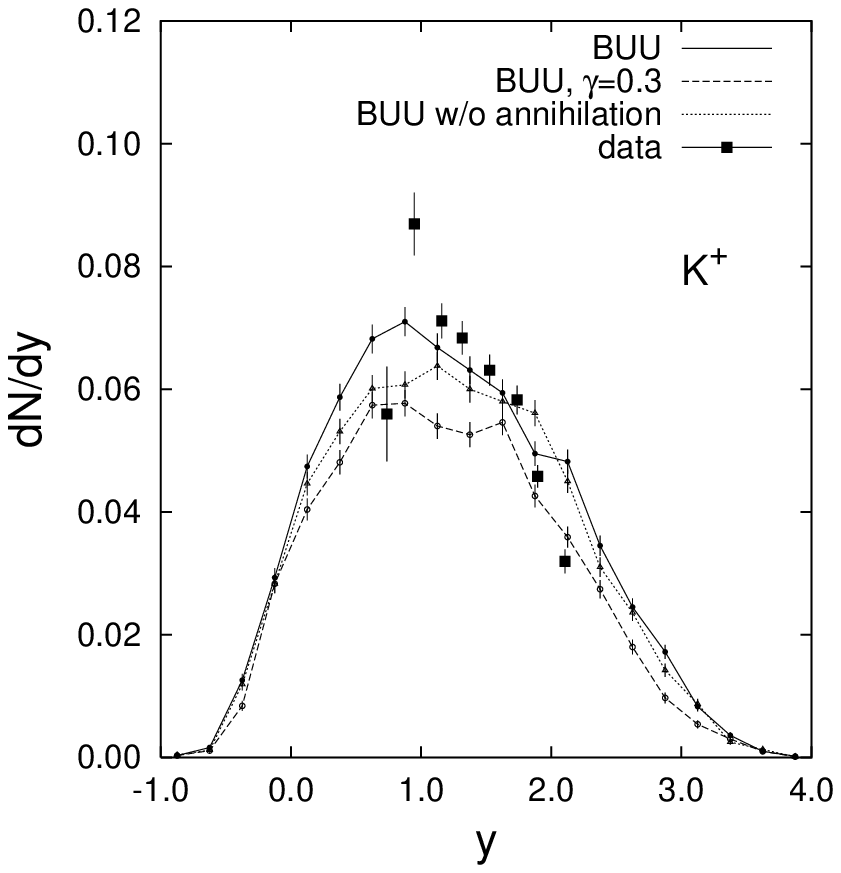} & 
\includegraphics[height=6.5cm]{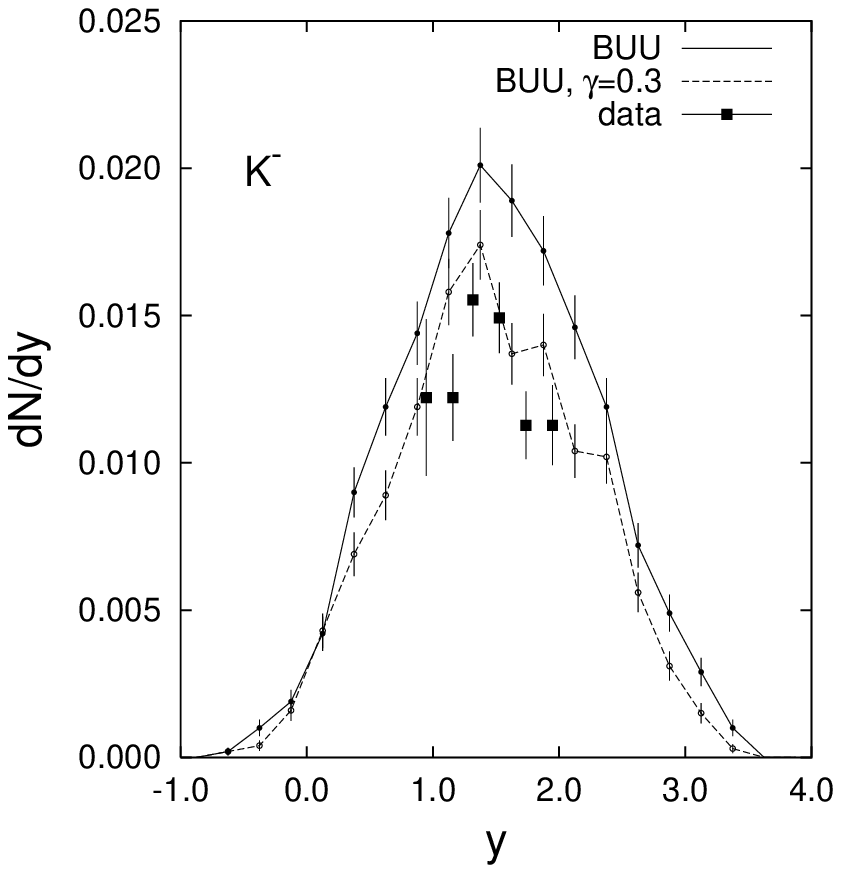} \\ 
\end{tabular} 
\caption{Rapidity distributions of $\pi^\pm$ and $K^\pm$ for $p$+Au at 14.6 GeV/c in comparison to data from \cite{pa1}. Solid and dashed lines show results with energy  dependent (see Eq.(\ref{eq1})) and constant ($\gamma=0.3$) strangeness suppression factor, respectively. The dotted line in the lower left panel shows a calculation without the $q\overline q$ annihilation channel in meson-baryon collisions. Errorbars on calculations are statistical.}\label{paurap}
\end{center}
\end{figure}

\begin{figure}
\begin{center}
\epsfig{file=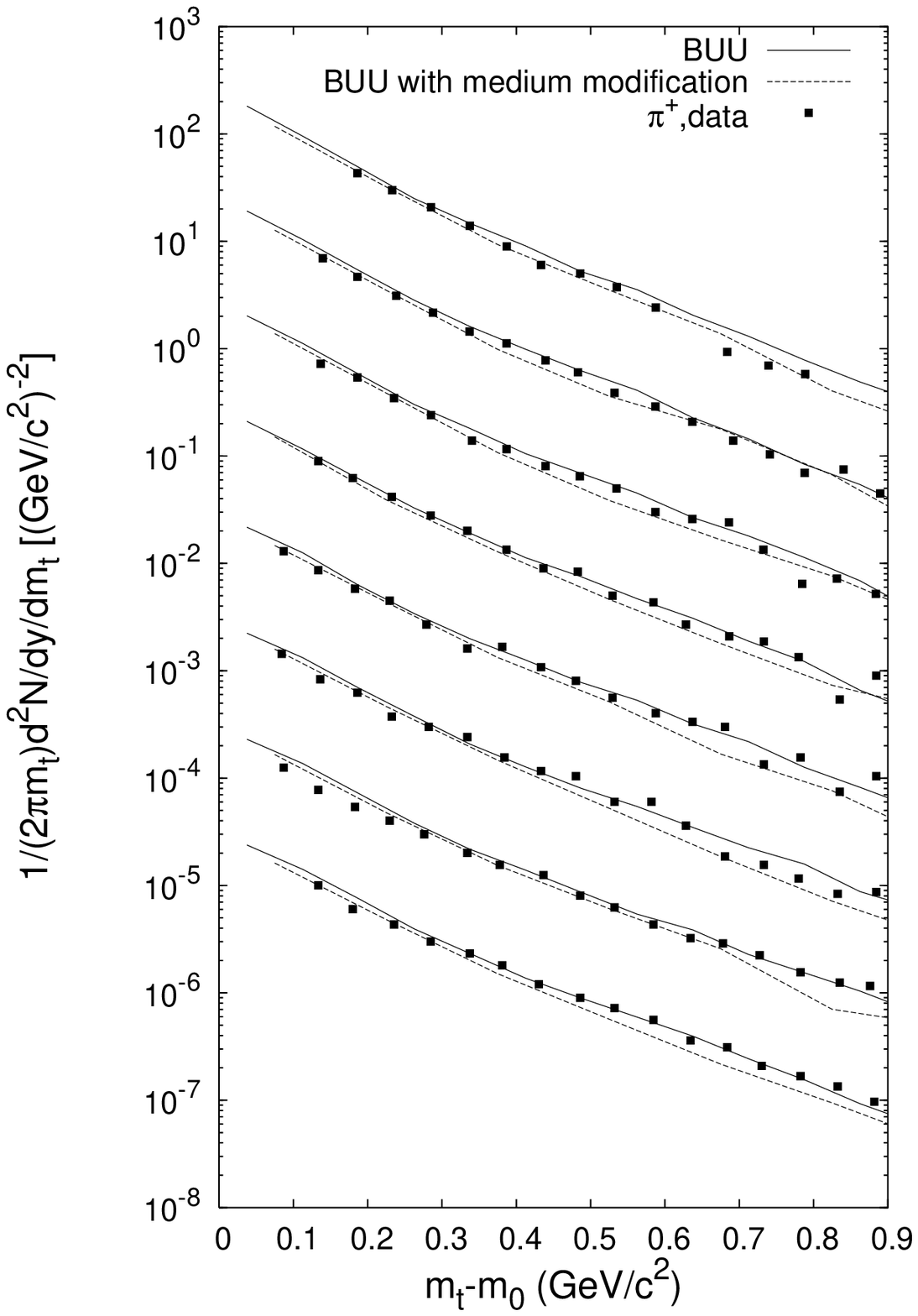,width=10cm}\caption{Transverse mass spectra for $\pi^+$ from Au+Au collisions at 10.7 A GeV for different slices of rapidity from the most central events selected by E$_{ZCAL}$. The rapidity slices range from 0.6-0.7 for the uppermost line to 1.3-1.4 for the lowermost line with a step 0.1. The spectra are multiplied by powers of 10: $10^0,10^{-1},...,10^{-7}$ from the uppermost to the lowermost line. Data are from \cite{hic1}.}
\label{auaupitr}
\end{center}
\end{figure}

\begin{figure}
\begin{center}
\epsfig{file=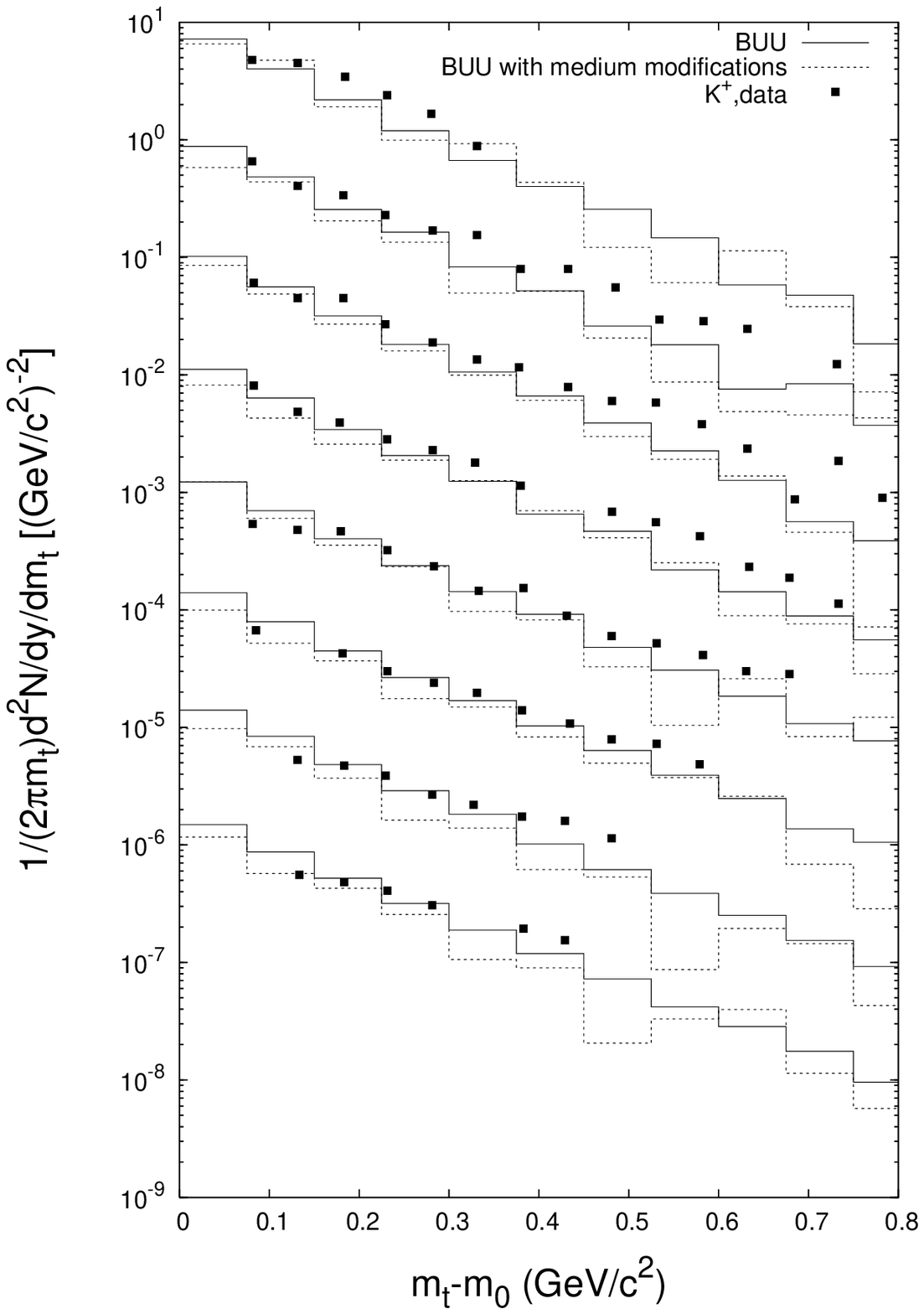,width=10cm}\caption{Transverse mass spectra for $K^+$ from Au+Au collisions at 10.7 A GeV for different slices of rapidity from the most central events selected by E$_{ZCAL}$. The rapidity slices range from 0.5-0.6 for the uppermost line to 1.2-1.3 for the lowermost line. The spectra are multiplied by powers of 10: $10^0,10^{-1},...,10^{-7}$ from the uppermost to the lowermost line. Data are from \cite{hic1}.}
\label{auauktr}
\end{center}
\end{figure}

\begin{figure}
\begin{center}
\begin{tabular}{cc}
\epsfig{file=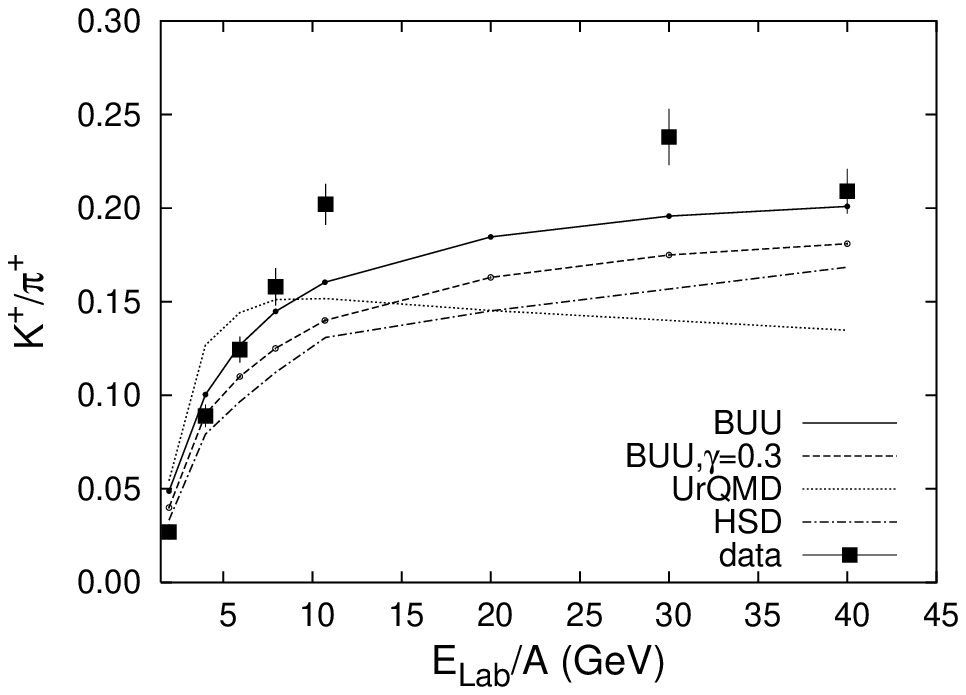,height=5cm} &
\epsfig{file=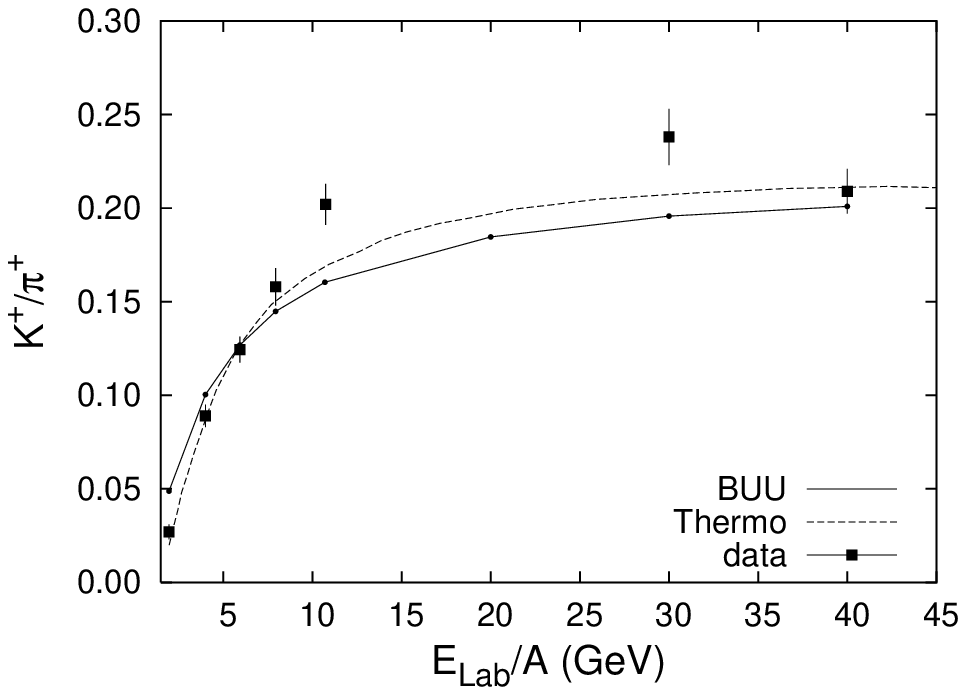,height=5cm}\\
\epsfig{file=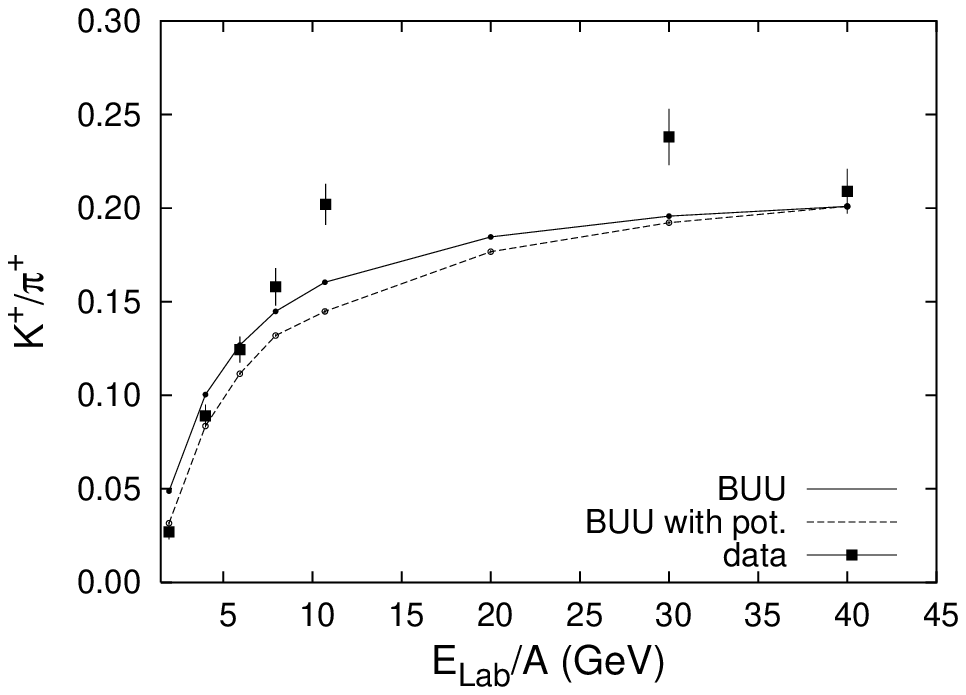,height=5cm} &
\epsfig{file=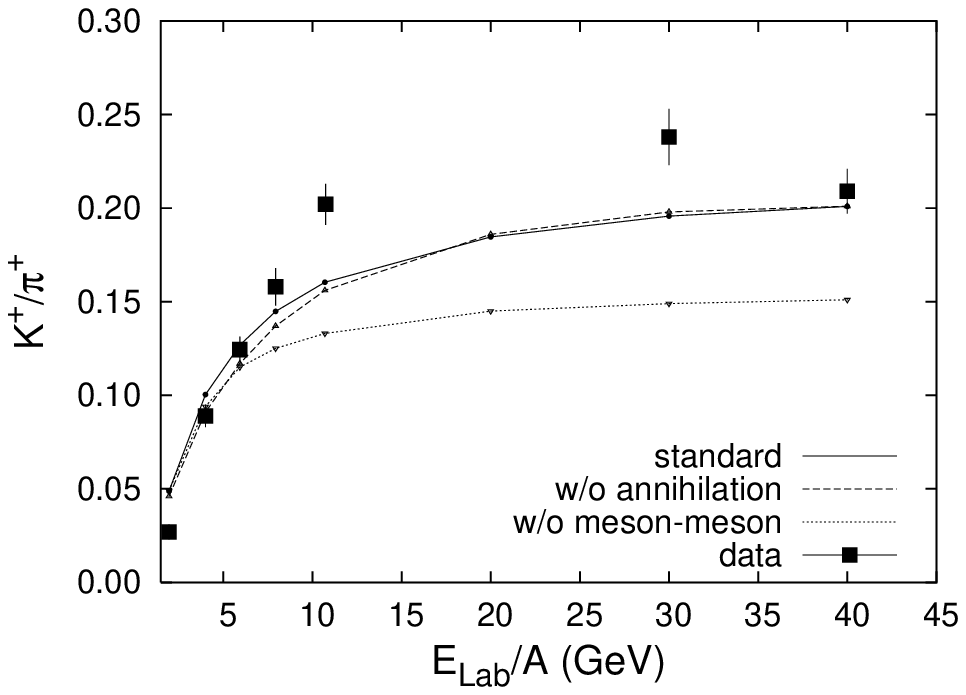,height=5cm}\\
\end{tabular}
\caption{$K^+/\pi^+$ ratio at midrapidity as a function of the beam energy. Upper left panel - comparison of the BUU results with the results of the UrQMD and HSD model \cite{strange}. Upper right panel - comparison between BUU and statistical model \cite{thermoref1}. Lower left panel - comparison of standard BUU (cascade mode) and BUU with nuclear mean field potential. Lower right panel - comparison of the standard BUU calculation with the calculations without $q\overline q$ annihilation and without meson-meson collisions. The data are from \cite{na491,ags1,na492}.}
\label{fexcit5}
\end{center}
\end{figure}

\begin{figure}
\begin{center}
\epsfig{file=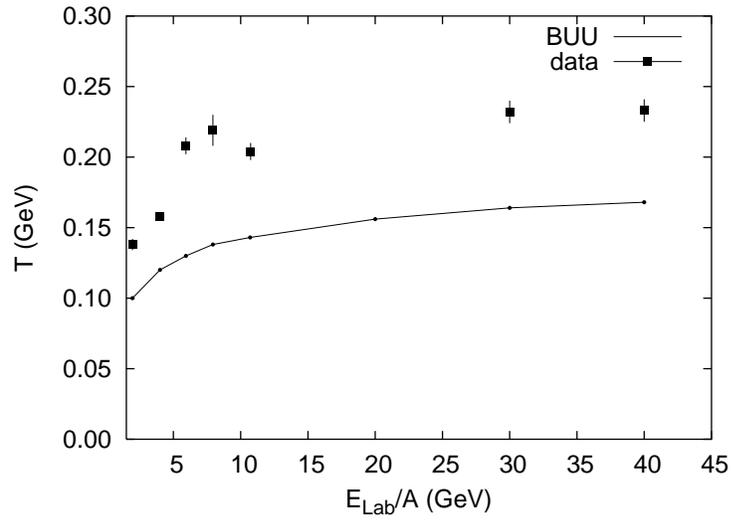,height=7cm}\caption{Inverse slope parameter T for 
$K^+$ as a function of energy in comparison to data from 
\cite{ags2,na491,na492}.}
\label{fexcit8}
\end{center}
\end{figure}

By performing the systematic study of the pion and kaon production for
various systems and collision energies we came to the following
conclusions:
\begin{enumerate}

\item The $\pi^\pm$ and $K^\pm$ rapidity spectra and the inverse slope
parameters of the transverse mass spectra from the proton-nucleus 
reactions are well described by BUU. The $q \bar q$-annihilation channel
improves the agreement with the data on the $K^+$ production in proton-nucleus 
collisions (Fig. \ref{paurap}).

\item In the peripheral Au+Au collisions at 10.7 A GeV the $\pi^+$ rapidity 
spectra are slightly underestimated, whereas the $K^+$ rapidity 
spectra are well described. This result is consistent
with a good agreement between BUU and the data on the proton-nucleus
reactions. The proton rapidity spectra are too narrow, i.e. the
stopping power is overestimated by BUU, for all collision centralities.

\item In the central nucleus-nucleus collisions (Si+Au at 14.6 A GeV/c,
Au+Au at 2-10.7 A GeV, Pb+Pb at 30 and 40 A GeV) the $K^+$ yields are,
overall, well described, the $K^-$ and hyperon yields are somewhat 
overestimated, the $\pi^+$ yields are overestimated in all central heavy-ion
collisions under study. The inverse slope parameters of the $K^+$ 
transverse mass spectra are strongly underestimated (Figs. \ref{auauktr},
\ref{fexcit8}). The pion slopes are well described (Fig. \ref{auaupitr}).
Our BUU results on $\pi^+$ and $K^+$ agree, generally, with results of the 
HSD and UrQMD calculations from Refs. \cite{strange,slopeel}.
The excitation function of the $K^+/\pi^+$ ratio is described, however,
better and closer to the thermal model results due to the introduction
of the new meson-meson channels of the strangeness production which 
increase the $K^+$ yield at the beam energy above 6 A GeV
(Fig. \ref{fexcit5}).

\item The in-medium modification of the FRITIOF model, which
reduces the particle production cross sections, was tested for the
system Au+Au at 10.7 A GeV. This leads to the better description of the 
pion production, whereas the $K^+$ production is overdamped by the in-medium
effects. The stopping power of the nuclear matter is reduced, which
results in the better description of the proton rapidity spectra.

\end{enumerate}

The $K^+/\pi^+$ ratio is quite sensitive to the meson-meson cross sections
and, within the reasonable choice of these cross sections, we have decreased the discrepancy between BUU and the data by about a factor of 2. 
The resulting disagreement $\sim 20$ \% can, thus, hardly to be considered
as a signal of the ``new physics''. The inverse slope parameter of
the $K^+$ transverse mass spectra is a more serious problem for BUU. 
It would be worthwhile to study this topic in more detail by introducing, e.g. string-string and many-body collisions, which could both enhance the hard part of the 
kaon spectra.

\vspace{1cm}

\noindent\begin{large}\textbf{Acknowledgements}\end{large} 

\vspace{0.5cm}

We thank W. Cassing for a careful reading of the manuscript and many helpful comments. We are also indebted to C. Greiner E. Bratkovskaya for useful discussions and thank the latter for making the results of HSD and UrQMD calculations available to us.

\bibliography{refs_new}

\newpage

\clearpage

\clearpage

\clearpage

\clearpage

\clearpage

\clearpage

\clearpage

\clearpage

\clearpage

\end{document}